\documentclass[%
reprint,
superscriptaddress,
nofootinbib,
 amsmath,amssymb,
 aps,prc
]{revtex4-2}

\usepackage{graphicx}
\usepackage{dcolumn}
\usepackage{bm}
\usepackage{subfig}
\usepackage{float}
\usepackage{xcolor}
\usepackage{ulem}

\usepackage{hyperref}

\begin{document}


\title{Odd-odd neutron-rich rhodium isotopes studied with\\ the double Penning trap JYFLTRAP}

\author{M.~Hukkanen}
\affiliation{University of Jyvaskyla, Department of Physics, Accelerator laboratory, P.O. Box 35(YFL) FI-40014 University of Jyvaskyla, Finland}
\affiliation{Universit\'e de Bordeaux, CNRS/IN2P3, LP2I Bordeaux, UMR 5797, F-33170 Gradignan, France}
\author{W.~Ryssens}
\affiliation{Institut d'Astronomie et d'Astrophysique, Universit\'e Libre de Bruxelles, Campus de la Plaine CP 226, 1050 Brussels, Belgium}
\author{P.~Ascher}
\affiliation{Universit\'e de Bordeaux, CNRS/IN2P3, LP2I Bordeaux, UMR 5797, F-33170 Gradignan, France}
\author{M.~Bender}
\affiliation{Universit\'e de Lyon, Universit\'e Claude Bernard Lyon 1, CNRS/IN2P3, IP2I Lyon, UMR 5822, F-69622 Villeurbanne, France}
\author{T.~Eronen}
\affiliation{University of Jyvaskyla, Department of Physics, Accelerator laboratory, P.O. Box 35(YFL) FI-40014 University of Jyvaskyla, Finland}
\author{S.~Gr\'evy}
\affiliation{Universit\'e de Bordeaux, CNRS/IN2P3, LP2I Bordeaux, UMR 5797, F-33170 Gradignan, France}
\author{A.~Kankainen}
\affiliation{University of Jyvaskyla, Department of Physics, Accelerator laboratory, P.O. Box 35(YFL) FI-40014 University of Jyvaskyla, Finland}
\author{M.~Stryjczyk}
\affiliation{University of Jyvaskyla, Department of Physics, Accelerator laboratory, P.O. Box 35(YFL) FI-40014 University of Jyvaskyla, Finland}
\author{L.~Al~Ayoubi}
\affiliation{University of Jyvaskyla, Department of Physics, Accelerator laboratory, P.O. Box 35(YFL) FI-40014 University of Jyvaskyla, Finland}
\affiliation{Universit\'e Paris Saclay, CNRS/IN2P3, IJCLab, 91405 Orsay, France}
\author{S.~Ayet}
\affiliation{II. Physikalisches Institut, Justus Liebig Universitat Gie{\ss}en, 35392 Gie{\ss}en, Germany}
\author{O.~Beliuskina}
\affiliation{University of Jyvaskyla, Department of Physics, Accelerator laboratory, P.O. Box 35(YFL) FI-40014 University of Jyvaskyla, Finland}
\author{C.~Delafosse}
 \altaffiliation[Present address: ]{Universit\'e Paris Saclay, CNRS/IN2P3, IJCLab, 91405 Orsay, France}
\affiliation{University of Jyvaskyla, Department of Physics, Accelerator laboratory, P.O. Box 35(YFL) FI-40014 University of Jyvaskyla, Finland}
\author{W.~Gins}
\affiliation{University of Jyvaskyla, Department of Physics, Accelerator laboratory, P.O. Box 35(YFL) FI-40014 University of Jyvaskyla, Finland}
\author{M.~Gerbaux}
\affiliation{Universit\'e de Bordeaux, CNRS/IN2P3, LP2I Bordeaux, UMR 5797, F-33170 Gradignan, France}
\author{A.~Husson}
\affiliation{Universit\'e de Bordeaux, CNRS/IN2P3, LP2I Bordeaux, UMR 5797, F-33170 Gradignan, France}
\author{A.~Jokinen}
\affiliation{University of Jyvaskyla, Department of Physics, Accelerator laboratory, P.O. Box 35(YFL) FI-40014 University of Jyvaskyla, Finland}
\author{D.A.~Nesterenko}
\affiliation{University of Jyvaskyla, Department of Physics, Accelerator laboratory, P.O. Box 35(YFL) FI-40014 University of Jyvaskyla, Finland}
\author{I.~Pohjalainen}
\affiliation{University of Jyvaskyla, Department of Physics, Accelerator laboratory, P.O. Box 35(YFL) FI-40014 University of Jyvaskyla, Finland}
\author{M.~Reponen}
\affiliation{University of Jyvaskyla, Department of Physics, Accelerator laboratory, P.O. Box 35(YFL) FI-40014 University of Jyvaskyla, Finland}
\author{S.~Rinta-Antila}
\affiliation{University of Jyvaskyla, Department of Physics, Accelerator laboratory, P.O. Box 35(YFL) FI-40014 University of Jyvaskyla, Finland}
\author{A.~de Roubin}
\altaffiliation[Present address: ]{KU Leuven, Instituut voor Kern- en Stralingsfysica, B-3001 Leuven, Belgium}
\affiliation{University of Jyvaskyla, Department of Physics, Accelerator laboratory, P.O. Box 35(YFL) FI-40014 University of Jyvaskyla, Finland}
\author{A.P.~Weaver}
\altaffiliation[Present address: ]{TRIUMF, 4004 Wesbrook Mall, Vancouver, British Columbia V6T 2A3, Canada}
\affiliation{School of Computing, Engineering and Mathematics, University of Brighton, Brighton BN2 4GJ, United Kingdom}

\date{\today}

\begin{abstract}
Precision mass measurements of neutron-rich rhodium isotopes have been performed at the JYFLTRAP Penning trap mass spectrometer at the Ion Guide Isotope Separator On-Line (IGISOL) facility. We report results on ground- and isomeric-state masses in $^{110,112,114,116,118}$Rh and the very first mass measurement of $^{120}$Rh. The isomeric states were separated and measured for the first time using the Phase-Imaging Ion-Cyclotron-Resonance (PI-ICR) technique. For $^{112}$Rh, we also report new half-lives for both the ground state and the isomer. The results are compared to theoretical predictions using the BSkG1 mass model and discussed in terms of triaxial deformation.

\end{abstract}

\maketitle


\section{\label{sec:intro} Introduction}

Neutron-rich rhodium isotopes (proton number $Z=45$) are located in a region known for rapid and drastic changes in nuclear structure and coexisting strongly deformed and spherical shapes \cite{Heyde2011}. At $Z\approx40$ and neutron number $N\approx60$, the strongly deformed configuration becomes the ground state. The atomic mass can be used to probe the evolution of nuclear structure along isotopic chains: the sudden onset of prolate ground-state deformation has been observed as a kink in two-neutron separation energies in the isotopic chains from Rb ($Z=37$) to Mo ($Z=42$)~\cite{Rahaman2007,Hager2006,Hager2007a,Hakala2011}. The same shape change has been observed as a steep increase in the mean-square charge radii ~\cite{Thibault1981,Lievens1996,Cheal2007,Campbell2002,Cheal2009,Charlwood2009}. This is one of the most dramatic shape changes on the nuclear chart.

Above $Z\approx42$, the shape transition becomes smoother but the situation remains complex. Coulomb excitation experiments of Ru and Pd isotopes with neutron numbers between $N = 60$ and 66~\cite{Svensson1995,Srebrny2006a,Doherty2017} indicate that these nuclei exhibit triaxial deformation. Their nuclear density in the intrinsic frame no longer has any rotational symmetry axis. In addition, the structure of neutron-rich Rh isotopes has been explored via prompt $\gamma$-ray spectroscopy of fission fragments \cite{Liu2011,Navin2017}. Comparison of the studied band structures with Triaxial Projected Shell Model (TPSM) calculations indicate a need for large, nearly constant, triaxial deformation in $^{116-119}$Rh \cite{Navin2017}. Several theoretical models~\cite{Hakala2011,Zhang2015,Abusara2017,Bucher2018} predict an additional contribution to the binding energy of triaxial nuclei compared to axially symmetric configurations. Most theoretical approaches predict the largest energy gains due to triaxial deformation in the vicinity of $^{106}$Ru~\cite{Goriely2009,Zhang2015,Moller2016,Scamps2021}.

The masses of neutron-rich odd-odd Rh isotopes have not been accurately known prior to this work due to the presence of low-lying long-living isomeric states \cite{Kondev2021}. The previous mass measurements \cite{Kolhinen2003,Hager2007b} on these isotopes at the JYFLTRAP double Penning trap \cite{Eronen2012} could not resolve the isomeric states, with an exception of $^{108}$Rh, because of the techniques used at the time.

The existence of two long-living states in neutron-rich odd-odd Rh isotopes has been well established via $\beta$-decay studies. The low-spin ($1^+$) states are strongly fed in the $\beta$-decay of even-even $^{110,112,114}$Ru nuclei~\cite{Jokinen1991,Jokinen1992} and have strong decay branches to the $0^+$ states in the Pd isotopes~\cite{Aysto1988, Lhersonneau1999,Lhersonneau2003}, whereas the high-spin Rh isomers $\beta$-decay dominantly to high-spin states in even $^{110-118}$Pd isotopes \cite{Aysto1988,Lhersonneau1999,Lhersonneau2003,Wang2001,Jokinen2000,Wang2006}. Because of the large spin difference, the internal transition is not favored and the excitation energies of these isomeric states have remained largely unknown \cite{Kondev2021}. Only in $^{104}$Rh, the excitation energy of the isomeric 5$^{+}$ state (129~keV) has been precisely established through $\gamma$-ray spectroscopy. In $^{106}$Rh, a 6$^{+}$ state is proposed at 137(13)~keV based on ${\beta}$-decay endpoint measurements \cite{Gurdal2012,Kondev2021} whereas a 5$^{+}$ state is proposed at 113.0(75)~keV in $^{108}$Rh based on a Time-Of-Flight Ion-Cyclotron-Resonance (TOF-ICR) mass measurement performed at JYFLTRAP \cite{Hager2007b}. For heavier Rh isotopes, the energies of the long-lived isomeric states have not been measured. Finally, it should be noted that the order of the two long-living states is established only in $^{104}$Rh and $^{106}$Rh, via $\gamma$-ray spectroscopy ($^{104}$Rh) or $\beta$-decay endpoint energies ($^{106}$Rh). For all the other Rh isotopes discussed in this work, the ordering of the two long-lived states is not firmly confirmed.

In this work, we studied odd-odd $^{110-120}$Rh isotopes via high-precision Penning-trap mass spectrometry. With the novel Phase-Imaging Ion-Cyclotron-Resonance (PI-ICR) technique at JYFLTRAP \cite{Eliseev2014,Nesterenko2018,Nesterenko2021}, we have resolved the ground and isomeric states in $^{110,112,114,116,118}$Rh isotopes for the first time. The reported measurements yield more accurate ground-state mass values as well as first direct determinations of the isomeric-state excitation energies. In $^{120}$Rh we report on the first ground-state mass measurement.

The measured masses provide essential data to improve future models of nuclear structure, particularly those that aim to describe nuclear masses. To benchmark existing models, we compare our experimental results to the recent global model BSkG1~\cite{Scamps2021}. This model is based on self-consistent Hartree-Fock-Bogoliubov (HFB) calculations using a Skyrme energy density functional (EDF) and aims at a global yet microscopic description of nuclear structure. Following the protocol developed for the earlier BSk-models (see Ref.~\cite{Goriely2016} and references therein), the model parameters have been adjusted to essentially all binding energies tabulated in Atomic Mass Evaluation 2016 (AME16)~\cite{Huang2017}, all known charge radii~\cite{Angeli2013} and realistic infinite nuclear matter properties. The resulting model achieves a root-mean-square deviation of 741 keV for the 2457 known masses of nuclei with $Z \geq 8$ included in Atomic Mass Evaluation 2020 (AME20)~\cite{Huang2021}.

\section{\label{sec:exp} Experimental methods}

Precision mass measurements of neutron-rich rhodium nuclei were performed at the Ion Guide Isotope Separator On-Line (IGISOL) facility \cite{Moore2013} utilizing the JYFLTRAP double Penning trap~\cite{Eronen2012}. The measured isotopes were produced in proton-induced fission, with 25-MeV protons from the K130 cyclotron impinging on a thin (15 mg/cm$^2$) natural uranium target. The proton beam intensity was varied between 1$-$10 $\mu$A depending on the studied case. The fission fragments were stopped in helium gas at about 300 mbar and extracted using a sextupole ion guide (SPIG)~\cite{Karvonen2008}. The secondary beam was accelerated to 30$q$~kV and mass-separated based on the mass-to-charge $m/q$ ratio with a $55^{\circ}$  dipole magnet, which has a mass resolving power of $m/\Delta m \approx 500$. The mass-separated beam was then transported to the buffer-gas filled radio-frequency quadrupole (RFQ) cooler-buncher~\cite{Nieminen2001}. To control the number of ions injected into the RFQ cooler-buncher, an electrode after the dipole magnet was connected to a fast switch to serve as a beamgate.

The cooled ion bunches from the RFQ were directed to the JYFLTRAP double Penning trap mass spectrometer~\cite{Eronen2012}, which consists of two cylindrical Penning traps housed inside a 7~T superconducting magnet. First, the ion bunch was injected into the purification trap where the ion sample was cooled, centered and cleaned utilizing the mass-selective buffer gas cooling technique~\cite{Savard1991}. Typically, a mass resolving power of $m/\Delta m \approx 10^5 $ is reached, which in most cases allowed removal of the isobaric contamination. 
The cleaned ion sample was transferred to the precision trap, where the mass determination of an ion was performed through the measurement of its cyclotron frequency $\nu_c$:
\begin{equation}
\nu_c = \frac{1}{2\pi}\frac{q}{m}B,
\end{equation}

where $q/m$ is the charge-to-mass ratio of the ion and $B$ is the magnetic field strength. To determine the magnetic field strength with enough precision, the cyclotron frequency of a reference ion with a well-known mass is measured. These references are measured alternatively with the ion of interest, in order to reduce the effect of the drift of the magnetic field. Utilizing the cyclotron frequency ratio $R= \nu_{\text{c,ref}}/\nu_{\text{c}}$, determined with singly-charged ions, the atomic mass of the nuclide of interest $m$ is then deduced as
\begin{equation}
m = R(m_{\text{ref}}-m_{\text{e}})+m_{\text{e}},
\end{equation}

where $m_e$ is the mass of an electron and $m_{\text{ref}}$ is the atomic mass of the reference nuclide.
In our experiment, $^{133}$Cs$^{+}$ ions from the IGISOL offline ion source station~\cite{Vilen2020} were used for most of the measurements. For $^{112}$Rh$^{2+}$ ions, $^{85}$Rb$^{+}$ ions from the offline ion source station were used as a reference. 

When one of the states, ground or isomeric state, had been measured against a well-known reference, it was later used as a reference for the other state, i.e. the ground state was measured against the isomeric state or vice versa. The reference for each measurement has been tabulated in Table \ref{tab:results}. The energy difference between the ground and the low-lying isomeric states, i.e. the excitation energy of the isomer $E_x$, was calculated utilizing the frequency ratio $R$ determined when these two states were measured against each other:

 \begin{equation}
    E_{x} = (R-1)[m_{\text{ref}} - m_{\text{e}}]c^2. 
\end{equation} 

\begin{figure}
\includegraphics[width=\columnwidth]{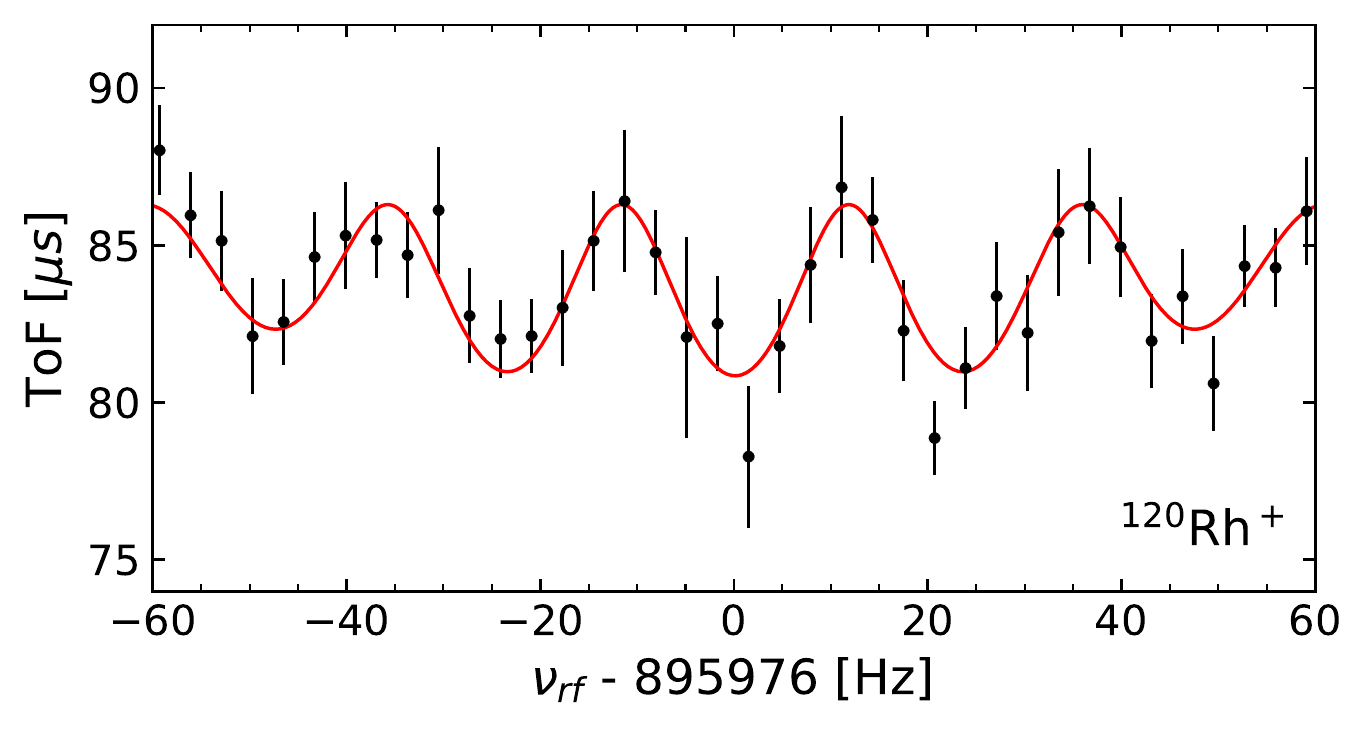}
\caption{\label{fig:120Rh_TOF-ICR} TOF-ICR resonance using a 10-30-10 ms (on-off-on) Ramsey excitation pattern  for $^{120}$Rh$^+$. The mean data points are shown in black, the fit of the theoretical curve \cite{Kretzschmar2007} in red.}
\end{figure}

In the precision trap, the measurement of the ion cyclotron frequency was performed utilizing either the Time-of-Flight Ion-Cyclotron-Resonance technique (TOF-ICR) \cite{Graff1980, Konig1995} or the Phase-Imaging Ion-Cyclotron-Resonance technique (PI-ICR) \cite{Eliseev2014, Nesterenko2018,Nesterenko2021}. In the TOF-ICR technique ion's radial energy is increased by a quadrupolar excitation and converted to axial energy when the ions travel through the magnetic field gradient after extraction from the precision trap. For the resonant ions this results in a shorter time-of-flight measured with a microchannel plate (MCP) detector. From the time-of-flight resonance the cyclotron frequency of the ion is then determined \cite{Konig1995}. To increase the precision, a technique called Ramsey method of time-separated oscillatory fields \cite{Kretzschmar2007, George2007} can be applied: a measurement scheme where the quadrupolar excitation is performed by two short pulses separated by a waiting time. The Ramsey technique with a pattern of 10-30-10 ms (on-off-on) was used for the most exotic isotope measured in this work, namely $^{120}$Rh (see Fig.~\ref{fig:120Rh_TOF-ICR}). The PI-ICR technique was utilized for all the other cases.

\begin{figure}[t]
\includegraphics[width=\columnwidth]{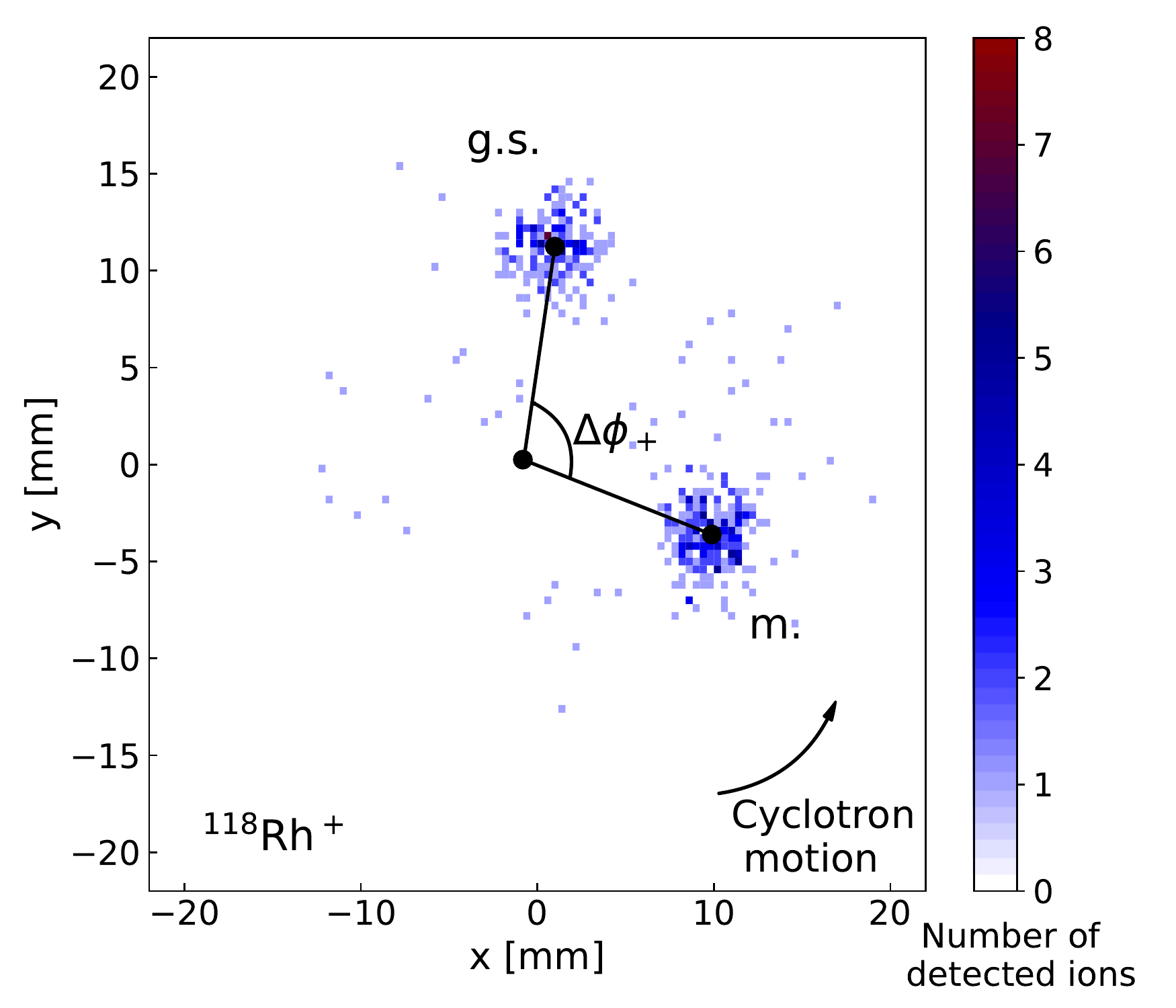}
\caption{\label{fig:118RhPI-ICR} Ion projection of the cyclotron phase spot of $^{118}$Rh$^+$ on the 2D MCP detector. The $^{118}$Rh ground state (g.s.) versus the isomeric state (m.) are separated with a 170 ms accumulation time. The phase difference $\Delta\phi_+$ corresponds to an excitation energy of 189(6) keV.}
\end{figure}

In the PI-ICR technique, the determination of the ion's cyclotron frequency is based on a measurement of the propagation of the ion's radial motions, magnetron motion $\nu_-$ and modified cyclotron motion $\nu_+$, projected onto a position sensitive MCP detector (2D MCP). For this purpose, first a dipole pulse at the ion's modified cyclotron frequency is applied.~The excitation scheme depends on which ion motion propagation is being measured. In the cyclotron phase $\phi_+$ measurement, the ion is let to accumulate phase for a time $t_{\text{acc}}$ on the modified cyclotron motion which is then converted with a $\pi$-pulse to magnetron motion. In the magnetron phase $\phi_-$ measurement, the modified cyclotron motion is first converted with a $\pi$-pulse to magnetron motion, which is then accumulated for a time $t_{\text{acc}}$ before extraction. The measurement pattern utilized at JYLFTRAP is described in more detail in Refs.~\cite{Nesterenko2018,Nesterenko2021} and the PI-ICR measurement technique in general in Ref.~\cite{Eliseev2014}. As an example the measurement of the cyclotron phase spot for $^{118}$Rh$^+$ is shown in Fig.~\ref{fig:118RhPI-ICR}, where the separation of the ground state from the isomeric state is seen with a phase accumulation time of $170$~ms. 

The measured phase difference of the ion's propagated radial motions $\phi_c = \phi_+ - \phi_-$ is related to the ion cyclotron frequency:

\begin{equation}
\nu_c = \nu_- + \nu_+ = \frac{\phi_c + 2\pi n}{2 \pi t_{\text{acc}}},
\label{eq:pi-icr_nc}
\end{equation}

where $n$ is the number of the full ion revolutions in the precision trap. The accumulation time $t_{\text{acc}}$ is selected to be a compromise where the half-life of the ion of interest, phase spot separation of different ion species, vacuum conditions and the stability of the electric field of the precision trap are considered. The phase accumulation times were 700~ms for the $^{110}$Rh$^+$ measurements, 1~s for $^{112}$Rh$^+$, 300~ms for $^{112}$Rh$^{2+}$, 450~ms for $^{114}$Rh$^+$ and $^{116}$Rh$^+$, and 170~ms for $^{118}$Rh$^+$. For $^{116}$Rh$^+$, $^{100}$Y$^{16}$O$^+$ ions were observed in the precision trap. Therefore an additional cleaning step using the Ramsey cleaning method~\cite{Eronen2008} was applied before the phase accumulation. To unambiguously determine $n$ in Eq.~(\ref{eq:pi-icr_nc}), the cyclotron frequency was determined with the TOF-ICR technique and at least two phase accumulation times were applied before the actual mass measurements. 

A count-rate class analysis \cite{Kellerbauer2003,Nesterenko2021} to account for the ion-ion interactions was performed when statistically feasible. Temporal magnetic field fluctuations of $\delta B/B = 2.01(25) \times 10^{-12}$ min$^{-1}$  $\times \delta t$ \cite{Nesterenko2021}, where $\delta t$ is the time between the measurements, were taken into account in the analysis. For measurements, where $^{133}$Cs$^+$ or $^{85}$Rb$^+$ ions were used as a reference, a mass-dependent shift of $\delta_m r/r = -2.35(81) \times 10^{-10} / \textnormal{u} \times (m_{\text{ref}} - m)$ \cite{Nesterenko2021} and a residual systematic uncertainty of $\delta_{\text{res}}r/r=9\times 10^{-9}$ \cite{Nesterenko2021} were added quadratically to the final frequency ratio uncertainty. For the mass doublets (i.e. ground state vs isomer), the mass-dependent and the residual systematic uncertainties are negligible. Systematic uncertainties of the magnetron phase advancement and systematic angle error \cite{Nesterenko2021} were also taken into account.

In addition to mass measurements, the  mass-selective buffer gas cooling technique \cite{Savard1991} was used to select $^{112}$Rh$^+$ and $^{112}$Rh$^{m, +}$ ions for post-trap spectroscopy in order to identify the states and determine their half-lives. 
After injection to the purification trap, the ions at $A=112$ were let to cool for roughly 60~ms, followed by a 10~ms dipolar excitation at the magnetron frequency. During this time, a fraction of the $^{112}$Ru$^+$ ions ($J^{\pi}=0^+, T_{1/2}=1.75(7)~s$) beta decay to $^{112}$Rh$^{2+}$, feeding only the low-spin (1$^{+}$) state in $^{112}$Rh, whereas fission dominantly produces $^{112}$Rh in the high-spin state. The ions of interest were selected using a 120~ms quadrupolar excitation either at the $^{112}$Rh$^+$ (from fission) or $^{112}$Rh$^{2+}$ (from in-trap decay) cyclotron frequency, extracted out of the trap, and implanted every 194~ms on a thin aluminum foil in front of a silicon detector after the Penning trap. The signals were read and time-stamped with the Nutaq data acquisition system with a 105~MHz clock. 

\section{\label{sec:results} Results}

\begin{figure}[h]
\includegraphics[width=0.49\textwidth]{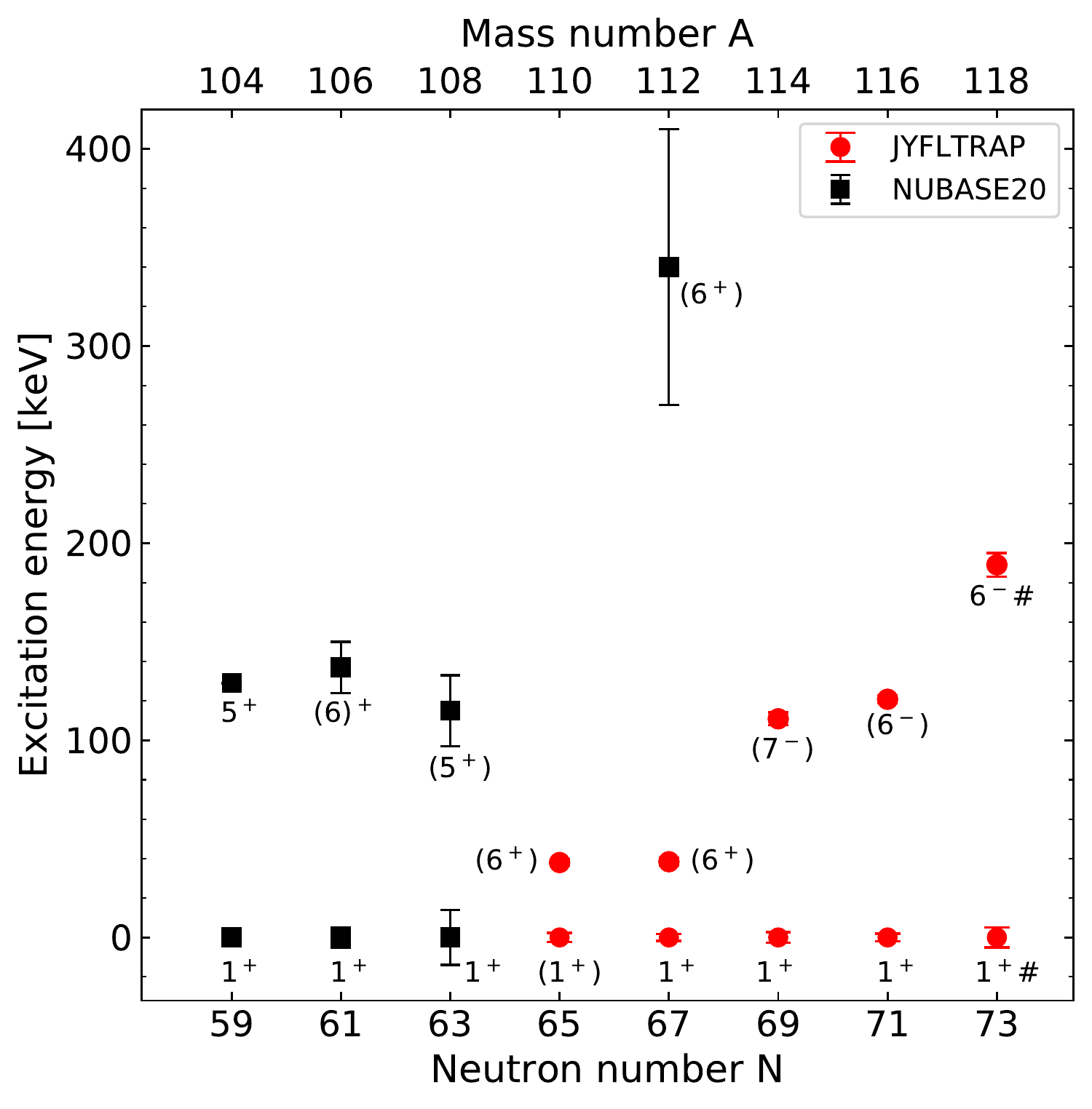}
\caption{\label{fig:RhExcitationEnergy} 
Systematics of the ground and isomeric states in $^{104-118}$Rh isotopes. The literature values from NUBASE20~\cite{Kondev2021} are shown in black, while the energies measured in this work at JYFLTRAP are shown in red. The spin/parities are adopted from NUBASE20 \cite{Kondev2021}. For $^{112}$Rh, we confirm the order of the states and show that it has been previously significantly overestimated.}
\end{figure}

\begin{figure*}[t]
\includegraphics[width=\textwidth]{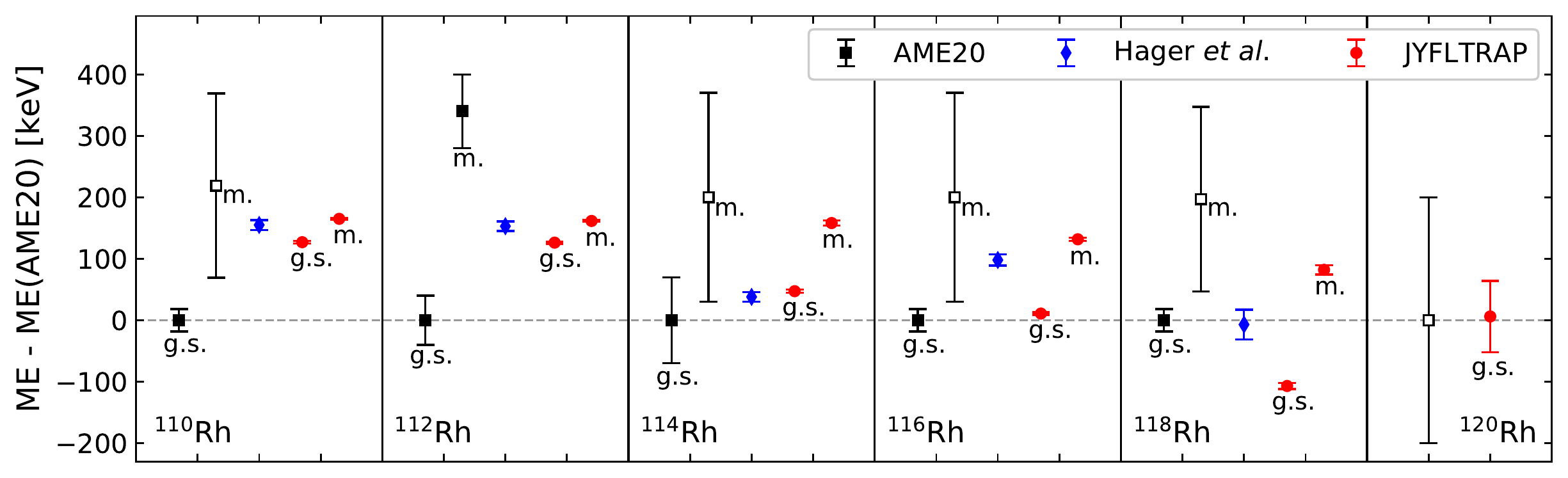}
\caption{\label{fig:Rh_AME2020_Hager} The mass-excess values (ME) for the Rh isotopes measured in this work at JYFLTRAP (red circles) compared to AME20~\cite{AME20} and NUBASE20~\cite{Kondev2021} (black squares) and the earlier JYFLTRAP results obtained with the TOF-ICR technique in Hager {\it et al.} (blue diamonds) \cite{Hager2007b}. Prior to this work, only the mass of the $^{112}$Rh isomer had been experimentally determined, whereas the rest of the isomeric states were based on extrapolations (shown with open black squares)~\cite{Kondev2021}.}
\end{figure*}

In total, we report six ground-state and five isomeric-state masses, which are summarized in Table~\ref{tab:results}. The mass of $^{120}$Rh and the isomeric states in $^{110,114,116,118}$Rh were measured for the first time. We have also determined the excitation energies of the isomers for the first time and show that the previously determined excitation energy of $^{112}$Rh isomer deviates significantly from our result~(see Table~\ref{tab:results} and Fig.~\ref{fig:RhExcitationEnergy}). Comparison to the previous JYFLTRAP mass measurement by Hager {\it et al.} \cite{Hager2007b} and to the AME20 \cite{AME20} values is given in Fig.~\ref{fig:Rh_AME2020_Hager}. Taking into account the ground- and isomeric-state yields, our results show a reasonable agreement with the values reported by Hager {\it et al.} \cite{Hager2007b}, where the measured masses were unresolved mixtures of the ground and isomeric states. Below, we report the results and compare to the literature for each of the isotopes separately.\\

\begin{table*}
\centering
\caption{\label{tab:results}Frequency ratios ($R$), mass excess values ($ME = (m-Au)c^2$) and isomer excitation energies ($E_x$) measured in this work. The masses of $^{133}$Cs and $^{85}$Rb reference ions were taken from the AME20~\cite{AME20}, whereas the masses determined in this work were used for the Rh reference ions (Ref.). The spin-parities $I^\pi$ and half-lives $T_{1/2}$ have been taken from NUBASE20 \cite{Kondev2021}, except for $^{112}$Rh, for which the half-lives are from this work. For comparison, the mass-excess values ME$_{lit.}$ from AME20~\cite{AME20} and excitation energies $E_{x,lit.}$ from NUBASE20 \cite{Kondev2021} are also provided, where \# marks an extrapolated value. }
\begin{ruledtabular}
\begin{tabular}{lllllllll}
 Nuclide & $I^\pi$ & $T_{1/2}$ & Ref. & $R=\nu_{\text{c,ref}}/\nu_{c}$ & ME [keV]  & ME$_{\text{lit.}}$ [keV]  & $E_x$ [keV]  & $E_{x,\text{lit.}}$ [keV] \vspace{0.08cm}\\ \hline
 \noalign{\vskip 1.3mm}   
 $^{110}$Rh & $(1^{+})$ & 3.35(12) s & $^{110}$Rh$^{m}$  & 0.999999629(15) & -82702.4(23) & -82829(18) & & \\
 $^{110}$Rh$^{m}$ & $(6^{+})$ & 28.5(13) s & $^{133}$Cs  & 0.826987604(15) & -82664.3(18) & -82610(150)\# & 38.0(15) & 220(150)\# \\
 \noalign{\vskip 1.3mm}
 $^{112}$Rh & $(1^{+})$ & 2.22(13) s\footnotemark[1] & $^{85}$Rb & 0.659002526(11)\footnotemark[2] & -79603.7(17) & -79730(40) & & \\
 $^{112}$Rh$^{m}$ & $(6^{+})$ & 6.96(8) s\footnotemark[1]  & $^{133}$Cs   & 0.842060991(16) & -79565.2(19) &  &  38.5(26) &\\
                  &           &             & $^{112}$Rh   & 1.000000368(27)  & -79565.4(33) & &  38.3(28) &\\
  &  &  &  &Weighted mean  & -79565.2(17) & -79390(60) & 38.4(19)  &  340(70) \\
  \noalign{\vskip 1.3mm}
 $^{114}$Rh & $1^{+}$ & 1.85(5) s & $^{133}$Cs         & 0.857140868(21) & -75662.7(26) & -75710(70) & &\\
 $^{114}$Rh$^{m}$ & $(7^{-})$ & 1.85(5) s & $^{114}$Rh   & 1.000001045(30) & -75551.8(41) & -75510(170)\# & 110.9(32) & 200(150)\#\\
 \noalign{\vskip 1.3mm}
 $^{116}$Rh & $1^{+}$ & 685(39) ms & $^{133}$Cs        & 0.872229073(17) & -70729(2) & -70740(70) & & \\
 $^{116}$Rh$^{m}$ & $(6^{-})$  & 570(50) ms & $^{116}$Rh   & 1.000001119(18) & -70608.3(28) & -70540(170)\# &  120.8(19)  & 200(150)\#\\
 \noalign{\vskip 1.3mm}
 $^{118}$Rh & $1^{+}$\# & 282(9) ms & $^{133}$Cs         & 0.887323752(40) & -64994(5) & -64887(24) & &  \\
 $^{118}$Rh$^{m}$ & $6^{-}\#$ & 310(30) ms & $^{118}$Rh  & 1.000001720(55) & -64804.9(78) & -64690(150)\#  & 189(6) & 200(150)\# \\
 \noalign{\vskip 1.3mm}
 $^{120}$Rh & $8^-$\# & 129.6(42) ms & $^{133}$Cs         & 0.902423642(470) & -58614(58) & -58620(200)\# & & \\
\end{tabular}
\footnotetext[1]{Half-life determined in this work, see Sect.~\ref{sec:halflives}.}
\footnotetext[2]{Measured with doubly charged $^{112}$Rh$^{2+}$ ions.}
\end{ruledtabular}
\end{table*}

\subsection{$^{110}$Rh}
\label{sec:110rh}

For $^{110}$Rh (N = 65), the measured mass excess of the ground state is $-82702.4(23)$~keV and  $-82664.3(18)$~keV for the isomer. The excitation energy of the isomer, $E_x = 38.0(15)$~keV, was directly extracted from the measured ground-state-to-isomer frequency ratio. 
The ground-state mass value determined in this work is 126(18)~keV larger than the AME20 value, which is based on $\beta$-decay endpoint energies from Jokinen~{\it et al.}~\cite{Jokinen1991} and a private communication~\cite{Kratz2000} (see Fig.~\ref{fig:Rh_AME2020_Hager}). However, our ground- and isomeric-state results are consistent with the earlier JYFLTRAP measurements done for a mixture of ground and isomeric states using the buffer-gas cooling technique in the first trap, $-82640(70)$~keV \cite{Kolhinen2003}, and with the TOF-ICR technique in the second trap, $-82674(8)$~keV \cite{Hager2007b}. Here we provide the first direct mass measurements of the ground and isomeric state of $^{110}$Rh. 

\begin{figure}
\begin{tabular}{l}
      \includegraphics[width=0.47\textwidth]{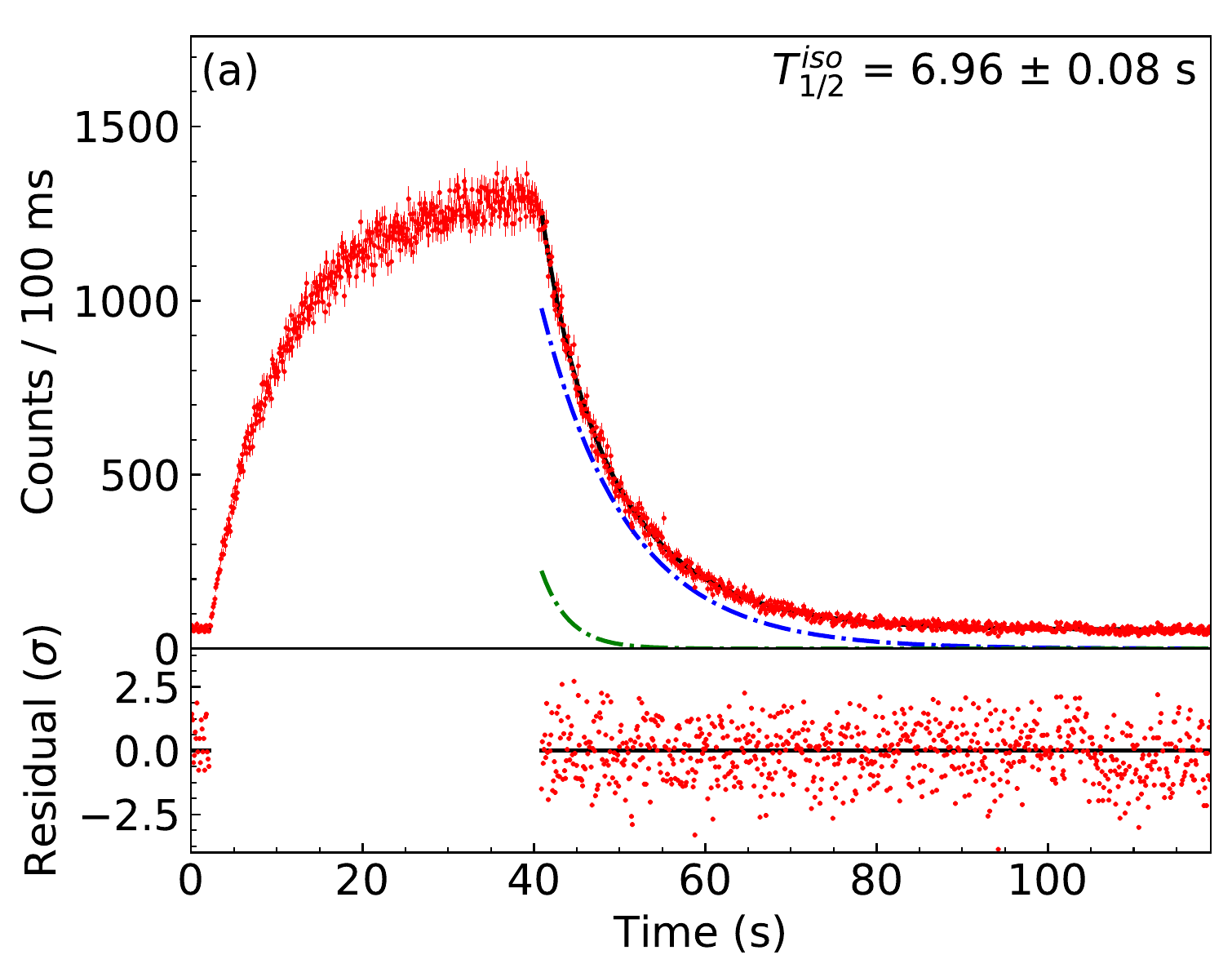}\\
    \includegraphics[width=0.47\textwidth]{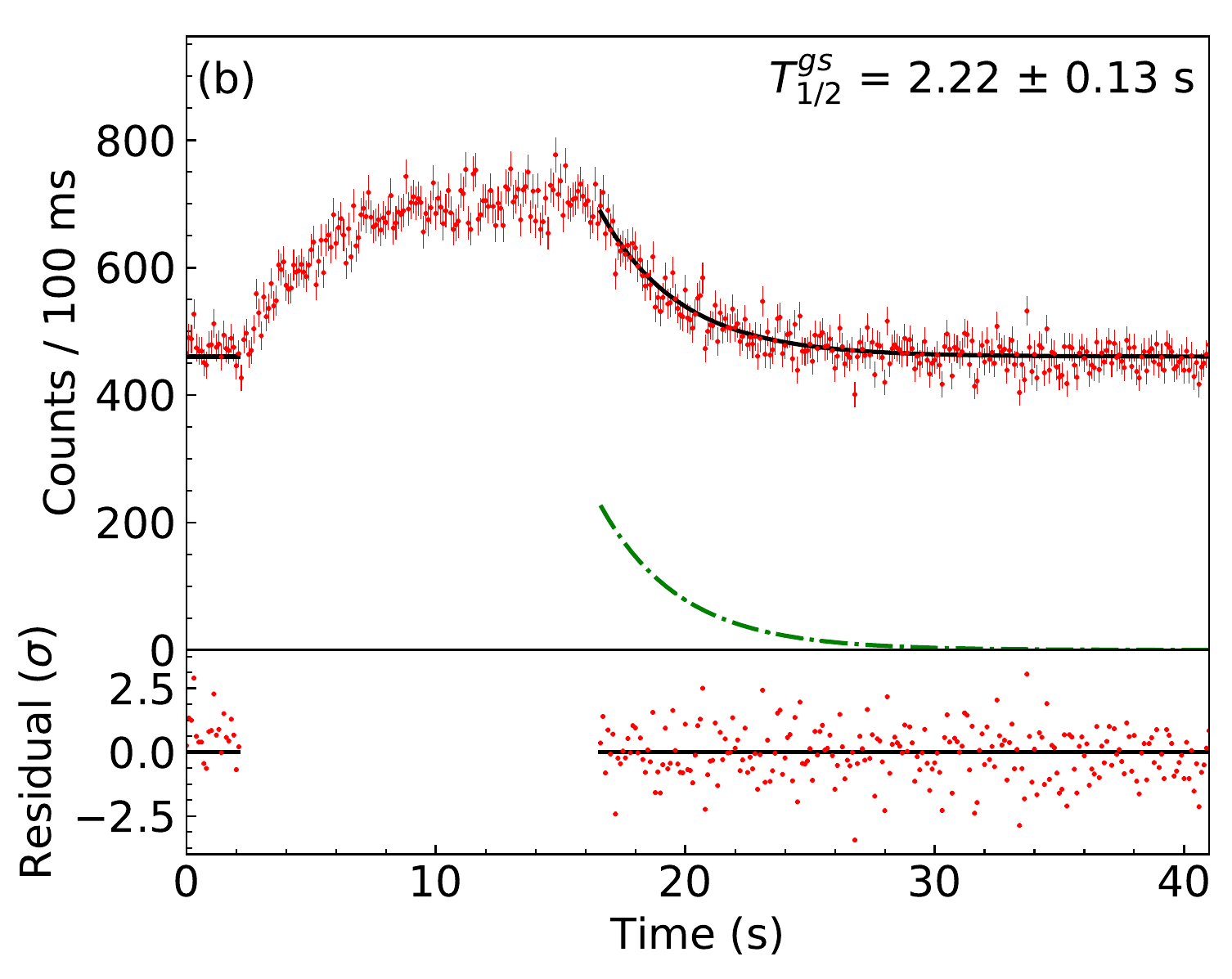}\\
        \end{tabular}
    \caption{Decay curves of $^{112}$Rh for the Penning-trap purified samples of (a) $^{112}$Rh$^+$ directly produced in fission and (b) $^{112}$Rh$^{2+}$produced via in-trap decay of $^{112}$Ru$^+$. The green and blue dash-dotted lines indicate the contribution of the ground state ($T_{1/2}=2.22(13)$~s) and the isomer ($T_{1/2}=6.96(8)$~s) decays, respectively. The lower panels show the residuals of the fit.}
    \label{fig:112Rh_halflife}
\end{figure}

\subsection{$^{112}$Rh}
\label{sec:112rh}
\subsubsection{Mass measurements}

The fission yield of $^{112}$Rh ($N = 67$) is largely dominated by the high-spin state,  and as explained in Sect.~\ref{sec:exp}, the low-spin state of $^{112}$Rh can be exclusively produced by the in-trap decay of the even-even $^{112}$Ru. The latter method was therefore used to determine the mass excess of the low-spin state of $^{112}$Rh in this work. The resulting mass-excess of $-79603.7(17)$~keV, is 27(9)~keV lower than reported for a mixture of the isomeric and ground states in Ref.~\cite{Hager2007b}. 
We note that our result is 127(44)~keV higher than the AME20 value, $-79730(40)$~keV \cite{AME20}, in which the previous JYFLTRAP measurements contributed by 16~\% \cite{Kolhinen2003} and 19~\% \cite{Hager2007b}, whereas a $\beta$-decay study based on private communication had a 66~\% contribution \cite{Kratz2000}.

For the high-spin state produced directly by fission, the final value for the mass excess, $-79565.2(17)$ keV, was taken as the weighted mean of the PI-ICR measurements performed with a 1~s accumulation time against (i) the $^{112}$Rh ground state ($-79565.4(33)$ keV) and (ii) $^{133}$Cs ($-79565.2(19)$ keV). In the measurement (i) the reference (ground state) statistics were rather low, thus the measurement was repeated with $^{133}$Cs as a reference. 
Thus, the $^{112}$Rh mass measurements now allow us to determine unambiguously that the low-spin state is the ground state of $^{112}$Rh whereas the high-spin state is the isomer. The excitation energy of the isomer was determined from the difference between the mass-excess values of the isomer and the ground-state, $E_x = 38.5(26)$~keV, as well as directly from the isomer against the ground state measurement, $E_x = 38.3(28)$~keV. The weighted mean, $E_x = 38.4(19)$~keV, was adopted as the final value. The excitation energy is significantly (more than $4\sigma$) lower than the NUBASE20~\cite{Kondev2021} value, $E_x = 340(70)$~keV, which is based on an unpublished $\beta$-decay study \cite{Kratz2000}. 

\subsubsection{\label{sec:halflives}Half-life measurements}

In order to further confirm the order of the low- and high-spin states, $\beta$-decay half-lives of the ground and isomeric states of $^{112}$Rh have been determined using a silicon detector located after the JYFLTRAP Penning trap. The first measurement, performed with an isomerically-mixed beam of $^{112}$Rh$^+$ produced directly in fission (see Fig.~\ref{fig:112Rh_halflife}(a)), was performed with a cycle of 2~s of waiting period, followed by 38.8~s of implantation and 80~s of decay. The second measurement was performed with the pure beam of the low-spin $^{112}$Rh$^{2+}$ ground-state ions produced by in-trap decay of $^{112}$Ru$^+$ (see Fig.~\ref{fig:112Rh_halflife}(b)). In this case, the cycle consisted of 2~s of waiting time, 14.6~s of implantation and 25~s of decay.

The data were fitted using the Markov chain Monte Carlo (MCMC) method. Depending on the dataset, the fitting models consisted of exponential functions, two for the isomerically-mixed $^{112}$Rh$^+$ and one for $^{112}$Rh$^{2+}$, as well as a constant background. The choice of this model is justified as the half-life of the daughter isotope $^{112}$Pd ($T_{1/2} \approx 21\mathrm{~h}$ \cite{ensdf112}) is much longer than the $^{112}$Rh half-lives. Both datasets were fitted simultaneously with the $^{112}$Rh ground-state half-life being a parameter shared by both models. 

The extracted half-lives are 2.22(13)~s for the ground state and 6.96(8)~s for the isomer. The latter is in a good agreement with 6.73(15)~s from Ref.~\cite{Lhersonneau1999} and 6.8(2)~s from Ref.~\cite{Aysto1988}. The ground-state half-life is more than $10\sigma$ away from the value reported in the ENSDF evaluation ($T_{1/2} = 3.6(3)$~s \cite{ensdf112}), however, it agrees with the result reported in Ref.~\cite{Jokinen1991} ($T_{1/2} = 2.1(3)$~s). It should be noted that the ENSDF value relies on Refs.~\cite{Aysto1988,Lhersonneau1999}. In all three publications, the beam had a significant $^{112}$Ru contamination which decays to the low-spin ground state of $^{112}$Rh. However, only in the work by Jokinen {\it et al.}~\cite{Jokinen1991}, this effect was taken into account.

Around $14(2)$\% of the $^{112}$Rh nuclei were in their ground state based on the PI-ICR measurement. This value was measured with 1~s accumulation time and only statistical uncertainty has been accounted for. The same ratio extracted from the half-life fits, 1~s after the implantation has been stopped, is equal to 16(2)\%. Both results are in an excellent agreement. The combination of independent mass measurements of the two states, together with the half-life measurements, allow us to unambiguously assign the 1$^+$ state as the ground state of $^{112}$Rh, and the high-spin $(6^+)$ state as the isomer.

\subsection{$^{114}$Rh}
\label{sec:114rh}

The measured ground-state mass excess of $^{114}$Rh ($N=69$), $-75662.7(26)$ keV, is in agreement with the AME20 \cite{AME20} value based on Hager {\it et al.} (59 \%) \cite{Hager2007b} and Kolhinen {\it et al.} ($41 \%$) \cite{Kolhinen2003}. The mass excess of the isomer $^{114}$Rh$^m$, $-75551.8(41)$ keV, was measured for the first time. The excitation energy of $^{114}$Rh$^m$, $E_x = 110.9(31)$~keV, was directly obtained from the measurement against the ground state and determined for the first time. Contrary to the other Rh isotopes studied in this work, the fission yield was dominated by the ground state. The ground state is presumably $1^+$ and the isomer $(7^-)$ \cite{Blachot2012b,Kondev2021}, however, the order of the low- and high-spin states has not yet been confirmed experimentally. Such an inversion in the observed isomeric fission yield ratio could be an indication of an inversion between the low-and high-spin states. Another explanation could be if the half-life for the high-spin state was much shorter than reported in literature, where the states have identical half-lives \cite{Kondev2021}.

\subsection{$^{116}$Rh}
\label{sec:116rh}

The measured mass-excess value for the ground state of $^{116}$Rh ($N=71$), $-70729(2)$ keV, agrees well with the AME20 \cite{AME20} value, $-70740(70)$~keV, but is 35 times more precise. The AME20 value is based 63~\% on the JYFLTRAP measurement by Hager {\it et al.}~\cite{Hager2007b} and 37~\% on an average between the earlier JYFLTRAP measurement~\cite{Kolhinen2003} and the storage-ring experiment using the Isochronous Mass Spectrometry (IMS) at GSI~\cite{Knoebel2008}. 

The mass of $^{116}$Rh$^m$, $-70608.3(28)$ keV, has been determined for the first time in this work. Its excitation energy, $E_x = 120.8(19)$ keV, was directly obtained from the measurement against the ground state. The isomer has been assumed to have spin-parity $(6^-)$ based on the $\beta$-decay feeding to a $(6^+)$ state in the daughter nucleus and no feeding to levels with $J\leq 4$ \cite{Blachot2010,Kondev2021}.  

\subsection{$^{118}$Rh}
\label{sec:118rh}

The measured mass-excess value for the ground state of $^{118}$Rh ($N=73$), $-64994(5)$~keV, is 107(25)~keV lower than the AME20 value \cite{AME20} based mainly on the previous study at JYFLTRAP~\cite{Hager2007b}, where a mass-excess of $-64894(24)$~keV was determined for a mixture of the ground and isomeric states. In addition, $^{118}$Rh has been measured via IMS at GSI, $-64830(270)$~keV~\cite{Matos2004}. 

The isomeric state $^{118}$Rh$^m$ was resolved from the ground state and measured for the first time (see Fig.~\ref{fig:118RhPI-ICR}). Its mass excess is $-64804.9(78)$~keV and the excitation energy, directly determined from the measurement against the ground state,  is $E_x = 189(6)$~keV. There is no experimental information on the spin-parities of the ground and isomeric states but they are assumed to be $1^+$ and $6^-$ based on systematics~\cite{Kondev2021}.

\subsection{$^{120}$Rh}
\label{sec:120rh}
The mass of $^{120}$Rh ($N = 75$) was measured experimentally for the first time (see Fig.~\ref{fig:120Rh_TOF-ICR}). The mass-excess value, $-58614(58)$~keV, is very close to the extrapolated value in AME20, $-58620(200)$\#~keV~\cite{AME20}. Whereas the ground state has a half-life of around 130~ms~\cite{Montes2006,Lorusso2015,Hall2021,Kondev2021}, the known isomeric state in $^{120}$Rh is too short-lived ($T_{1/2}=295(16)$~ns~\cite{Kameda2012}) for a Penning-trap measurement. The spin-parities of the ground and isomeric states of $^{120}$Rh are unknown. \\

\section{\label{sec:disc} Discussion}

We start by introducing mean-field modeling of nuclear masses and the BSkG1 model in Sect.~\ref{sec:BSkG1lmodel}. The experimental mass values obtained in this work are compared to the predictions from the BSkG1 model in Sect.~\ref{sec:BSkG1comparison}, whereas two-neutron separation energies, as well as other mass differences are discussed in Sect.~\ref{sec:derivables}. The quadrupole deformation, its evolution and their role in the region are further discussed in Sect.~\ref{sec:triaxiality} and the case of $^{112}$Rh in Sect.~\ref{sec:112rh-theor}.

\subsection{Mean-field modeling of nuclear masses}
\label{sec:BSkG1lmodel}
The BSkG1 model is based on self-consistent HFB calculations using a Skyrme EDF \cite{Scamps2021}. In such approaches one searches for the nuclear configuration that minimizes the total energy in a large variational space, which generally includes symmetry-broken Bogoliubov states that correspond to nuclei that are not spherically symmetric. 
In this way, the shape of the nucleus in its ground state arises naturally as a prediction of the model. It is usually characterized in terms of the multipole moments $Q_{\ell m}$ of the nuclear density, which we define in Sec.~\ref{sec:triaxiality},~and whose relative importance generally quickly decreases with increasing $\ell$ \cite{Scamps2021}.
Not all possible multipole moments are explored by all nuclei, since many of them take shapes that are symmetric in some way. This fact is often exploited by practical implementations that enforce commonly found symmetries to reduce the computational cost. For example, the triaxial shapes we study here combine non-zero values for both quadrupole moments $Q_{20}$ and $Q_{22}$ but are only encountered in specific regions of the nuclear chart~\cite{Scamps2021}. By assuming axial symmetry, i.e. by considering only shapes with one rotational symmetry axis, a significant amount of computational effort can be saved at the cost of possibly missing some physics.

\begin{table*}[t!]
\begin{tabular}{lclllll}
\hline
\hline
& & \multicolumn{5}{c}{Mass excess (keV)} \\
\cline{3-5} \cline{6-7} 
Nucleus & $N$ &  \multicolumn{1}{c}{JYFLTRAP} & \multicolumn{1}{c}{BSkG1} & \multicolumn{1}{c}{Diff.}   & \multicolumn{1}{c}{BSkG1$_{\text{ax}}$} & \multicolumn{1}{c}{Diff.$_{\text{ax}}$} \\
 \hline
 $^{110}$Rh & 65 & $-82702.4$ (23)  & $-83408$ & $-706$     & $-81545$         & $+1157$  \\
 $^{112}$Rh & 67 & $-79603.7$(17)  & $-80263$ & $-660$     & $-78386$         & $+1218$  \\
 $^{114}$Rh & 69 & $-75662.7$(26)  & $-76401$ & $-739$    & $-74538$         & $+1125$  \\
 $^{116}$Rh & 71 & $-70729$ (2)     & $-71538$ & $-809$       & $-70168$         & $+561$      \\
 $^{118}$Rh & 73 & $-64994$(5)      & $-65776$ & $-783$       & $-65090$         & $-96$      \\
 $^{120}$Rh & 75 & $-58613.8$(58) & $-59412$ & $-799$    & $-59373$         & $-763$ \\
\hline
\hline
\end{tabular}
\caption{Mass excesses of the odd-odd Rh isotopes measured in this work (JYFLTRAP) and those obtained with the BSkG1 model~\cite{Scamps2021} and differences (Diff.) between both, all expressed in keV. To illustrate the impact of non-axial shapes, results from calculations restricted to axial symmetry BSkG1$_{\text{ax}}$
and their difference from experiment (Diff.$_{\text{ax}}$) are also given. We remind the reader that BSkG1 describes essentially all known masses in AME20 with an rms error of 741 keV.}
\label{tab:thvsexp_absmass}
\end{table*}

An EDF-based model is characterized by a sizeable number of parameters, which need to be fitted to experimental data using one of many different possible strategies. The standard procedure constructs the objective function for this parameter adjustment from the masses and charge radii of a few, usually spherically symmetric, even-even nuclei, nuclear matter 
properties, and some spectroscopic information. With only a few exceptions, the parameterisations constructed this way are nevertheless applicable to the modelling of all nuclei, independent of their shape.
Some of these parameterisations have been widely used for local studies of nuclear structure in small regions all over the nuclear chart; examples that studied the appearance of triaxial shapes in the mass region of interest here are Refs.~\cite{Hakala2011,Zhang2015,Abusara2017,Bucher2018}.
There also exist more global mass-table calculations with such parameterisations, but in the majority of cases these are limited to even-even nuclei with axial shapes ~\cite{Erler2012a,Agbemava2014a}, with some of them also including non-axial shapes \cite{Lu2015,Yang2021}. Large-scale calculations of this kind including odd- and odd-odd nuclei are even more sparse, an exception being Ref.~\cite{Hilaire2007a}, however, these calculations are restricted to axial shapes only.\footnote{The update of the database also considers non-axial shapes for even-even nuclei.} 

Despite their usefulness for the study of differential quantities in small regions of the nuclear chart, such 'standard' models usually do not describe \textit{absolute binding energies} well: their objective function includes but a handful of nuclei, resulting in root-mean-square (rms) deviations with respect to the entirety of known masses that can reach 10 MeV~\cite{Scamps2021} and are generally not smaller than 1 MeV. This former strategy has to be contrasted with that adopted by what we refer as \textit{global models}: their adjustment includes, among other things, the binding energies and charge radii of almost \textit{all} nuclei for which data are available, be they even-even, odd or odd-odd. Including this large amount of data in the fit typically results in rms deviation on all masses that lie between 600-800 keV~\cite{Goriely2016}. The lowest rms deviations, for microscopic and microscopic-macroscopic models alike, are generally not lower than 500 keV~\cite{Goriely2013,Moller2016}.

However, including more than a handful of nuclei requires allowing for nuclear deformation during the parameter adjustment, which is computationally costly. Most global models, EDF-based or not, were adjusted for this reason with the assumption of axial symmetry. The first global model to overcome this limitation is the 2012 version of the semi-microscopic finite-range droplet model \cite{Moller2016}. The authors of Ref.~\cite{Goriely2009} achieved a first global EDF fit that allowed for triaxial shapes, although some of the model ingredients for odd- and odd-odd nuclei were interpolated from adjacent even-even nuclei.

Profiting from recent numerical and algorithmic developments, BSkG1 is the first global model based on a Skyrme EDF that allowed for all nuclei to take non-axial shapes during the parameter adjustment. This was achieved with the MOCCa code~\cite{RyssensPhD}, a tool that represents the single-particle wave functions on a three-dimensional coordinate-space mesh. It offers both an easily-controlled numerical accuracy~\cite{Ryssens15} and also advanced algorithms to achieve a rapid and stable solution of the self-consistent Skyrme-HFB equations~\cite{Ryssens2019}.

The calculations we report on here were performed using the same tool, in identical numerical conditions as in Ref.~\cite{Scamps2021}. We repeat here only the main features of our treatment of nuclei with an odd neutron number, an odd proton number or both: for each nucleus we performed multiple self-consistent blocking calculations using the equal-filling approximation~\cite{Perez-Martin2008}. Among such a set of calculations, we selected the ground state as the one with lowest energy after convergence.

\subsection{Comparison of mass values to the BSkG1 model}
\label{sec:BSkG1comparison}

The measured mass excesses, the BSkG1 values~\cite{Scamps2021} and the differences between both are shown for the Rh isotopes in Table~\ref{tab:thvsexp_absmass}.
The model overbinds all isotopes with a remarkably constant energy difference on the order of 700 keV. 
This value is close to the overall rms deviation BSkG1 achieves on all nuclear masses, indicating that the Rh isotopes are, on average, neither better nor worse described than other isotopic chains. 
The difference between theory and experiment is nevertheless several orders of magnitude larger than the experimental uncertainty, but this reflects the current state-of-the-art of global mass models such as BSkG1. 

To illustrate the additional energy gain due to triaxiality, Table~\ref{tab:thvsexp_absmass} also lists the mass excesses obtained when restricting the calculations to axial shapes (BSkG1$_{\text{ax}}$), as well as its difference 
(Diff.$_{\text{ax}}$) from experiment. The full calculation leads to binding energies that are larger than those obtained 
from an axial calculation, as expected from the variational principle. With the exception of $^{120}$Rh, the energy gain due to triaxiality is large, up to 1.9 MeV for $^{112}$Rh, and depends on $N$. Subtracting this energy gain results in calculated mass excesses that increase faster with neutron number than the measured ones, as is reflected also in the mass differences we discuss below.

Calculations spanning the nuclear chart that account for triaxiality are scarce, but all of them predict triaxial deformations in the vicinity of $^{106}$Ru~\cite{Moller2006,Moller2016,Lu2015,Goriely2009,Scamps2021,Yang2021}. The strength of the effect, as evaluated by the theoretical energy gain due to triaxial deformation, is strongly model-dependent. The microscopic-macroscopic model of Ref.~\cite{Moller2006}
\footnote{We note that the latest model in this series includes the effect of triaxiality, but Ref.~\cite{Moller2016} provides no information on triaxial deformation or its associated energy gain. } 
predicts the largest energy effect for $^{108}$Ru at about $500$ keV. They obtain triaxial deformations for several isotopes in the Rh chain, but the energy gain is often smaller than 300~keV. These values are much smaller than the ones found for BSkG1, which in this mass region are often significantly larger than one MeV, see Table~\ref{tab:thvsexp_absmass}, and can even reach 2.3~MeV for $^{112}$Tc. The larger BSkG1 values are in qualitative agreement with the results of Ref.~\cite{Goriely2009} obtained with the D1M Gogny-type EDF. This model predicts an energy gain of up to 1.6~MeV for $^{110}$Pd, but the region of triaxial nuclei is limited to nuclei with $N \leq 70$. The authors of Ref.~\cite{Yang2021} do not report on calculations for odd-mass or odd-odd nuclei, but find a modest effect on the order of a few hundred keV~\cite{Lu2015} for $^{102, 104, 106, 108}$Ru; they find the smallest amount of nuclei with ground state triaxial deformation among the models considered here.

\subsection{Comparison of mass differences\\ to the BSkG1 model}
\label{sec:derivables}

The mass differences are not affected by the global offset of about 700 keV between the experimental and the BSkG1 model values. The two-neutron separation energy is a sensitive probe for structural changes, such as shell closures or the onset of deformation as a function of neutron number \cite{Lunney2003}. It is defined as 

\begin{equation}
S_{2n}(Z,N) = BE(Z,N) - BE(Z,N-2) \, , 
\end{equation}
where $BE$ is the nuclear binding energy of a nucleus. In terms of mass values, it can be written as $S_{2n}=[m(Z,N-2)+2m_n-m(Z,N)]c^2$, where $m$ denotes the masses for the nuclides $(Z,N)$, $(Z,N-2)$ and the neutron, and $c$ is the speed of light in vacuum. 

\begin{figure}[t]
\includegraphics[width=\columnwidth]{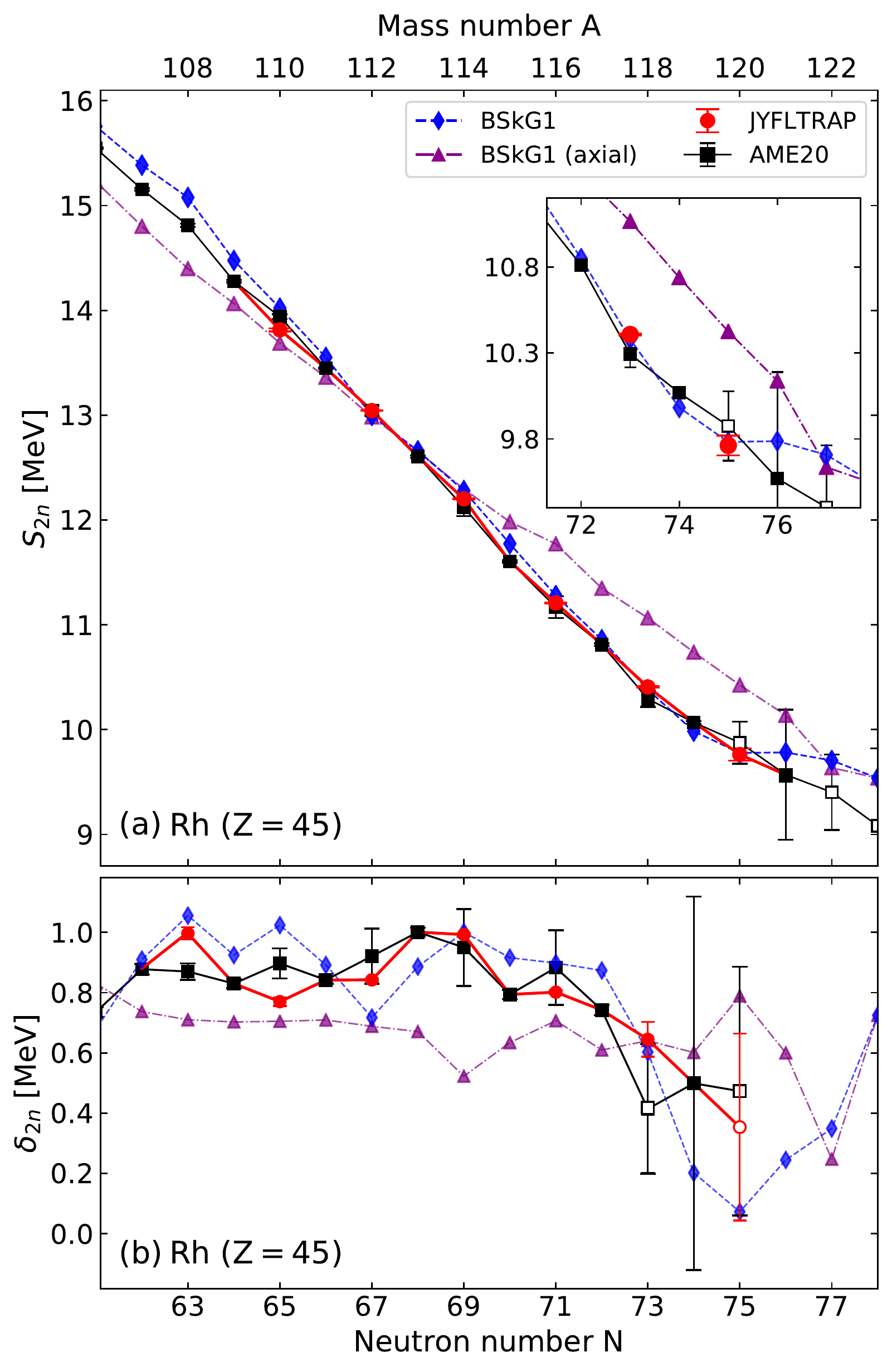}
\caption{(a) Comparison of two-neutron-separation energies $S_{2n}$ of the Rh isotopes affected by 
         our new measurements (red circles), AME20 values (full and open
          black square) and BSkG1 results, as tabulated in 
          Ref.~\cite{Scamps2021} (blue diamonds) or restricted to axially
          symmetric shapes (purple triangles), see text. Open markers represent values at least partially based on extrapolated mass values from AME20~\cite{AME20}. (b) Two-neutron shell-gap energies $\delta_{2n}$.}
\label{fig:S2N_BSkG1_JYFLTRAP} 
\end{figure}

\begin{figure*}[t]
\includegraphics[width=\textwidth]{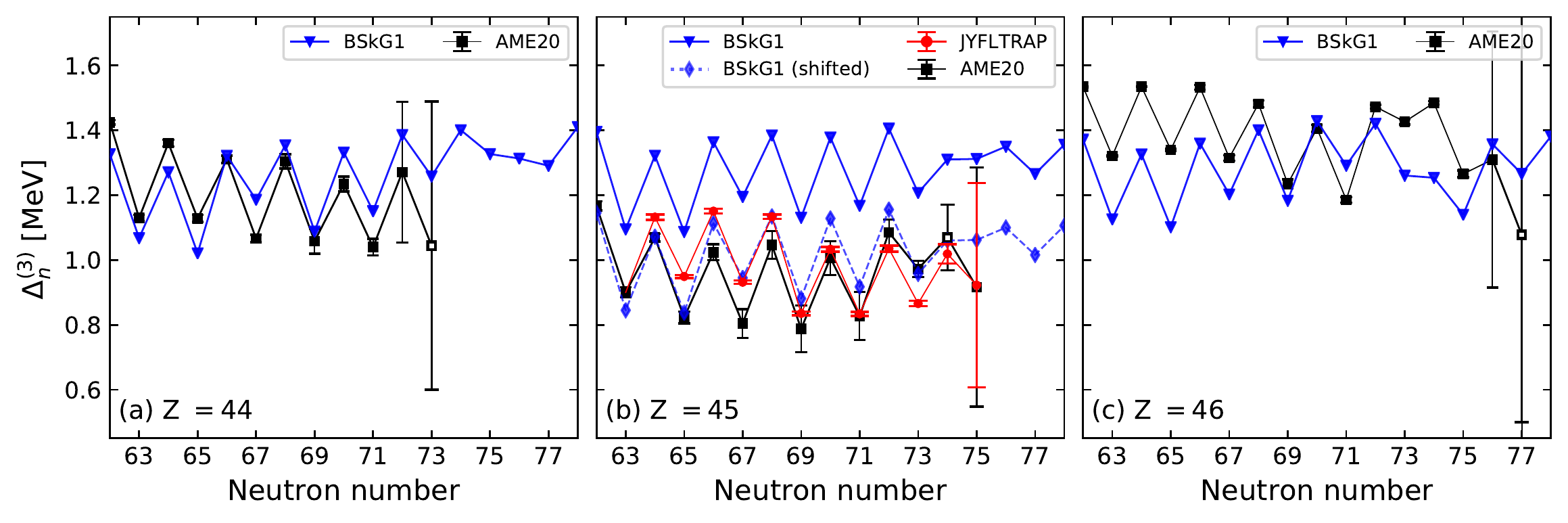}
\caption{Three-point neutron gaps $\Delta^{(3)}_{n}$ for (a) Ru $(Z = 44)$, (b) Rh $(Z = 45)$ and (c) Pd $(Z = 46)$ isotopes. We compare values obtained from BSkG1 (blue triangles with solid line), AME20 (black squares) and from this work at JYFLTRAP (red circles). As illustration, we also show the BSkG1 values with an additional shift of $-250$ keV for Rh isotopes (Z = 45) (blue diamonds with dotted line). \label{fig:Delta3N}}
\end{figure*}

Figure~\ref{fig:S2N_BSkG1_JYFLTRAP}(a) shows the experimental and calculated two-neutron separation energies. As before, we compare complete BSkG1 calculations allowing for non-axial shapes and results from calculations restricted to axial shapes.
The BSkG1 model reproduces the experimental data remarkably well when triaxial deformation is allowed for, as could be expected from the near constant offset between the calculated and experimental mass-excess values (seen Table~\ref{tab:thvsexp_absmass}). 
The $S_{2n}$ values from axial calculations deviate significantly from the experimental ones. Their slope does not follow that of the experimental data, which is a consequence of the gradual evolution of the energy gain from triaxial shapes with $N$. For the heaviest Rh isotopes the $S_{2n}$ values from the full and axial calculations become very similar again as the energy gain due to triaxiality is small.~We note that older global Skyrme EDF-based models that do not allow for triaxial deformation, such as those of Ref.~\cite{Goriely2016}, do not match the smooth trend of the experimental $S_{2n}$ as well as BSkG1. They typically predict one or more sudden transitions between prolate and oblate shapes between $N=70$ and $N=75$ with associated non-smooth features of the calculated two-neutron separation energies.


In order to highlight the changes in the evolution of the two-neutron separation energies as a function of $N$, Fig.~\ref{fig:S2N_BSkG1_JYFLTRAP}(b) shows a quantity often called the two-neutron shell gap:
\begin{align}
\delta_{2n}(Z,N) 
& = S_{2n}(Z,N) - S_{2n}(Z,N+2) \, ,
\end{align}

which quantifies the changes in slope of the $S_{2n}$ and filters out discontinuities. It is also often used to analyse the evolution of magicity in spherical nuclei \cite{Bender08a}, but can also indicate a change of shape. The experimental $\delta_{2n}$ values stay rather constant until $N=71$, implying a near-constant slope of the $S_{2n}$ caused by near-parabolic behaviour of the masses of odd and odd-odd Rh isotopes, respectively. From $N=71$ with added neutron numbers the $\delta_{2n}$ values have a decreasing trend signalling a change in the slope of the $S_{2n}$ values, or, equivalently, a discontinuity in the $N$-dependence of the masses. At $N = 73$, the slope decreases less with newly measured $^{120}$Rh $(N = 75)$ mass than compared to the AME20 based on an extrapolated mass value of $^{120}$Rh.
The $\delta_{2n}$ from the full BSkG1 model follow the experimental values up to $N=73$, as expected from the good reproduction of the experimental $S_{2n}$ values. Beyond, the BSkG1 values fall to even lower values of $\delta_{2n}$, reflecting the nearly flat trend of the $S_{2n}$ and signalling a structural change in the isotopic chain associated with the disappearance of triaxial deformation, see Sect.~\ref{sec:triaxiality}. The experimental data for $N=74$ and $N=75$ indicate a less dramatic change, but their precision is degraded by the large uncertainty on the masses of $^{121,122}$Rh. Of these, $^{121}$Rh is based on a storage-ring measurement \cite{Knoebel2016}, and only an extrapolated value for $^{122}$Rh is given in AME20 \cite{AME20}. To better constrain the trend with neutron number, masses of more exotic rhodium nuclides have to be measured to high precision.


The masses of odd-even and odd-odd Rh isotopes fall on two distinct curves that smoothly evolve with $N$ in a similar way, but that are separated by an energy gap. The distance, or the gap, between the two curves can be estimated by a three-point formula

\begin{equation}
\begin{aligned}
\Delta^{(3)}_{n}(Z,N) 
=& \frac{(-1)^{N+1}}{2} \,  \big[ BE(Z, N+1)\\ 
&-2BE(Z,N) + BE(Z, N-1) \big] \, .
\label{eq:Delta3}
\end{aligned}
\end{equation}
The gap is usually associated with the size of the neutron pairing gap, but is also affected by discontinuities such as shell closures and shape transitions~\cite{Bender00a,Duguet01a}. We compare the $\Delta^{(3)}_{n}$ values for the Rh chain obtained from our measurements to those derived from AME20 and to calculated values in Fig.~\ref{fig:Delta3N}(b). 
\footnote{
Note that the $\Delta^{(3)}_{n}$ values in Fig.~\ref{fig:Delta3N} exhibit an odd-even staggering, but this does not indicate a difference in neutron pairing correlations between isotopes with odd and even neutron number. 
Compared to finite difference formulas of higher order, a three-point formula does not perfectly eliminate mean-field, i.e. non-pairing, contributions to the evolution of the nuclear binding energy with $N$~\cite{Bender00a,Duguet01a}. We stick to Eq.~\eqref{eq:Delta3} however, as it allows us to extend the experimental curves Rh (Z = 45) on Fig.~\ref{fig:Delta3N} to $N=75$.} We have drastically reduced the uncertainties of the experimental $\Delta^{(3)}_n$ values of neutron-rich Rh isotopes in this work. Interestingly, the new $\Delta^{(3)}_{n}$ values do not agree with AME20 for many of the studied Rh isotopes. This is a consequence of the differences between the mass-excess values determined in this work and AME20 (see Table \ref{tab:results} and Fig.~\ref{fig:Rh_AME2020_Hager}). For instance $N=65$ and 67, the $\Delta^{(3)}_{n}$ value increases by slightly more than 100 keV, whereas for $N=73$ it decreases by a similar amount.

The amplitude of the odd-even staggering of the $\Delta^{(3)}_{n}$ is well reproduced by the BSkG1 model, while the absolute size of the $\Delta^{(3)}_{n}$ is not. It turns out that the model's systematic overestimation of the neutron pairing gaps in the Rh chain that is visible in Fig.~\ref{fig:Delta3N}(b) is a deficiency particular to isotopic chains of elements with odd proton number $Z$. For isotopic chains with even $Z$, the $\Delta^{(3)}_{n}$ values are in general reasonably well described in this region of the nuclear chart; we illustrate this for the adjacent Ru ($Z=44$) isotopes in the Fig.~\ref{fig:Delta3N}(a) but have also checked this for other isotopic chains. The experimental $\Delta^{(3)}_{n}$ values of the  Rh isotopes are systematically smaller than the values in the adjacent Pd and Ru isotopes with same $N$, while the calculated ones are of same size.

The differences in the size of calculated and experimental gaps appear systematically all across the chart of nuclei, which points to missing physics in the BSkG1 model. 
The effect in question is usually interpreted as an interaction between the unpaired proton and neutron in odd-odd nuclei that produces additional binding energy~\cite{Moller2016,Wu16a}. 
Roughly speaking, the offset between experimental $\Delta^{(3)}_{n}$ values of the Rh isotopes and those in the adjacent Pd and Ru chains is about 250 keV. 
If we subtract this estimate from the BSkG1 masses of the odd-odd Rh isotopes, the $\Delta_{n}^{(3)}$ values shift down along the entire chain. As an illustration 
we show the BSkG1 values for the Rh isotopes shifted this way for the Rh isotopes on Fig.~\ref{fig:Delta3N}(b): the shifted BSkG1 three-point gaps agree quite well with the experimental values. 

To further clarify the systematic difference between odd-$Z$ and even-$Z$ chains, we also show values for the three-point gaps for Ru isotopes in Fig.~\ref{fig:Delta3N}(a) and for Pd isotopes in Fig.~\ref{fig:Delta3N}(c).
In any event, this discussion implies that for nuclides with odd $Z$, the $\Delta_{n}^{(3)}$ values systematically contain a sizable contribution that has nothing to do with neutron pairing correlations. 
The same applies of course also to three-point formula for protons, $\Delta_{p}^{(3)}(Z)$, for nuclides with odd $N$.
The pattern of the calculated $\Delta^{(3)}_{n}(N)$ becomes irregular for Rh and Pd beyond $N=73$ while for the Ru beyond $N=74$. This can be attributed  to the same structural change that is at the origin of the decrease of the $\delta_{2n}$ in panel (b) of Fig.~\ref{fig:S2N_BSkG1_JYFLTRAP}. We discuss the change in the shape of the nuclides with $N$ in the next section.


\subsection{Evolution of quadrupole deformation}
\label{sec:triaxiality}
\begin{figure}[t]
\includegraphics[width=0.45\textwidth]{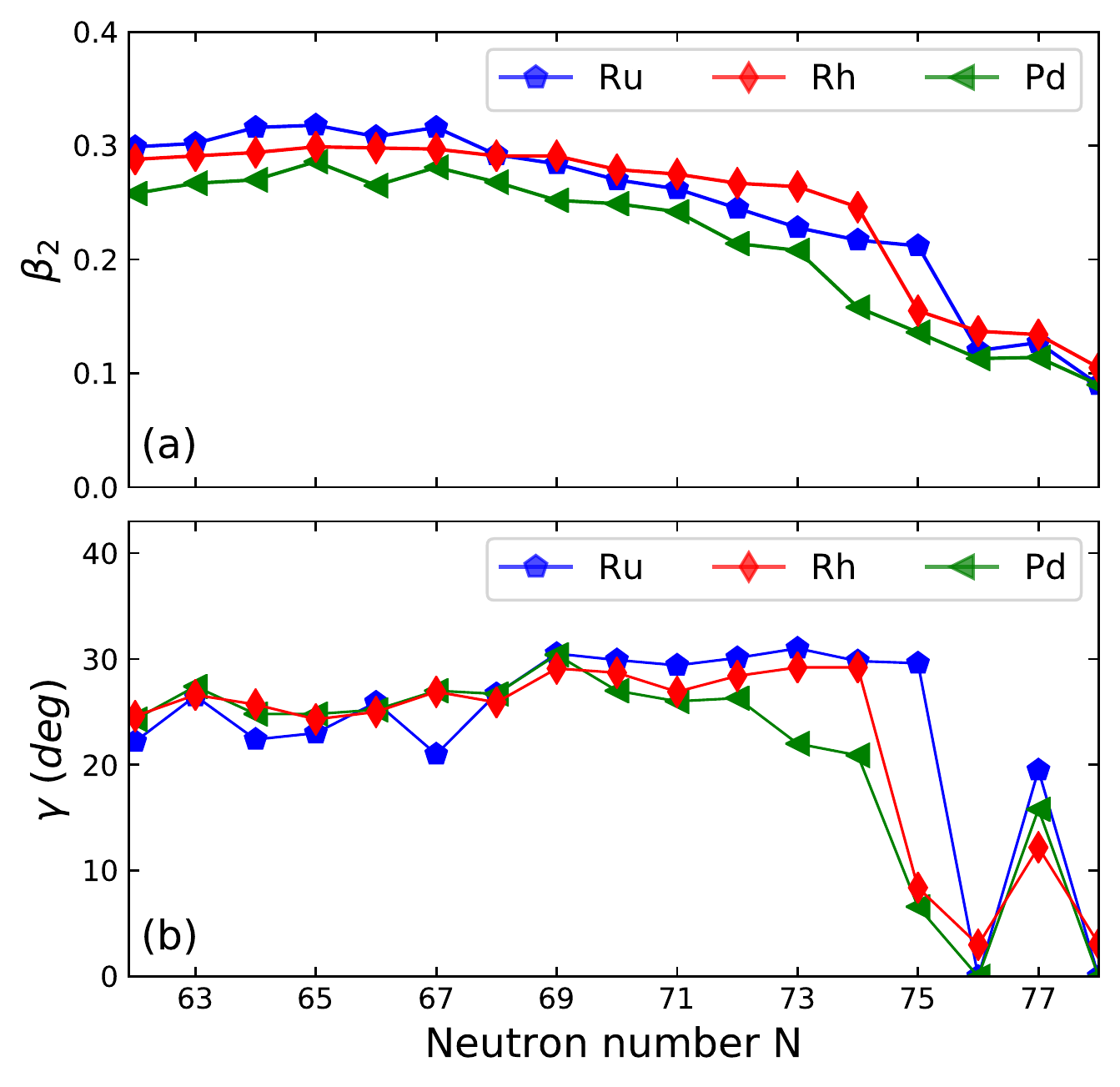}
\caption{ (a) Deformation parameter $\beta_2$ and (b) the triaxiality angle 
          $\gamma$ for neutron-rich ruthenium ($Z = 44)$, rhodium ($Z = 45)$ and 
          palladium ($Z = 46)$ isotopes as calculated from the BSkG1 model. 
          }          
\label{fig:triaxialityBSkG1}
\end{figure}

To better understand the experimental results and the role of deformation and triaxiality for the structure of neutron-rich Rh isotopes, in what follows we will analyze the evolution of 
the ground-state shape as predicted by BSkG1 for this region. To this end, we will discuss the quadrupole deformation of a nucleus of mass $A$ in terms of the dimensionless deformation $\beta_{2}$ and the triaxiality angle $\gamma$ as
\begin{align}
\beta_2 &= \frac{4 \pi}{ 3 R^2 A }\sqrt{Q_{20}^2 + 2 Q_{22}^2}\, , \\ 
\gamma &= \text{atan} \left( \sqrt{2} Q_{22}/Q_{20} \right)\, .
\end{align}
where $R = 1.2 A^{1/3}$ fm. The quadrupole moments $Q_{20}$ and $Q_{22}$ are defined in terms of integrals 
of the nuclear density distribution and spherical harmonics $Y_{2m}$ as $Q_{2 m} \equiv \int d^3 r \, r^2 \, \rho(\bold{r}) \, \Re \{ Y_{2 m}(\bold{r}) \}$
, where $\rho(\bold{r})$ is the nuclear matter density and $\Re \{Y_{2 m}(\bold{r)}\}$ is the real part of the relevant spherical harmonic, for $m=0$ and 2~\cite{Scamps2021}.
The quantity $\beta_2$ characterizes the total size of the quadrupole deformation and is always positive such that its sign cannot  be used to distinguish between prolate and oblate shapes. The triaxiality angle $\gamma$ can make the distinction: values of $0^{\circ}$ and $60^{\circ}$ correspond to an axially symmetric prolate or oblate shape, respectively. Values in between these extremes indicate triaxial shapes that no longer posses
a rotational symmetry axis. 

Figure~\ref{fig:triaxialityBSkG1}(a) shows the calculated $\beta_{2}$ and Fig.~\ref{fig:triaxialityBSkG1}(b) $\gamma$ values for the neutron-rich odd-$Z$ Rh isotopes, as well as for the
neighbouring Ru ($Z=44$) and Pd ($Z=46$) isotopic chains.  From $N=63$ up to $N=74$, Rh nuclei exhibit a fairly large deformation ($\beta_{2} \simeq 0.25-0.3$)
which evolves smoothly. For these neutron numbers, BSkG1 predicts $\gamma$-values close to $30^{\circ}$, i.e.\ the nuclear shapes are close to being maximally triaxial. 
For the heaviest Rh isotopes, the model predicts a return to axial shapes: beyond $N=74$, the calculated values for $\gamma$ drop significantly as does the total 
quadrupole deformation $\beta$. This change in deformation coincides with the reduction of the additional binding from triaxial shapes in Table~\ref{tab:thvsexp_absmass} that 
is responsible for different changes in Figs.~\ref{fig:S2N_BSkG1_JYFLTRAP} and \ref{fig:Delta3N} around $N=75$. More precisely, it leads to
(i) the flattening of the $S_{2n}$ in panel (a) of Fig.~\ref{fig:S2N_BSkG1_JYFLTRAP}, (ii) the drop in $\delta_{2n}$ in panel (b) of Fig.~\ref{fig:S2N_BSkG1_JYFLTRAP}, and 
(iii) the irregularities in 
the odd-even staggering of the $\Delta^{(3)}_{n}$ in Fig.~\ref{fig:Delta3N}.

The trends in even-$Z$ Ru and Pd chains follow the same general lines: strong deformation that evolves smoothly for the lighter isotopes, and smaller deformation and values of $\gamma$ for neutron-rich isotopes. The details differ between chains: the transition between both regimes happens one neutron later ($N=75$) in the Ru chain, while the calculated $\beta$-values in the Pd chain vary more smoothly than those in the other two chains. 

Direct experimental information on triaxial deformation in nuclear ground states
is scarce, especially for odd-mass and odd-odd nuclei. For even-even nuclei, 
information on quadrupole deformation can be extracted from experimental data
in a model-independent way through the use of rotational invariants~\cite{Kumar1972,Cline1986}. 
The authors of Ref.~\cite{Svensson1995}, the only such study in this region of 
the nuclear chart that we are aware of, report $\gamma$-values for the stable
$^{106, 108, 110}$Pd isotopes ($Z=46$, $N=60$, 62, 64) as $20^{+2}_{-2}, 19^{+4}_{-5}$
and $16^{+1}_{-1}$ degrees, respectively. These values are somewhat smaller than 
the calculated triaxiality angles for these nuclei \cite{Scamps2021}, but we consider this a satisfactory level of description for a global model. Coulomb excitation data for $^{104}$Ru ($Z=44$, $N=60$) \cite{Srebrny2006a} and $^{110}$Ru ($N=66$) \cite{Doherty2017} also point towards triaxiality of these two nuclides.

Also, as already mentioned in the introduction, the interpretation of the 
available information about rotational bands of neutron-rich Rh isotopes and 
nuclides in adjacent isotopic chains \cite{Liu2011,Liu2013,Navin2017,Fotiades2003,Zhang2015,Soderstrom2013,Luo2004,Hagen2018} 
consistently requires the assumption of triaxial shapes within the various approaches used to 
model them.

\subsection{More detailed look at $^{112}$Rh}
\label{sec:112rh-theor}

As illustrated in Table~\ref{tab:thvsexp_absmass} and Fig.~\ref{fig:S2N_BSkG1_JYFLTRAP}, the 
effect of triaxiality on the BSkG1 mass values for the Rh isotopes is large, and the largest effect is seen for $^{112}$Rh. The trend in the isomeric-state excitation energies (see Fig.~\ref{fig:RhExcitationEnergy}) has a minimum around the midshell at $N=66$, i.e. at $^{110}$Rh ($N=65$) and $^{112}$Rh ($N=67$). Thus, $^{112}$Rh is a special case among the studied Rh isotopes. In the following, we will discuss its structure in detail. 

\begin{figure}
\includegraphics[width=.48\textwidth]{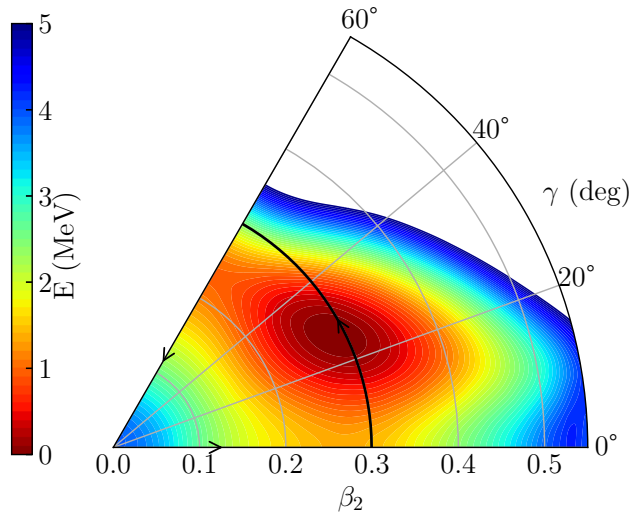}
\caption{Potential energy surface in the ($\beta$,$\gamma$)-plane for 
         false-vacuum calculations (see text) of $^{112}$Rh. The shape-trajectory
         followed by the Nilsson diagram in 
         Fig.~\ref{fig:nilssonRh112} is indicated by black arrows.
         }
\label{fig:surfaceRh112}
\end{figure}

To obtain a correct description of an odd-odd nucleus in a mean-field calculations, one has to construct a quasiparticle excitation for each nucleon species on top of a reference state that describes an even-even nucleus~\cite{RingSchuck}. 
Among a multitude of possible choices for such \textit{blocked} states, the combination leading to the lowest total energy varies with deformation in a discontinous way rendering it impossible to draw consistent potential energy surfaces (PES) for such nuclei.
Instead, one can forego the construction of quasiparticle excitations and perform a false-vacuum calculation, which fixes the average number of protons and neutrons to be odd but otherwise treats the nucleus as even-even~\cite{Duguet2001a,Duguet2001b}. We show the PES for such a calculation for $^{112}$Rh in Fig.~\ref{fig:surfaceRh112}.
While we obtain for this nucleus the largest difference between the triaxial minimum and the axially symmetric saddle among all Rh isotopes, the topography of the surface is representative for $^{108-119}$Rh. A peculiar aspect of the PES in this region is that the triaxial minimum is at a slightly larger total deformation $\beta_{2}$ than either of the prolate and oblate saddle-points.

\begin{figure*}
\includegraphics[width=\textwidth]{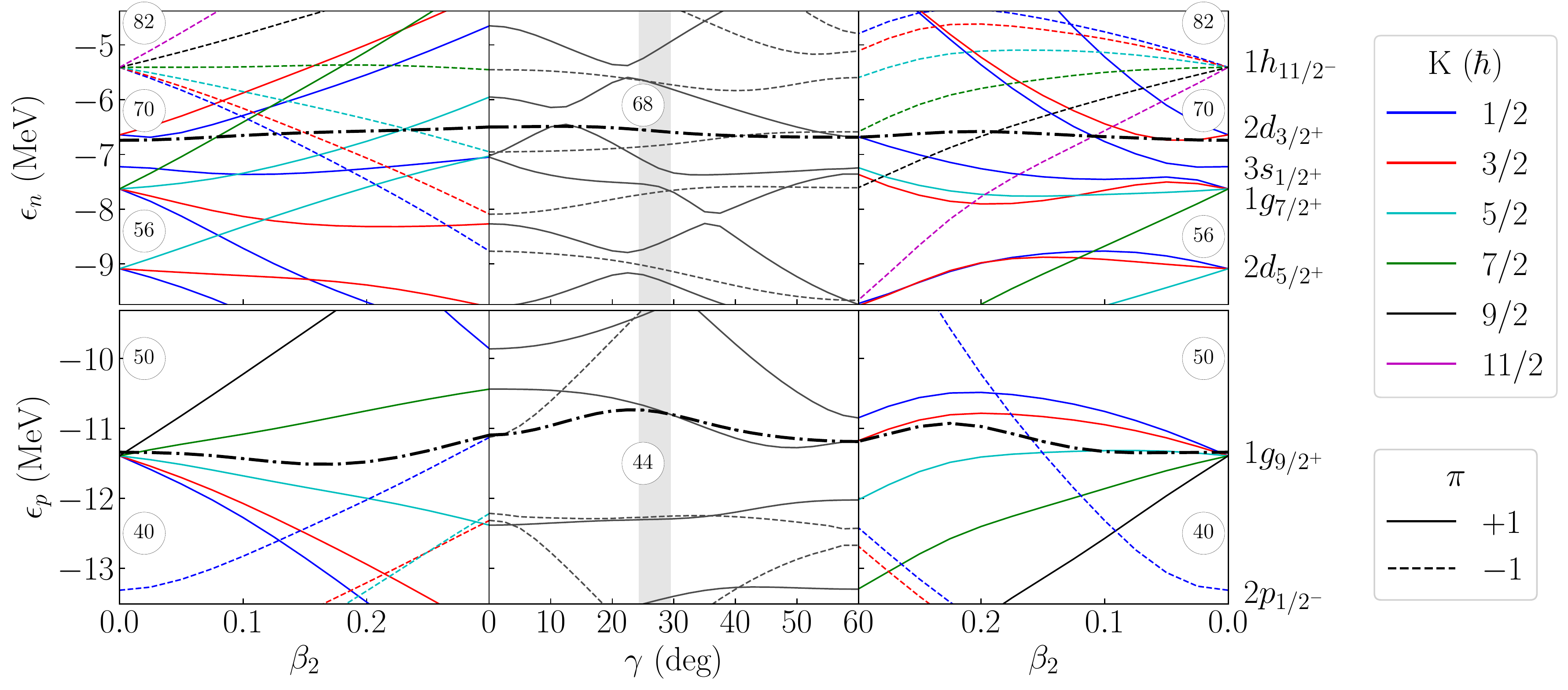}
\caption{(Color online) Nilsson diagram of the eigenvalues of the single-particle Hamiltonian for neutrons (top row) and protons (bottom row) along the path in the $\beta$-$\gamma$ plane indicated by arrows in Fig.~\ref{fig:surfaceRh112} for $^{112}$Rh (see text for details). The Fermi energy is drawn as a dash-dotted line, while full (dashed) lines indicate single-particle levels of positive (negative) parity. The three indicated regions correspond to axially symmetric prolate shape with $\gamma = 0^{\circ}$ (left column), fixed quadrupole deformation $\beta_2 = 0.3$ with varying $\gamma$ (center column) and axially oblate shape with $\gamma = 60^{\circ}$ (right column). The vertical gray band in the center panels is centered at $\gamma = 27^{\circ}$, the value obtained in a complete calculation of $^{112}$Rh. The quantum numbers of the shells at sphericity are indicated on the right-hand-side. }
\label{fig:nilssonRh112}
\end{figure*}

The appearance of triaxial deformation in this region can be understood 
qualitatively by inspecting the single-particle spectra as a function of
deformation. Figure~\ref{fig:nilssonRh112} shows the eigenvalues of 
the single-particle Hamiltonian of neutrons (top row) and protons (bottom 
row)  for $^{112}$Rh ($Z=45$, $N=67$)
along the path indicated by black arrows in Fig.~\ref{fig:surfaceRh112}. 
The left column explores the variation of the single-particle spectrum 
as a function of quadrupole deformation $\beta$ at $\gamma = 0^{\circ}$, 
i.e.\ for axially symmetric prolate shapes. The right-most column shows the same for 
oblate shapes with $\gamma = 60^{\circ}$, while the center column is 
drawn for fixed quadrupole deformation $\beta_2 = 0.3$ while varying $\gamma$.
For axially symmetric shapes it is possible to assign the single-particle 
states the quantum number $K$ according to the projection of their angular
momentum on the symmetry axis. This quantum number is indicated by colors
in the left and right columns of Fig.~\ref{fig:nilssonRh112}, but cannot
be used in the center column where configurations are not axially symmetric.

The origin of the drive towards triaxial shapes can be found in the proton single-particle spectrum. According to Strutinsky's theorem~\cite{Strutinsky1967,Brack1972}, local minima (maxima) in deformation energy surfaces correspond to regions where the bunching of single-particle levels around the Fermi energy is very low (high). At spherical shape, the proton Fermi energy is in the middle of the highly degenerate $1g_{9/2^+}$ shell, leading to large level density. Through prolate deformation, the substates of the spherical proton $1g_{9/2^+}$ spread out strongly and produce a very sparse spectrum with large gaps among the positive-parity states. The upsloping negative-parity states spoil these gaps though, particularly the $K^{\pi} = 1/2^-$ level coming from the spherical $2p_{1/2^-}$ shell. However, this state curves up strongly for increasing values of $\gamma$, such that triaxial deformation allows the nucleus to decrease the level density near the Fermi energy significantly, opening up a gap at $Z=44$. Similarly large gaps open up as well for $Z=42$
and $Z=46$, explaining the preference for non-axial deformation in this region. If a given nucleus takes a triaxial shape, however, depends also on the neutron spectrum at these deformations.

For the mid-shell nucleus $^{112}$Rh, the overall density of neutron levels around the Fermi energy is much larger than the one of protons. The former is also not visibly decreased when going from spherical to prolate or oblate shapes with $\beta_{2} \simeq 0.3$. Still, for $N=68$ and some other adjacent neutron numbers, modest single-particle gaps open up for finite values of $\gamma$. When increasing the neutron number however, the neutron Fermi energy rises and encounters larger shell gaps at smaller prolate deformations. Sufficiently close to $N=82$, this effect trumps the preference of the protons for triaxial deformation. 

The staggering of the triaxiality angle at $N=77$ in Fig.~\ref{fig:triaxialityBSkG1} can be understood in a similar way. The mean-field minimum for $ N \geq 76$  is axially symmetric for all three isotopic chains, as neutrons close to the shell closure strongly prefer axially symmetric shapes. Nevertheless, the potential energy surface of a false-vacuum calculation at $N=77$ remains sufficiently soft with respect to $\gamma$ such that the creation of a neutron quasiparticle excitation is sufficient to break the symmetry for $^{121}$Ru and $^{123}$Pd. As the level density of neutrons is much larger than that of protons, the polarising effect of a blocked neutron is larger than that of a blocked protons; the latter cannot generate triaxial deformation in $^{121}$Rh.

Without symmetry restoration techniques, the symmetry-broken mean-field calculations we report on here cannot produce definite spin assignments for ground or excited states. For even-$N$ Rh isotopes however, we can make tentative assignments due to the sparsity of the single-particle proton spectrum. The picture that emerges is consistent with experimental spectroscopic information: a positive parity proton state, linked to the $K^{\pi}=7/2^+$ state on the prolate side, is located close to the Fermi energy in the center of Fig.~\ref{fig:nilssonRh112}. This matches well the $J^{\pi}=7/2^+$ ground-states of the $^{105,107,109}$Rh isotopes
\footnote{The ground states of the neutron-rich even-$N$ Rh isotopes with  $111 \leq A \leq 125$ have also been tentatively assigned $J^{\rm \pi} = 7/2^+$~\cite{Kondev2021}.}.
Without triaxial deformation, theory would not reproduce the spin-parities: on the prolate side the $K^{\pi}=1/2^-$ state suggests a negative-parity ground state, whereas an oblate shape would result in a state with a lower spin. These considerations seem to be independent of the type of EDF that is employed: calculations with the Gogny D1S parameterization lead to a similar single-particle spectrum for the protons~\cite{Bucher2018}.

For odd-$N$ Rh isotopes however, extracting spin-parity assignments for the ground and isomeric states is particularly difficult due to the large number of states near the Fermi energy that are all $K$-mixed. 
Based on the (often tentative) experimental information on spin-parities in this mass region, we can however offer some observations. For both $^{110}$Rh and $^{112}$Rh, the ground and isomeric states have been assigned $(1)^+$ and $(6)^+$, i.e.\ states of identical parity that differ by a large amount of angular momentum. In the strong-coupling limit, these assignments can be naturally explained by combining the angular momentum of a $K^{\pi}=7/2^+$ proton and a $K^{\pi}=5/2^+$ neutron in two different ways: a parallel (anti-parallel) coupling results in a high (low) spin. Such spin assignment for the odd nucleons would make the ground and isomeric state a so-called Gallagher-Moszkowski pair~\cite{Gallagher1958}.
The appearance of triaxial deformation stops us from relying on the $K$-quantum number, but we observe that the central panel of the top row of Fig.~\ref{fig:nilssonRh112} shows two neutron single-particle states near the Fermi energy
that link to $K^{\pi}=7/2^+$ states on the prolate side. Although the negative parity neutron state linked to the $K^{\pi}=7/2^-$ state on the prolate side could lead to a negative parity ground state in a symmetry-restored calculation, the BSkG1 Nilsson scheme in Fig.~\ref{fig:nilssonRh112} is thus not incompatible with this scenario for the ground and isomeric states in $^{110,112}$Rh.

Assuming that the isomeric states in $^{110,112}$Rh are the Gallagher-Moszkowski partners of their respective ground states also offers a possible, albeit tentative, interpretation for the extremely low excitation energy of their isomers of $^{110,112}$Rh.
For other nuclei in this mass region, the excitation energy of 
spin-isomers is often of the order of 100 keV, see for example 
Fig.~\ref{fig:RhExcitationEnergy}. This splitting is generally 
interpreted as being the result of the relative 
orientation of the intrinsic spins
of the odd neutron ($\boldsymbol{s}_n$) and odd proton ($\boldsymbol{s}_p$): depending on the relative orientation of spin, the proton-neutron spin-spin interaction is either attractive or repulsive,  leading to a small difference in total binding energy. The state with lowest energy has spins that are parallel ($\boldsymbol{s}_n \cdot \boldsymbol{s}_p > 0$) while the excited partner has anti-parallel spins  ($\boldsymbol{s}_n \cdot \boldsymbol{s}_p < 0$). 
The situation of $^{110}$Rh and $^{112}$Rh is somewhat peculiar due to the presence of \textit{two} positive parity neutron levels with similar
angular momenta near the Fermi energy that are candidates for the observed
states. These two single-particle states differ, at least for axial prolate shapes, in their relative orientation between their orbital ($\boldsymbol{\ell}_n$) and spin angular momentum ($\boldsymbol{s}_n$), although they have roughly similar total angular momentum ($\boldsymbol{j}_n = \boldsymbol{\ell}_n + \boldsymbol{s}_n $).
The neutron level originating from the spherical $1g_{7/2^+}$ has both mostly anti-parallel ($ \boldsymbol{\ell}_n \cdot \boldsymbol{s}_n \leq 0$), while the one connected to the $2d_{5/2^+}$ shell has both mostly parallel  ($ \boldsymbol{\ell}_n \cdot \boldsymbol{s}_n \geq 0$).  
For triaxial shapes these two single-particle states mix, and in the many-body wave 
functions of the two partner states the quasiparticle configurations in which either is 
blocked might also be mixed. This could lead to a situation where the 
spins of the odd nucleons are almost perpendicular 
($\boldsymbol{s}_n \cdot \boldsymbol{s}_p \approx 0$), resulting in a very
small Gallagher-Moszkowski splitting and hence a very low excitation 
energy of the isomers in $^{110}$Rh and $^{112}$Rh. This interpretation also naturally explains why such low-lying isomers are only observed only for a small  number (two) of the Rh isotopes; as $N$ increases (decreases), the neutron Fermi energy moves up (down) in Fig.~\ref{fig:nilssonRh112}, further away from the $K^{\pi} = 7/2^+$ neutron state with highest (lowest) energy and removing the possibility of this mixing. Although we cannot provide definite predictions, it is natural to expect a larger, i.e. normal, Gallagher-Moszkowski splitting for neutron numbers sufficiently far from $N=65,67$.

We emphasize this discussion is no substitute for more advanced many-body calculations capable of constructing spectra with the associated quantum numbers for these nuclei. Performing such calculations with predictive power based on nuclear EDFs might however not be possible in the near future. On top of the technical challenges inherent in employing symmetry-restoration techniques for odd-odd nuclei, it has been argued that virtually all existing EDF-based models are not suited to describe the attraction between the odd nucleons in such nuclei~\cite{Robledo2014}. 



\section{Conclusions}
We have determined the ground- and isomeric-state masses of $^{110,112,114,116,118}$Rh and determined the isomeric state excitation energies accurately for the first time. Also the mass of $^{120}$Rh was measured for the first time. The new ground-state mass values revealed deviations up to around 100 keV from the adopted mass values of $^{110,112,118}$Rh in AME20 \cite{AME20}. We have unambiguously determined the masses and half-lives of the ground and isomeric states in $^{112}$Rh and confirm that the low-spin $1^+$ state is the ground state. The new ground-state half-life was found to be more than $10\sigma$ shorter than the value reported in the ENSDF evaluation \cite{ensdf112}, but to agree with the result reported in Ref.~\cite{Jokinen1991} which had taken into account $^{112}$Ru contamination. 
The experimental results have been compared to the results of the global BSkG1 mass model \cite{Scamps2021} that was adjusted allowing nuclei to take triaxial shapes. The trends of mass excesses and two-neutron separation energies of the Rh isotopes are very well reproduced with the full BSkG1 model, whereas limiting the variational space to axial shapes leads to substantial deviations from the experimental values. This result underlines the important role of triaxiality in the region of studied Rh isotopes. 
The predictions of BSkG1 for the potential energy surface as well as the 
 deformation-dependence of the single-particle energies have been studied in detail for $^{112}$Rh, the nucleus for which the effect of triaxiality is largest in the BSkG1
 model. The results indicate that the proton shell effects drive the Rh nuclides to triaxial 
 shapes, and that neutron shell effects moderate for which isotopes this happens. A drastic change in the deformation is predicted to take place at $N=75$ ($^{120}$Rh), which is imprinted on several mass differences. In the future, mass measurements of the more exotic Rh isotopes are needed to explore if this effect can be seen as a change in the slope of the two-neutron separation energies beyond $N=75$.

\begin{acknowledgments}
The authors would like to thank L. Bonneau for the fruitful discussions. The present research benefited from computational resources made available on the Tier-1 supercomputer of the F\'ed\'eration Wallonie-Bruxelles, infrastructure funded by the Walloon Region under the grant agreement No 1117545. W.R. acknowledges financial support from the FNRS (Belgium). Work by M.B.\ has been supported by the Agence Nationale de la Recherche, France, Grant No.~19-CE31-0015-01 (NEWFUN). Funding from the European Union’s Horizon 2020 research and innovation programme under grant agreement No 771036 (ERC CoG MAIDEN) is gratefully acknowledged. M.H. acknowledges financial support from the Ellen \& Artturi Nyyss\"onen foundation. We are grateful for the mobility support from Projet International de Coop\'eration Scientifique Manipulation of Ions in Traps and Ion sourCes for Atomic and Nuclear Spectroscopy (MITICANS) of CNRS. T.E. and A.R. acknowledge support from the Academy of Finland project No. 295207, 306980 and 327629.
\end{acknowledgments}

\pagebreak
\bibliographystyle{apsrev}
\bibliography{main_submitted_v2}

\begin{thebibliography}{95}
\expandafter\ifx\csname natexlab\endcsname\relax\def\natexlab#1{#1}\fi
\expandafter\ifx\csname bibnamefont\endcsname\relax
  \def\bibnamefont#1{#1}\fi
\expandafter\ifx\csname bibfnamefont\endcsname\relax
  \def\bibfnamefont#1{#1}\fi
\expandafter\ifx\csname citenamefont\endcsname\relax
  \def\citenamefont#1{#1}\fi
\expandafter\ifx\csname url\endcsname\relax
  \def\url#1{\texttt{#1}}\fi
\expandafter\ifx\csname urlprefix\endcsname\relax\def\urlprefix{URL }\fi
\providecommand{\bibinfo}[2]{#2}
\providecommand{\eprint}[2][]{\url{#2}}

\bibitem[{\citenamefont{Heyde and Wood}(2011)}]{Heyde2011}
\bibinfo{author}{\bibfnamefont{K.}~\bibnamefont{Heyde}} \bibnamefont{and}
  \bibinfo{author}{\bibfnamefont{J.~L.} \bibnamefont{Wood}},
  \bibinfo{journal}{Rev. Mod. Phys.} \textbf{\bibinfo{volume}{83}},
  \bibinfo{pages}{1467} (\bibinfo{year}{2011}),
  \urlprefix\url{https://doi.org/10.1103/RevModPhys.83.1467}.

\bibitem[{\citenamefont{{Rahaman, S.} et~al.}(2007)\citenamefont{{Rahaman, S.},
  {Hager, U.}, {Elomaa, V. -V.}, {Eronen, T.}, {Hakala, J.}, {Jokinen, A.},
  {Kankainen, A.}, {Karvonen, P.}, {Moore, I. D.}, {Penttil\"a, H.}
  et~al.}}]{Rahaman2007}
\bibinfo{author}{\bibnamefont{{Rahaman, S.}}},
  \bibinfo{author}{\bibnamefont{{Hager, U.}}},
  \bibinfo{author}{\bibnamefont{{Elomaa, V. -V.}}},
  \bibinfo{author}{\bibnamefont{{Eronen, T.}}},
  \bibinfo{author}{\bibnamefont{{Hakala, J.}}},
  \bibinfo{author}{\bibnamefont{{Jokinen, A.}}},
  \bibinfo{author}{\bibnamefont{{Kankainen, A.}}},
  \bibinfo{author}{\bibnamefont{{Karvonen, P.}}},
  \bibinfo{author}{\bibnamefont{{Moore, I. D.}}},
  \bibinfo{author}{\bibnamefont{{Penttil\"a, H.}}}, \bibnamefont{et~al.},
  \bibinfo{journal}{Eur. Phys. J. A} \textbf{\bibinfo{volume}{32}},
  \bibinfo{pages}{87} (\bibinfo{year}{2007}),
  \urlprefix\url{https://doi.org/10.1140/epja/i2006-10297-y}.

\bibitem[{\citenamefont{Hager et~al.}(2006)\citenamefont{Hager, Eronen, Hakala,
  Jokinen, Kolhinen, Kopecky, Moore, Nieminen, Oinonen, Rinta-Antila
  et~al.}}]{Hager2006}
\bibinfo{author}{\bibfnamefont{U.}~\bibnamefont{Hager}},
  \bibinfo{author}{\bibfnamefont{T.}~\bibnamefont{Eronen}},
  \bibinfo{author}{\bibfnamefont{J.}~\bibnamefont{Hakala}},
  \bibinfo{author}{\bibfnamefont{A.}~\bibnamefont{Jokinen}},
  \bibinfo{author}{\bibfnamefont{V.}~\bibnamefont{Kolhinen}},
  \bibinfo{author}{\bibfnamefont{S.}~\bibnamefont{Kopecky}},
  \bibinfo{author}{\bibfnamefont{I.}~\bibnamefont{Moore}},
  \bibinfo{author}{\bibfnamefont{A.}~\bibnamefont{Nieminen}},
  \bibinfo{author}{\bibfnamefont{M.}~\bibnamefont{Oinonen}},
  \bibinfo{author}{\bibfnamefont{S.}~\bibnamefont{Rinta-Antila}},
  \bibnamefont{et~al.}, \bibinfo{journal}{Phys. Rev. Lett.}
  \textbf{\bibinfo{volume}{96}}, \bibinfo{pages}{042504}
  (\bibinfo{year}{2006}),
  \urlprefix\url{https://doi.org/10.1103/PhysRevLett.96.042504}.

\bibitem[{\citenamefont{Hager et~al.}(2007{\natexlab{a}})\citenamefont{Hager,
  Jokinen, Elomaa, Eronen, Hakala, Kankainen, Rahaman, Rissanen, Moore,
  Rinta-Antila et~al.}}]{Hager2007a}
\bibinfo{author}{\bibfnamefont{U.}~\bibnamefont{Hager}},
  \bibinfo{author}{\bibfnamefont{A.}~\bibnamefont{Jokinen}},
  \bibinfo{author}{\bibfnamefont{V.-V.} \bibnamefont{Elomaa}},
  \bibinfo{author}{\bibfnamefont{T.}~\bibnamefont{Eronen}},
  \bibinfo{author}{\bibfnamefont{J.}~\bibnamefont{Hakala}},
  \bibinfo{author}{\bibfnamefont{A.}~\bibnamefont{Kankainen}},
  \bibinfo{author}{\bibfnamefont{S.}~\bibnamefont{Rahaman}},
  \bibinfo{author}{\bibfnamefont{J.}~\bibnamefont{Rissanen}},
  \bibinfo{author}{\bibfnamefont{I.}~\bibnamefont{Moore}},
  \bibinfo{author}{\bibfnamefont{S.}~\bibnamefont{Rinta-Antila}},
  \bibnamefont{et~al.}, \bibinfo{journal}{Nucl. Phys. A}
  \textbf{\bibinfo{volume}{793}}, \bibinfo{pages}{20}
  (\bibinfo{year}{2007}{\natexlab{a}}), ISSN \bibinfo{issn}{0375-9474},
  \urlprefix\url{https://www.sciencedirect.com/science/article/pii/S0375947407006070}.

\bibitem[{\citenamefont{Hakala et~al.}(2011)\citenamefont{Hakala,
  Rodriguez-Guzman, Elomaa, Eronen, Jokinen, Kolhinen, Moore, Penttil\"a,
  Reponen, Rissanen et~al.}}]{Hakala2011}
\bibinfo{author}{\bibfnamefont{J.}~\bibnamefont{Hakala}},
  \bibinfo{author}{\bibfnamefont{R.}~\bibnamefont{Rodriguez-Guzman}},
  \bibinfo{author}{\bibfnamefont{V.~V.} \bibnamefont{Elomaa}},
  \bibinfo{author}{\bibfnamefont{T.}~\bibnamefont{Eronen}},
  \bibinfo{author}{\bibfnamefont{A.}~\bibnamefont{Jokinen}},
  \bibinfo{author}{\bibfnamefont{V.~S.} \bibnamefont{Kolhinen}},
  \bibinfo{author}{\bibfnamefont{I.~D.} \bibnamefont{Moore}},
  \bibinfo{author}{\bibfnamefont{H.}~\bibnamefont{Penttil\"a}},
  \bibinfo{author}{\bibfnamefont{M.}~\bibnamefont{Reponen}},
  \bibinfo{author}{\bibfnamefont{J.}~\bibnamefont{Rissanen}},
  \bibnamefont{et~al.}, \bibinfo{journal}{Eur. Phys. J. A}
  \textbf{\bibinfo{volume}{47}}, \bibinfo{pages}{129} (\bibinfo{year}{2011}),
  \urlprefix\url{https://doi.org/10.1140/epja/i2011-11129-9}.

\bibitem[{\citenamefont{Thibault et~al.}(1981)\citenamefont{Thibault, Touchard,
  B\"uttgenbach, Klapisch, de~Saint~Simon, Duong, Jacquinot, Juncar, Liberman,
  Pillet et~al.}}]{Thibault1981}
\bibinfo{author}{\bibfnamefont{C.}~\bibnamefont{Thibault}},
  \bibinfo{author}{\bibfnamefont{F.}~\bibnamefont{Touchard}},
  \bibinfo{author}{\bibfnamefont{S.}~\bibnamefont{B\"uttgenbach}},
  \bibinfo{author}{\bibfnamefont{R.}~\bibnamefont{Klapisch}},
  \bibinfo{author}{\bibfnamefont{M.}~\bibnamefont{de~Saint~Simon}},
  \bibinfo{author}{\bibfnamefont{H.~T.} \bibnamefont{Duong}},
  \bibinfo{author}{\bibfnamefont{P.}~\bibnamefont{Jacquinot}},
  \bibinfo{author}{\bibfnamefont{P.}~\bibnamefont{Juncar}},
  \bibinfo{author}{\bibfnamefont{S.}~\bibnamefont{Liberman}},
  \bibinfo{author}{\bibfnamefont{P.}~\bibnamefont{Pillet}},
  \bibnamefont{et~al.}, \bibinfo{journal}{Phys. Rev. C}
  \textbf{\bibinfo{volume}{23}}, \bibinfo{pages}{2720} (\bibinfo{year}{1981}),
  \urlprefix\url{https://doi.org/10.1103/PhysRevC.23.2720}.

\bibitem[{\citenamefont{Lievens et~al.}(1996)\citenamefont{Lievens, Arnold,
  Borchers, Georg, Keim, Klein, Neugart, Vermeeren, and
  Silverans}}]{Lievens1996}
\bibinfo{author}{\bibfnamefont{P.}~\bibnamefont{Lievens}},
  \bibinfo{author}{\bibfnamefont{E.}~\bibnamefont{Arnold}},
  \bibinfo{author}{\bibfnamefont{W.}~\bibnamefont{Borchers}},
  \bibinfo{author}{\bibfnamefont{U.}~\bibnamefont{Georg}},
  \bibinfo{author}{\bibfnamefont{M.}~\bibnamefont{Keim}},
  \bibinfo{author}{\bibfnamefont{A.}~\bibnamefont{Klein}},
  \bibinfo{author}{\bibfnamefont{R.}~\bibnamefont{Neugart}},
  \bibinfo{author}{\bibfnamefont{L.}~\bibnamefont{Vermeeren}},
  \bibnamefont{and} \bibinfo{author}{\bibfnamefont{R.~E.}
  \bibnamefont{Silverans}}, \bibinfo{journal}{Europhysics Letters ({EPL})}
  \textbf{\bibinfo{volume}{33}}, \bibinfo{pages}{11} (\bibinfo{year}{1996}),
  \urlprefix\url{https://doi.org/10.1209/epl/i1996-00296-0}.

\bibitem[{\citenamefont{Cheal et~al.}(2007)\citenamefont{Cheal, Gardner,
  Avgoulea, Billowes, Bissell, Campbell, Eronen, Flanagan, Forest, Huikari
  et~al.}}]{Cheal2007}
\bibinfo{author}{\bibfnamefont{B.}~\bibnamefont{Cheal}},
  \bibinfo{author}{\bibfnamefont{M.}~\bibnamefont{Gardner}},
  \bibinfo{author}{\bibfnamefont{M.}~\bibnamefont{Avgoulea}},
  \bibinfo{author}{\bibfnamefont{J.}~\bibnamefont{Billowes}},
  \bibinfo{author}{\bibfnamefont{M.}~\bibnamefont{Bissell}},
  \bibinfo{author}{\bibfnamefont{P.}~\bibnamefont{Campbell}},
  \bibinfo{author}{\bibfnamefont{T.}~\bibnamefont{Eronen}},
  \bibinfo{author}{\bibfnamefont{K.}~\bibnamefont{Flanagan}},
  \bibinfo{author}{\bibfnamefont{D.}~\bibnamefont{Forest}},
  \bibinfo{author}{\bibfnamefont{J.}~\bibnamefont{Huikari}},
  \bibnamefont{et~al.}, \bibinfo{journal}{Physics Letters B}
  \textbf{\bibinfo{volume}{645}}, \bibinfo{pages}{133} (\bibinfo{year}{2007}),
  ISSN \bibinfo{issn}{0370-2693},
  \urlprefix\url{https://www.sciencedirect.com/science/article/pii/S037026930601608X}.

\bibitem[{\citenamefont{Campbell et~al.}(2002)\citenamefont{Campbell, Thayer,
  Billowes, Dendooven, Flanagan, Forest, Griffith, Huikari, Jokinen, Moore
  et~al.}}]{Campbell2002}
\bibinfo{author}{\bibfnamefont{P.}~\bibnamefont{Campbell}},
  \bibinfo{author}{\bibfnamefont{H.~L.} \bibnamefont{Thayer}},
  \bibinfo{author}{\bibfnamefont{J.}~\bibnamefont{Billowes}},
  \bibinfo{author}{\bibfnamefont{P.}~\bibnamefont{Dendooven}},
  \bibinfo{author}{\bibfnamefont{K.~T.} \bibnamefont{Flanagan}},
  \bibinfo{author}{\bibfnamefont{D.~H.} \bibnamefont{Forest}},
  \bibinfo{author}{\bibfnamefont{J.~A.~R.} \bibnamefont{Griffith}},
  \bibinfo{author}{\bibfnamefont{J.}~\bibnamefont{Huikari}},
  \bibinfo{author}{\bibfnamefont{A.}~\bibnamefont{Jokinen}},
  \bibinfo{author}{\bibfnamefont{R.}~\bibnamefont{Moore}},
  \bibnamefont{et~al.}, \bibinfo{journal}{Phys. Rev. Lett.}
  \textbf{\bibinfo{volume}{89}}, \bibinfo{pages}{082501}
  (\bibinfo{year}{2002}),
  \urlprefix\url{https://link.aps.org/doi/10.1103/PhysRevLett.89.082501}.

\bibitem[{\citenamefont{Cheal et~al.}(2009)\citenamefont{Cheal, Baczynska,
  Billowes, Campbell, Charlwood, Eronen, Forest, Jokinen, Kessler, Moore
  et~al.}}]{Cheal2009}
\bibinfo{author}{\bibfnamefont{B.}~\bibnamefont{Cheal}},
  \bibinfo{author}{\bibfnamefont{K.}~\bibnamefont{Baczynska}},
  \bibinfo{author}{\bibfnamefont{J.}~\bibnamefont{Billowes}},
  \bibinfo{author}{\bibfnamefont{P.}~\bibnamefont{Campbell}},
  \bibinfo{author}{\bibfnamefont{F.~C.} \bibnamefont{Charlwood}},
  \bibinfo{author}{\bibfnamefont{T.}~\bibnamefont{Eronen}},
  \bibinfo{author}{\bibfnamefont{D.~H.} \bibnamefont{Forest}},
  \bibinfo{author}{\bibfnamefont{A.}~\bibnamefont{Jokinen}},
  \bibinfo{author}{\bibfnamefont{T.}~\bibnamefont{Kessler}},
  \bibinfo{author}{\bibfnamefont{I.~D.} \bibnamefont{Moore}},
  \bibnamefont{et~al.}, \bibinfo{journal}{Phys. Rev. Lett.}
  \textbf{\bibinfo{volume}{102}}, \bibinfo{pages}{222501}
  (\bibinfo{year}{2009}),
  \urlprefix\url{https://link.aps.org/doi/10.1103/PhysRevLett.102.222501}.

\bibitem[{\citenamefont{Charlwood et~al.}(2009)\citenamefont{Charlwood,
  Baczynska, Billowes, Campbell, Cheal, Eronen, Forest, Jokinen, Kessler, Moore
  et~al.}}]{Charlwood2009}
\bibinfo{author}{\bibfnamefont{F.}~\bibnamefont{Charlwood}},
  \bibinfo{author}{\bibfnamefont{K.}~\bibnamefont{Baczynska}},
  \bibinfo{author}{\bibfnamefont{J.}~\bibnamefont{Billowes}},
  \bibinfo{author}{\bibfnamefont{P.}~\bibnamefont{Campbell}},
  \bibinfo{author}{\bibfnamefont{B.}~\bibnamefont{Cheal}},
  \bibinfo{author}{\bibfnamefont{T.}~\bibnamefont{Eronen}},
  \bibinfo{author}{\bibfnamefont{D.}~\bibnamefont{Forest}},
  \bibinfo{author}{\bibfnamefont{A.}~\bibnamefont{Jokinen}},
  \bibinfo{author}{\bibfnamefont{T.}~\bibnamefont{Kessler}},
  \bibinfo{author}{\bibfnamefont{I.}~\bibnamefont{Moore}},
  \bibnamefont{et~al.}, \bibinfo{journal}{Phys. Lett. B}
  \textbf{\bibinfo{volume}{674}}, \bibinfo{pages}{23} (\bibinfo{year}{2009}),
  ISSN \bibinfo{issn}{0370-2693},
  \urlprefix\url{https://www.sciencedirect.com/science/article/pii/S0370269309002470}.

\bibitem[{\citenamefont{Svensson et~al.}(1995)\citenamefont{Svensson,
  Fahlander, Hasselgren, B, Westerberg, Cline, Czosnyka, Wu, Diamond, and
  Kluge}}]{Svensson1995}
\bibinfo{author}{\bibfnamefont{L.~E.} \bibnamefont{Svensson}},
  \bibinfo{author}{\bibfnamefont{C.}~\bibnamefont{Fahlander}},
  \bibinfo{author}{\bibfnamefont{L.}~\bibnamefont{Hasselgren}},
  \bibinfo{author}{\bibfnamefont{A.}~\bibnamefont{B}},
  \bibinfo{author}{\bibfnamefont{L.}~\bibnamefont{Westerberg}},
  \bibinfo{author}{\bibfnamefont{D.}~\bibnamefont{Cline}},
  \bibinfo{author}{\bibfnamefont{T.}~\bibnamefont{Czosnyka}},
  \bibinfo{author}{\bibfnamefont{C.~Y.} \bibnamefont{Wu}},
  \bibinfo{author}{\bibfnamefont{R.~M.} \bibnamefont{Diamond}},
  \bibnamefont{and} \bibinfo{author}{\bibfnamefont{H.}~\bibnamefont{Kluge}},
  \bibinfo{journal}{Nucl. Phys. A} \textbf{\bibinfo{volume}{584}},
  \bibinfo{pages}{547} (\bibinfo{year}{1995}).

\bibitem[{\citenamefont{Srebrny et~al.}(2006)\citenamefont{Srebrny, Czosnyka,
  Droste, Rohozi{\'n}ski, Pr{\'o}chniak, Zajac, Pomorski, Cline, Wu,
  B{\"a}cklin et~al.}}]{Srebrny2006a}
\bibinfo{author}{\bibfnamefont{J.}~\bibnamefont{Srebrny}},
  \bibinfo{author}{\bibfnamefont{T.}~\bibnamefont{Czosnyka}},
  \bibinfo{author}{\bibfnamefont{C.}~\bibnamefont{Droste}},
  \bibinfo{author}{\bibfnamefont{S.}~\bibnamefont{Rohozi{\'n}ski}},
  \bibinfo{author}{\bibfnamefont{L.}~\bibnamefont{Pr{\'o}chniak}},
  \bibinfo{author}{\bibfnamefont{K.}~\bibnamefont{Zajac}},
  \bibinfo{author}{\bibfnamefont{K.}~\bibnamefont{Pomorski}},
  \bibinfo{author}{\bibfnamefont{D.}~\bibnamefont{Cline}},
  \bibinfo{author}{\bibfnamefont{C.~Y.} \bibnamefont{Wu}},
  \bibinfo{author}{\bibfnamefont{A.}~\bibnamefont{B{\"a}cklin}},
  \bibnamefont{et~al.}, \bibinfo{journal}{Nuclear Physics A}
  \textbf{\bibinfo{volume}{766}}, \bibinfo{pages}{25} (\bibinfo{year}{2006}),
  ISSN \bibinfo{issn}{0375-9474},
  \urlprefix\url{https://www.sciencedirect.com/science/article/pii/S0375947405012029}.

\bibitem[{\citenamefont{Doherty et~al.}(2017)\citenamefont{Doherty, Allmond,
  Janssens, Korten, Zhu, Zieli{\'{n}}ska, Radford, Ayangeakaa, Bucher,
  Batchelder et~al.}}]{Doherty2017}
\bibinfo{author}{\bibfnamefont{D.}~\bibnamefont{Doherty}},
  \bibinfo{author}{\bibfnamefont{J.}~\bibnamefont{Allmond}},
  \bibinfo{author}{\bibfnamefont{R.}~\bibnamefont{Janssens}},
  \bibinfo{author}{\bibfnamefont{W.}~\bibnamefont{Korten}},
  \bibinfo{author}{\bibfnamefont{S.}~\bibnamefont{Zhu}},
  \bibinfo{author}{\bibfnamefont{M.}~\bibnamefont{Zieli{\'{n}}ska}},
  \bibinfo{author}{\bibfnamefont{D.}~\bibnamefont{Radford}},
  \bibinfo{author}{\bibfnamefont{A.}~\bibnamefont{Ayangeakaa}},
  \bibinfo{author}{\bibfnamefont{B.}~\bibnamefont{Bucher}},
  \bibinfo{author}{\bibfnamefont{J.}~\bibnamefont{Batchelder}},
  \bibnamefont{et~al.}, \bibinfo{journal}{Phys. Lett. B}
  \textbf{\bibinfo{volume}{766}}, \bibinfo{pages}{334} (\bibinfo{year}{2017}),
  ISSN \bibinfo{issn}{03702693},
  \urlprefix\url{https://linkinghub.elsevier.com/retrieve/pii/S0370269317300497}.

\bibitem[{\citenamefont{Liu et~al.}(2011)\citenamefont{Liu, Hamilton, Ramayya,
  Chen, Gao, Zhu, Gu, Yeoh, Brewer, Hwang et~al.}}]{Liu2011}
\bibinfo{author}{\bibfnamefont{S.~H.} \bibnamefont{Liu}},
  \bibinfo{author}{\bibfnamefont{J.~H.} \bibnamefont{Hamilton}},
  \bibinfo{author}{\bibfnamefont{A.~V.} \bibnamefont{Ramayya}},
  \bibinfo{author}{\bibfnamefont{Y.~S.} \bibnamefont{Chen}},
  \bibinfo{author}{\bibfnamefont{Z.~C.} \bibnamefont{Gao}},
  \bibinfo{author}{\bibfnamefont{S.~J.} \bibnamefont{Zhu}},
  \bibinfo{author}{\bibfnamefont{L.}~\bibnamefont{Gu}},
  \bibinfo{author}{\bibfnamefont{E.~Y.} \bibnamefont{Yeoh}},
  \bibinfo{author}{\bibfnamefont{N.~T.} \bibnamefont{Brewer}},
  \bibinfo{author}{\bibfnamefont{J.~K.} \bibnamefont{Hwang}},
  \bibnamefont{et~al.}, \bibinfo{journal}{Phys. Rev. C}
  \textbf{\bibinfo{volume}{83}}, \bibinfo{pages}{064310}
  (\bibinfo{year}{2011}),
  \urlprefix\url{https://link.aps.org/doi/10.1103/PhysRevC.83.064310}.

\bibitem[{\citenamefont{Navin et~al.}(2017)\citenamefont{Navin, Rejmund,
  Bhattacharyya, Palit, Bhat, Sheikh, Lemasson, Caama\~{n}o, Cl\'ement, Delaune
  et~al.}}]{Navin2017}
\bibinfo{author}{\bibfnamefont{A.}~\bibnamefont{Navin}},
  \bibinfo{author}{\bibfnamefont{M.}~\bibnamefont{Rejmund}},
  \bibinfo{author}{\bibfnamefont{S.}~\bibnamefont{Bhattacharyya}},
  \bibinfo{author}{\bibfnamefont{R.}~\bibnamefont{Palit}},
  \bibinfo{author}{\bibfnamefont{G.}~\bibnamefont{Bhat}},
  \bibinfo{author}{\bibfnamefont{J.~A.} \bibnamefont{Sheikh}},
  \bibinfo{author}{\bibfnamefont{A.}~\bibnamefont{Lemasson}},
  \bibinfo{author}{\bibfnamefont{M.}~\bibnamefont{Caama\~{n}o}},
  \bibinfo{author}{\bibfnamefont{E.}~\bibnamefont{Cl\'ement}},
  \bibinfo{author}{\bibfnamefont{O.}~\bibnamefont{Delaune}},
  \bibnamefont{et~al.}, \bibinfo{journal}{Phys. Lett. B}
  \textbf{\bibinfo{volume}{767}}, \bibinfo{pages}{480} (\bibinfo{year}{2017}),
  \urlprefix\url{https://doi.org/10.1016/j.physletb.2016.11.020}.

\bibitem[{\citenamefont{Zhang et~al.}(2015)\citenamefont{Zhang, Bhat,
  Nazarewicz, Sheikh, and Shi}}]{Zhang2015}
\bibinfo{author}{\bibfnamefont{C.~L.} \bibnamefont{Zhang}},
  \bibinfo{author}{\bibfnamefont{G.~H.} \bibnamefont{Bhat}},
  \bibinfo{author}{\bibfnamefont{W.}~\bibnamefont{Nazarewicz}},
  \bibinfo{author}{\bibfnamefont{J.~A.} \bibnamefont{Sheikh}},
  \bibnamefont{and} \bibinfo{author}{\bibfnamefont{Y.}~\bibnamefont{Shi}},
  \bibinfo{journal}{Phys. Rev. C} \textbf{\bibinfo{volume}{92}},
  \bibinfo{pages}{034307} (\bibinfo{year}{2015}), ISSN
  \bibinfo{issn}{0556-2813}.

\bibitem[{\citenamefont{Abusara et~al.}(2017)\citenamefont{Abusara, Ahmad, and
  Othman}}]{Abusara2017}
\bibinfo{author}{\bibfnamefont{H.}~\bibnamefont{Abusara}},
  \bibinfo{author}{\bibfnamefont{S.}~\bibnamefont{Ahmad}}, \bibnamefont{and}
  \bibinfo{author}{\bibfnamefont{S.}~\bibnamefont{Othman}},
  \bibinfo{journal}{Physical Review C} \textbf{\bibinfo{volume}{95}},
  \bibinfo{pages}{054302} (\bibinfo{year}{2017}), ISSN
  \bibinfo{issn}{2469-9985, 2469-9993},
  \urlprefix\url{http://link.aps.org/doi/10.1103/PhysRevC.95.054302}.

\bibitem[{\citenamefont{Bucher et~al.}(2018)\citenamefont{Bucher, MacH,
  Aprahamian, Robledo, Simpson, Rissanen, Ghi\c{t}\v{a}, Olaizola, Kurcewicz,
  {\"{A}}yst{\"{o}} et~al.}}]{Bucher2018}
\bibinfo{author}{\bibfnamefont{B.}~\bibnamefont{Bucher}},
  \bibinfo{author}{\bibfnamefont{H.}~\bibnamefont{MacH}},
  \bibinfo{author}{\bibfnamefont{A.}~\bibnamefont{Aprahamian}},
  \bibinfo{author}{\bibfnamefont{L.~M.} \bibnamefont{Robledo}},
  \bibinfo{author}{\bibfnamefont{G.~S.} \bibnamefont{Simpson}},
  \bibinfo{author}{\bibfnamefont{J.}~\bibnamefont{Rissanen}},
  \bibinfo{author}{\bibfnamefont{D.~G.} \bibnamefont{Ghi\c{t}\v{a}}},
  \bibinfo{author}{\bibfnamefont{B.}~\bibnamefont{Olaizola}},
  \bibinfo{author}{\bibfnamefont{W.}~\bibnamefont{Kurcewicz}},
  \bibinfo{author}{\bibfnamefont{J.}~\bibnamefont{{\"{A}}yst{\"{o}}}},
  \bibnamefont{et~al.}, \bibinfo{journal}{Phys. Rev. C}
  \textbf{\bibinfo{volume}{98}}, \bibinfo{pages}{064320}
  (\bibinfo{year}{2018}), ISSN \bibinfo{issn}{24699993}.

\bibitem[{\citenamefont{Goriely et~al.}(2009)\citenamefont{Goriely, Hilaire,
  Girod, and P{\'{e}}ru}}]{Goriely2009}
\bibinfo{author}{\bibfnamefont{S.}~\bibnamefont{Goriely}},
  \bibinfo{author}{\bibfnamefont{S.}~\bibnamefont{Hilaire}},
  \bibinfo{author}{\bibfnamefont{M.}~\bibnamefont{Girod}}, \bibnamefont{and}
  \bibinfo{author}{\bibfnamefont{S.}~\bibnamefont{P{\'{e}}ru}},
  \bibinfo{journal}{Phys. Rev. Lett.} \textbf{\bibinfo{volume}{102}},
  \bibinfo{pages}{242501} (\bibinfo{year}{2009}), ISSN
  \bibinfo{issn}{0031-9007}.

\bibitem[{\citenamefont{M{\"{o}}ller et~al.}(2016)\citenamefont{M{\"{o}}ller,
  Sierk, Ichikawa, and Sagawa}}]{Moller2016}
\bibinfo{author}{\bibfnamefont{P.}~\bibnamefont{M{\"{o}}ller}},
  \bibinfo{author}{\bibfnamefont{A.}~\bibnamefont{Sierk}},
  \bibinfo{author}{\bibfnamefont{T.}~\bibnamefont{Ichikawa}}, \bibnamefont{and}
  \bibinfo{author}{\bibfnamefont{H.}~\bibnamefont{Sagawa}},
  \bibinfo{journal}{At. Data Nucl. Data Tables}
  \textbf{\bibinfo{volume}{109-110}} (\bibinfo{year}{2016}), ISSN
  \bibinfo{issn}{0092640X}.

\bibitem[{\citenamefont{Scamps et~al.}(2021)\citenamefont{Scamps, Goriely,
  Olsen, Bender, and Ryssens}}]{Scamps2021}
\bibinfo{author}{\bibfnamefont{G.}~\bibnamefont{Scamps}},
  \bibinfo{author}{\bibfnamefont{S.}~\bibnamefont{Goriely}},
  \bibinfo{author}{\bibfnamefont{E.}~\bibnamefont{Olsen}},
  \bibinfo{author}{\bibfnamefont{M.}~\bibnamefont{Bender}}, \bibnamefont{and}
  \bibinfo{author}{\bibfnamefont{W.}~\bibnamefont{Ryssens}},
  \bibinfo{journal}{Eur. Phys. J. A} \textbf{\bibinfo{volume}{57}},
  \bibinfo{pages}{333} (\bibinfo{year}{2021}),
  \urlprefix\url{https://doi.org/10.1140/epja/s10050-021-00642-1}.

\bibitem[{\citenamefont{Kondev et~al.}(2021)\citenamefont{Kondev, Wang, Huang,
  Naimi, and Audi}}]{Kondev2021}
\bibinfo{author}{\bibfnamefont{F.}~\bibnamefont{Kondev}},
  \bibinfo{author}{\bibfnamefont{M.}~\bibnamefont{Wang}},
  \bibinfo{author}{\bibfnamefont{W.}~\bibnamefont{Huang}},
  \bibinfo{author}{\bibfnamefont{S.}~\bibnamefont{Naimi}}, \bibnamefont{and}
  \bibinfo{author}{\bibfnamefont{G.}~\bibnamefont{Audi}},
  \bibinfo{journal}{Chin. Phys. C} \textbf{\bibinfo{volume}{45}},
  \bibinfo{pages}{030001} (\bibinfo{year}{2021}),
  \urlprefix\url{https://doi.org/10.1088/1674-1137/abddae}.

\bibitem[{\citenamefont{Kolhinen}(2003)}]{Kolhinen2003}
\bibinfo{author}{\bibfnamefont{V.}~\bibnamefont{Kolhinen}}, Ph.D. thesis,
  \bibinfo{school}{University of Jyv\"askyl\"a} (\bibinfo{year}{2003}),
  \urlprefix\url{http://urn.fi/URN:ISBN:978-951-39-3141-4}.

\bibitem[{\citenamefont{Hager et~al.}(2007{\natexlab{b}})\citenamefont{Hager,
  Elomaa, Eronen, Hakala, Jokinen, Kankainen, Rahaman, Rinta-Antila,
  Saastamoinen, Sonoda et~al.}}]{Hager2007b}
\bibinfo{author}{\bibfnamefont{U.}~\bibnamefont{Hager}},
  \bibinfo{author}{\bibfnamefont{V.-V.} \bibnamefont{Elomaa}},
  \bibinfo{author}{\bibfnamefont{T.}~\bibnamefont{Eronen}},
  \bibinfo{author}{\bibfnamefont{J.}~\bibnamefont{Hakala}},
  \bibinfo{author}{\bibfnamefont{A.}~\bibnamefont{Jokinen}},
  \bibinfo{author}{\bibfnamefont{A.}~\bibnamefont{Kankainen}},
  \bibinfo{author}{\bibfnamefont{S.}~\bibnamefont{Rahaman}},
  \bibinfo{author}{\bibfnamefont{S.}~\bibnamefont{Rinta-Antila}},
  \bibinfo{author}{\bibfnamefont{A.}~\bibnamefont{Saastamoinen}},
  \bibinfo{author}{\bibfnamefont{T.}~\bibnamefont{Sonoda}},
  \bibnamefont{et~al.}, \bibinfo{journal}{Phys. Rev. C}
  \textbf{\bibinfo{volume}{75}}, \bibinfo{pages}{064302}
  (\bibinfo{year}{2007}{\natexlab{b}}),
  \urlprefix\url{https://link.aps.org/doi/10.1103/PhysRevC.75.064302}.

\bibitem[{\citenamefont{Eronen et~al.}(2012)\citenamefont{Eronen, Kolhinen,
  Elomaa, Gorelov, Hager, Hakala, Jokinen, Kankainen, Karvonen, Kopecky
  et~al.}}]{Eronen2012}
\bibinfo{author}{\bibfnamefont{T.}~\bibnamefont{Eronen}},
  \bibinfo{author}{\bibfnamefont{V.}~\bibnamefont{Kolhinen}},
  \bibinfo{author}{\bibfnamefont{V.-V.} \bibnamefont{Elomaa}},
  \bibinfo{author}{\bibfnamefont{D.}~\bibnamefont{Gorelov}},
  \bibinfo{author}{\bibfnamefont{U.}~\bibnamefont{Hager}},
  \bibinfo{author}{\bibfnamefont{J.}~\bibnamefont{Hakala}},
  \bibinfo{author}{\bibfnamefont{A.}~\bibnamefont{Jokinen}},
  \bibinfo{author}{\bibfnamefont{A.}~\bibnamefont{Kankainen}},
  \bibinfo{author}{\bibfnamefont{P.}~\bibnamefont{Karvonen}},
  \bibinfo{author}{\bibfnamefont{S.}~\bibnamefont{Kopecky}},
  \bibnamefont{et~al.}, \bibinfo{journal}{Eur. Phys. J. A}
  \textbf{\bibinfo{volume}{48}}, \bibinfo{pages}{46} (\bibinfo{year}{2012}),
  \urlprefix\url{https://doi.org/10.1140/epja/i2012-12046-1}.

\bibitem[{\citenamefont{Jokinen et~al.}(1991)\citenamefont{Jokinen, \"Ayst\"o,
  Dendooven, Eskola, Janas, Jauho, Leino, Parmonen, Penttil\"a, Rykaczewski
  et~al.}}]{Jokinen1991}
\bibinfo{author}{\bibfnamefont{A.}~\bibnamefont{Jokinen}},
  \bibinfo{author}{\bibfnamefont{J.}~\bibnamefont{\"Ayst\"o}},
  \bibinfo{author}{\bibfnamefont{P.}~\bibnamefont{Dendooven}},
  \bibinfo{author}{\bibfnamefont{K.}~\bibnamefont{Eskola}},
  \bibinfo{author}{\bibfnamefont{Z.}~\bibnamefont{Janas}},
  \bibinfo{author}{\bibfnamefont{P.~P.} \bibnamefont{Jauho}},
  \bibinfo{author}{\bibfnamefont{M.~E.} \bibnamefont{Leino}},
  \bibinfo{author}{\bibfnamefont{J.~M.} \bibnamefont{Parmonen}},
  \bibinfo{author}{\bibfnamefont{H.}~\bibnamefont{Penttil\"a}},
  \bibinfo{author}{\bibfnamefont{K.}~\bibnamefont{Rykaczewski}},
  \bibnamefont{et~al.}, \bibinfo{journal}{Z. Phys. A}
  \textbf{\bibinfo{volume}{340}}, \bibinfo{pages}{21 } (\bibinfo{year}{1991}),
  \urlprefix\url{https://doi.org/10.1007/BF01284476}.

\bibitem[{\citenamefont{Jokinen et~al.}(1992)\citenamefont{Jokinen,
  \"Ayst{\"o}, Jauho, Leino, Parmonen, Penttil\"a, Eskola, and
  Janas}}]{Jokinen1992}
\bibinfo{author}{\bibfnamefont{A.}~\bibnamefont{Jokinen}},
  \bibinfo{author}{\bibfnamefont{J.}~\bibnamefont{\"Ayst{\"o}}},
  \bibinfo{author}{\bibfnamefont{P.}~\bibnamefont{Jauho}},
  \bibinfo{author}{\bibfnamefont{M.}~\bibnamefont{Leino}},
  \bibinfo{author}{\bibfnamefont{J.}~\bibnamefont{Parmonen}},
  \bibinfo{author}{\bibfnamefont{H.}~\bibnamefont{Penttil\"a}},
  \bibinfo{author}{\bibfnamefont{K.}~\bibnamefont{Eskola}}, \bibnamefont{and}
  \bibinfo{author}{\bibfnamefont{Z.}~\bibnamefont{Janas}},
  \bibinfo{journal}{Nuclear Physics A} \textbf{\bibinfo{volume}{549}},
  \bibinfo{pages}{420} (\bibinfo{year}{1992}), ISSN \bibinfo{issn}{0375-9474},
  \urlprefix\url{https://www.sciencedirect.com/science/article/pii/0375947492900882}.

\bibitem[{\citenamefont{\"Ayst{\"o} et~al.}(1988)\citenamefont{\"Ayst{\"o},
  Davids, Hattula, Honkanen, Honkanen, Jauho, Julin, Juutinen, Kumpulainen,
  L{\"o}nnroth et~al.}}]{Aysto1988}
\bibinfo{author}{\bibfnamefont{J.}~\bibnamefont{\"Ayst{\"o}}},
  \bibinfo{author}{\bibfnamefont{C.}~\bibnamefont{Davids}},
  \bibinfo{author}{\bibfnamefont{J.}~\bibnamefont{Hattula}},
  \bibinfo{author}{\bibfnamefont{J.}~\bibnamefont{Honkanen}},
  \bibinfo{author}{\bibfnamefont{K.}~\bibnamefont{Honkanen}},
  \bibinfo{author}{\bibfnamefont{P.}~\bibnamefont{Jauho}},
  \bibinfo{author}{\bibfnamefont{R.}~\bibnamefont{Julin}},
  \bibinfo{author}{\bibfnamefont{S.}~\bibnamefont{Juutinen}},
  \bibinfo{author}{\bibfnamefont{J.}~\bibnamefont{Kumpulainen}},
  \bibinfo{author}{\bibfnamefont{T.}~\bibnamefont{L{\"o}nnroth}},
  \bibnamefont{et~al.}, \bibinfo{journal}{Nucl. Phys. A}
  \textbf{\bibinfo{volume}{480}}, \bibinfo{pages}{104} (\bibinfo{year}{1988}),
  \urlprefix\url{https://doi.org/10.1016/0375-9474(88)90387-9}.

\bibitem[{\citenamefont{Lhersonneau et~al.}(1999)\citenamefont{Lhersonneau,
  Wang, Hankonen, Dendooven, Jones, Julin, and \"Ayst\"o}}]{Lhersonneau1999}
\bibinfo{author}{\bibfnamefont{G.}~\bibnamefont{Lhersonneau}},
  \bibinfo{author}{\bibfnamefont{J.~C.} \bibnamefont{Wang}},
  \bibinfo{author}{\bibfnamefont{S.}~\bibnamefont{Hankonen}},
  \bibinfo{author}{\bibfnamefont{P.}~\bibnamefont{Dendooven}},
  \bibinfo{author}{\bibfnamefont{P.}~\bibnamefont{Jones}},
  \bibinfo{author}{\bibfnamefont{R.}~\bibnamefont{Julin}}, \bibnamefont{and}
  \bibinfo{author}{\bibfnamefont{J.}~\bibnamefont{\"Ayst\"o}},
  \bibinfo{journal}{Phys. Rev. C} \textbf{\bibinfo{volume}{60}},
  \bibinfo{pages}{014315} (\bibinfo{year}{1999}),
  \urlprefix\url{https://link.aps.org/doi/10.1103/PhysRevC.60.014315}.

\bibitem[{\citenamefont{Lhersonneau et~al.}(2003)\citenamefont{Lhersonneau,
  Wang, Capote, Suhonen, Dendooven, Huikari, Per\"aj\"arvi, and
  Wang}}]{Lhersonneau2003}
\bibinfo{author}{\bibfnamefont{G.}~\bibnamefont{Lhersonneau}},
  \bibinfo{author}{\bibfnamefont{Y.}~\bibnamefont{Wang}},
  \bibinfo{author}{\bibfnamefont{R.}~\bibnamefont{Capote}},
  \bibinfo{author}{\bibfnamefont{J.}~\bibnamefont{Suhonen}},
  \bibinfo{author}{\bibfnamefont{P.}~\bibnamefont{Dendooven}},
  \bibinfo{author}{\bibfnamefont{J.}~\bibnamefont{Huikari}},
  \bibinfo{author}{\bibfnamefont{K.}~\bibnamefont{Per\"aj\"arvi}},
  \bibnamefont{and} \bibinfo{author}{\bibfnamefont{J.~C.} \bibnamefont{Wang}},
  \bibinfo{journal}{Phys. Rev. C} \textbf{\bibinfo{volume}{67}},
  \bibinfo{pages}{024303} (\bibinfo{year}{2003}),
  \urlprefix\url{https://link.aps.org/doi/10.1103/PhysRevC.67.024303}.

\bibitem[{\citenamefont{Wang et~al.}(2001)\citenamefont{Wang, Dendooven,
  Huikari, Jokinen, Kolhinen, Lhersonneau, Nieminen, Nummela, Penttil\"a,
  Per\"aj\"arvi et~al.}}]{Wang2001}
\bibinfo{author}{\bibfnamefont{Y.}~\bibnamefont{Wang}},
  \bibinfo{author}{\bibfnamefont{P.}~\bibnamefont{Dendooven}},
  \bibinfo{author}{\bibfnamefont{J.}~\bibnamefont{Huikari}},
  \bibinfo{author}{\bibfnamefont{A.}~\bibnamefont{Jokinen}},
  \bibinfo{author}{\bibfnamefont{V.~S.} \bibnamefont{Kolhinen}},
  \bibinfo{author}{\bibfnamefont{G.}~\bibnamefont{Lhersonneau}},
  \bibinfo{author}{\bibfnamefont{A.}~\bibnamefont{Nieminen}},
  \bibinfo{author}{\bibfnamefont{S.}~\bibnamefont{Nummela}},
  \bibinfo{author}{\bibfnamefont{H.}~\bibnamefont{Penttil\"a}},
  \bibinfo{author}{\bibfnamefont{K.}~\bibnamefont{Per\"aj\"arvi}},
  \bibnamefont{et~al.}, \bibinfo{journal}{Phys. Rev. C}
  \textbf{\bibinfo{volume}{63}}, \bibinfo{pages}{024309}
  (\bibinfo{year}{2001}),
  \urlprefix\url{https://link.aps.org/doi/10.1103/PhysRevC.63.024309}.

\bibitem[{\citenamefont{Jokinen et~al.}(2000)\citenamefont{Jokinen, Wang,
  \"Ayst\"o, Dendooven, Nummela, Huikari, Kolhinen, Nieminen, Per\"aj\"arvi,
  and Rinta-Antila}}]{Jokinen2000}
\bibinfo{author}{\bibfnamefont{A.}~\bibnamefont{Jokinen}},
  \bibinfo{author}{\bibfnamefont{J.}~\bibnamefont{Wang}},
  \bibinfo{author}{\bibfnamefont{J.}~\bibnamefont{\"Ayst\"o}},
  \bibinfo{author}{\bibfnamefont{P.}~\bibnamefont{Dendooven}},
  \bibinfo{author}{\bibfnamefont{S.}~\bibnamefont{Nummela}},
  \bibinfo{author}{\bibfnamefont{J.}~\bibnamefont{Huikari}},
  \bibinfo{author}{\bibfnamefont{V.}~\bibnamefont{Kolhinen}},
  \bibinfo{author}{\bibfnamefont{A.}~\bibnamefont{Nieminen}},
  \bibinfo{author}{\bibfnamefont{K.}~\bibnamefont{Per\"aj\"arvi}},
  \bibnamefont{and}
  \bibinfo{author}{\bibfnamefont{S.}~\bibnamefont{Rinta-Antila}},
  \bibinfo{journal}{Eur. Phys. J. A} \textbf{\bibinfo{volume}{9}},
  \bibinfo{pages}{9 } (\bibinfo{year}{2000}),
  \urlprefix\url{https://doi.org/10.1007/s100500070049}.

\bibitem[{\citenamefont{{Wang You-Bao, P. Dendooven, J. Huikari, A. Jokinen, V.
  S. Kolhinen, G. Lhersonneau, A. Nieminen, S. Nummela, H. Penttil{\"{a}}, K.
  Per{\"{a}}j{\"{a}}rvi, S. Rinta-Antila, J. Szerypo} and
  {\"{A}}yst{\"{o}}}(2006)}]{Wang2006}
\bibinfo{author}{\bibfnamefont{J.~C.~W.} \bibnamefont{{Wang You-Bao, P.
  Dendooven, J. Huikari, A. Jokinen, V. S. Kolhinen, G. Lhersonneau, A.
  Nieminen, S. Nummela, H. Penttil{\"{a}}, K. Per{\"{a}}j{\"{a}}rvi, S.
  Rinta-Antila, J. Szerypo}}} \bibnamefont{and}
  \bibinfo{author}{\bibfnamefont{J.}~\bibnamefont{{\"{A}}yst{\"{o}}}},
  \bibinfo{journal}{Chinese Phys. Lett.} \textbf{\bibinfo{volume}{23}},
  \bibinfo{pages}{808} (\bibinfo{year}{2006}).

\bibitem[{\citenamefont{G\"urdal and Kondev}(2012)}]{Gurdal2012}
\bibinfo{author}{\bibfnamefont{G.}~\bibnamefont{G\"urdal}} \bibnamefont{and}
  \bibinfo{author}{\bibfnamefont{F.}~\bibnamefont{Kondev}},
  \bibinfo{journal}{Nuclear Data Sheets} \textbf{\bibinfo{volume}{113}},
  \bibinfo{pages}{1315} (\bibinfo{year}{2012}), ISSN \bibinfo{issn}{0090-3752},
  \urlprefix\url{https://www.sciencedirect.com/science/article/pii/S0090375212000415}.

\bibitem[{\citenamefont{Eliseev et~al.}(2014)\citenamefont{Eliseev, Blaum,
  Block, D{\"o}rr, Droese, Eronen, Goncharov, H{\"o}cker, Ketter,
  Minaya~Ramirez et~al.}}]{Eliseev2014}
\bibinfo{author}{\bibfnamefont{S.}~\bibnamefont{Eliseev}},
  \bibinfo{author}{\bibfnamefont{K.}~\bibnamefont{Blaum}},
  \bibinfo{author}{\bibfnamefont{M.}~\bibnamefont{Block}},
  \bibinfo{author}{\bibfnamefont{A.}~\bibnamefont{D{\"o}rr}},
  \bibinfo{author}{\bibfnamefont{C.}~\bibnamefont{Droese}},
  \bibinfo{author}{\bibfnamefont{T.}~\bibnamefont{Eronen}},
  \bibinfo{author}{\bibfnamefont{M.}~\bibnamefont{Goncharov}},
  \bibinfo{author}{\bibfnamefont{M.}~\bibnamefont{H{\"o}cker}},
  \bibinfo{author}{\bibfnamefont{J.}~\bibnamefont{Ketter}},
  \bibinfo{author}{\bibfnamefont{E.}~\bibnamefont{Minaya~Ramirez}},
  \bibnamefont{et~al.}, \bibinfo{journal}{Appl. Phys. B}
  \textbf{\bibinfo{volume}{114}}, \bibinfo{pages}{107} (\bibinfo{year}{2014}),
  \urlprefix\url{https://doi.org/10.1007/s00340-013-5621-0}.

\bibitem[{\citenamefont{Nesterenko et~al.}(2018)\citenamefont{Nesterenko,
  Eronen, Kankainen, Canete, Jokinen, Moore, Penttil\"a, Rinta-Antila,
  de~Roubin, and Vilen}}]{Nesterenko2018}
\bibinfo{author}{\bibfnamefont{D.~A.} \bibnamefont{Nesterenko}},
  \bibinfo{author}{\bibfnamefont{T.}~\bibnamefont{Eronen}},
  \bibinfo{author}{\bibfnamefont{A.}~\bibnamefont{Kankainen}},
  \bibinfo{author}{\bibfnamefont{L.}~\bibnamefont{Canete}},
  \bibinfo{author}{\bibfnamefont{A.}~\bibnamefont{Jokinen}},
  \bibinfo{author}{\bibfnamefont{I.~D.} \bibnamefont{Moore}},
  \bibinfo{author}{\bibfnamefont{H.}~\bibnamefont{Penttil\"a}},
  \bibinfo{author}{\bibfnamefont{S.}~\bibnamefont{Rinta-Antila}},
  \bibinfo{author}{\bibfnamefont{A.}~\bibnamefont{de~Roubin}},
  \bibnamefont{and} \bibinfo{author}{\bibfnamefont{M.}~\bibnamefont{Vilen}},
  \bibinfo{journal}{Eur. Phys. J. A.} \textbf{\bibinfo{volume}{54}},
  \bibinfo{pages}{154} (\bibinfo{year}{2018}),
  \urlprefix\url{https://doi.org/10.1140/epja/i2018-12589-y}.

\bibitem[{\citenamefont{Nesterenko et~al.}(2021)\citenamefont{Nesterenko,
  Eronen, Ge, Kankainen, and Vil\'en}}]{Nesterenko2021}
\bibinfo{author}{\bibfnamefont{D.}~\bibnamefont{Nesterenko}},
  \bibinfo{author}{\bibfnamefont{T.}~\bibnamefont{Eronen}},
  \bibinfo{author}{\bibfnamefont{Z.}~\bibnamefont{Ge}},
  \bibinfo{author}{\bibfnamefont{A.}~\bibnamefont{Kankainen}},
  \bibnamefont{and} \bibinfo{author}{\bibfnamefont{M.}~\bibnamefont{Vil\'en}},
  \bibinfo{journal}{Eur. Phys. Jour. A} \textbf{\bibinfo{volume}{57}},
  \bibinfo{pages}{302} (\bibinfo{year}{2021}),
  \urlprefix\url{https://doi.org/10.1140/epja/s10050-021-00608-3}.

\bibitem[{\citenamefont{Goriely et~al.}(2016)\citenamefont{Goriely, Chamel, and
  Pearson}}]{Goriely2016}
\bibinfo{author}{\bibfnamefont{S.}~\bibnamefont{Goriely}},
  \bibinfo{author}{\bibfnamefont{N.}~\bibnamefont{Chamel}}, \bibnamefont{and}
  \bibinfo{author}{\bibfnamefont{J.~M.} \bibnamefont{Pearson}},
  \bibinfo{journal}{Phys. Rev. C} \textbf{\bibinfo{volume}{93}},
  \bibinfo{pages}{034337} (\bibinfo{year}{2016}), ISSN
  \bibinfo{issn}{2469-9985}.

\bibitem[{\citenamefont{Huang et~al.}(2017)\citenamefont{Huang, Audi, Wang,
  Kondev, Naimi, and Xu}}]{Huang2017}
\bibinfo{author}{\bibfnamefont{W.~J.} \bibnamefont{Huang}},
  \bibinfo{author}{\bibfnamefont{G.}~\bibnamefont{Audi}},
  \bibinfo{author}{\bibfnamefont{M.}~\bibnamefont{Wang}},
  \bibinfo{author}{\bibfnamefont{F.~G.} \bibnamefont{Kondev}},
  \bibinfo{author}{\bibfnamefont{S.}~\bibnamefont{Naimi}}, \bibnamefont{and}
  \bibinfo{author}{\bibfnamefont{X.}~\bibnamefont{Xu}},
  \bibinfo{journal}{Chinese Phys. C} \textbf{\bibinfo{volume}{41}}
  (\bibinfo{year}{2017}), ISSN \bibinfo{issn}{16741137}.

\bibitem[{\citenamefont{Angeli and Marinova}(2013)}]{Angeli2013}
\bibinfo{author}{\bibfnamefont{I.}~\bibnamefont{Angeli}} \bibnamefont{and}
  \bibinfo{author}{\bibfnamefont{K.~P.} \bibnamefont{Marinova}},
  \bibinfo{journal}{At. Data Nucl. Data Tables} \textbf{\bibinfo{volume}{99}},
  \bibinfo{pages}{69} (\bibinfo{year}{2013}), ISSN \bibinfo{issn}{0092640X}.

\bibitem[{\citenamefont{Huang et~al.}(2021)\citenamefont{Huang, Wang, Kondev,
  Audi, and Naimi}}]{Huang2021}
\bibinfo{author}{\bibfnamefont{W.}~\bibnamefont{Huang}},
  \bibinfo{author}{\bibfnamefont{M.}~\bibnamefont{Wang}},
  \bibinfo{author}{\bibfnamefont{F.}~\bibnamefont{Kondev}},
  \bibinfo{author}{\bibfnamefont{G.}~\bibnamefont{Audi}}, \bibnamefont{and}
  \bibinfo{author}{\bibfnamefont{S.}~\bibnamefont{Naimi}},
  \bibinfo{journal}{Chinese Phys. C} \textbf{\bibinfo{volume}{45}},
  \bibinfo{pages}{030002} (\bibinfo{year}{2021}), ISSN
  \bibinfo{issn}{1674-1137},
  \urlprefix\url{https://iopscience.iop.org/article/10.1088/1674-1137/abddb0}.

\bibitem[{\citenamefont{Moore et~al.}(2013)\citenamefont{Moore, Eronen,
  Gorelov, Hakala, Jokinen, Kankainen, Kolhinen, Koponen, Penttil\"a,
  Pohjalainen et~al.}}]{Moore2013}
\bibinfo{author}{\bibfnamefont{I.~D.} \bibnamefont{Moore}},
  \bibinfo{author}{\bibfnamefont{T.}~\bibnamefont{Eronen}},
  \bibinfo{author}{\bibfnamefont{D.}~\bibnamefont{Gorelov}},
  \bibinfo{author}{\bibfnamefont{J.}~\bibnamefont{Hakala}},
  \bibinfo{author}{\bibfnamefont{A.}~\bibnamefont{Jokinen}},
  \bibinfo{author}{\bibfnamefont{A.}~\bibnamefont{Kankainen}},
  \bibinfo{author}{\bibfnamefont{V.}~\bibnamefont{Kolhinen}},
  \bibinfo{author}{\bibfnamefont{J.}~\bibnamefont{Koponen}},
  \bibinfo{author}{\bibfnamefont{H.}~\bibnamefont{Penttil\"a}},
  \bibinfo{author}{\bibfnamefont{I.}~\bibnamefont{Pohjalainen}},
  \bibnamefont{et~al.}, \bibinfo{journal}{Nuc. Inst. and Meth. in Physics
  Research Section B: Beam Interactions with Materials and Atoms}
  \textbf{\bibinfo{volume}{317}}, \bibinfo{pages}{208} (\bibinfo{year}{2013}),
  \urlprefix\url{https://doi.org/10.1016/j.nimb.2013.06.036}.

\bibitem[{\citenamefont{Karvonen et~al.}(2008)\citenamefont{Karvonen, Moore,
  Sonoda, Kessler, Penttil\"a, Per\"aj\"arvi, Ronkanen, and
  \"Ayst{\"o}}}]{Karvonen2008}
\bibinfo{author}{\bibfnamefont{P.}~\bibnamefont{Karvonen}},
  \bibinfo{author}{\bibfnamefont{I.~D.} \bibnamefont{Moore}},
  \bibinfo{author}{\bibfnamefont{T.}~\bibnamefont{Sonoda}},
  \bibinfo{author}{\bibfnamefont{T.}~\bibnamefont{Kessler}},
  \bibinfo{author}{\bibfnamefont{H.}~\bibnamefont{Penttil\"a}},
  \bibinfo{author}{\bibfnamefont{K.}~\bibnamefont{Per\"aj\"arvi}},
  \bibinfo{author}{\bibfnamefont{P.}~\bibnamefont{Ronkanen}}, \bibnamefont{and}
  \bibinfo{author}{\bibfnamefont{J.}~\bibnamefont{\"Ayst{\"o}}},
  \bibinfo{journal}{Nucl. Instrum. Meth. Phys. Res. B}
  \textbf{\bibinfo{volume}{266}}, \bibinfo{pages}{4794} (\bibinfo{year}{2008}),
  \urlprefix\url{https://doi.org/10.1016/j.nimb.2008.07.022}.

\bibitem[{\citenamefont{Nieminen et~al.}(2001)\citenamefont{Nieminen, Huikari,
  Jokinen, \"Ayst{\"o}, Campbell, and Cochrane}}]{Nieminen2001}
\bibinfo{author}{\bibfnamefont{A.}~\bibnamefont{Nieminen}},
  \bibinfo{author}{\bibfnamefont{J.}~\bibnamefont{Huikari}},
  \bibinfo{author}{\bibfnamefont{A.}~\bibnamefont{Jokinen}},
  \bibinfo{author}{\bibfnamefont{J.}~\bibnamefont{\"Ayst{\"o}}},
  \bibinfo{author}{\bibfnamefont{P.}~\bibnamefont{Campbell}}, \bibnamefont{and}
  \bibinfo{author}{\bibfnamefont{E.~C.~A.} \bibnamefont{Cochrane}},
  \bibinfo{journal}{Nucl. Instrum. Meth. Phys. Res. A}
  \textbf{\bibinfo{volume}{469}}, \bibinfo{pages}{244} (\bibinfo{year}{2001}),
  \urlprefix\url{https://doi.org/10.1016/S0168-9002(00)00750-6}.

\bibitem[{\citenamefont{Savard et~al.}(1991)\citenamefont{Savard, Becker,
  Bollen, Kluge, Moore, Otto, Schweikhard, Stolzenberg, and
  Wiess}}]{Savard1991}
\bibinfo{author}{\bibfnamefont{G.}~\bibnamefont{Savard}},
  \bibinfo{author}{\bibfnamefont{S.}~\bibnamefont{Becker}},
  \bibinfo{author}{\bibfnamefont{G.}~\bibnamefont{Bollen}},
  \bibinfo{author}{\bibfnamefont{H.-J.} \bibnamefont{Kluge}},
  \bibinfo{author}{\bibfnamefont{R.~B.} \bibnamefont{Moore}},
  \bibinfo{author}{\bibfnamefont{T.}~\bibnamefont{Otto}},
  \bibinfo{author}{\bibfnamefont{L.}~\bibnamefont{Schweikhard}},
  \bibinfo{author}{\bibfnamefont{H.}~\bibnamefont{Stolzenberg}},
  \bibnamefont{and} \bibinfo{author}{\bibfnamefont{U.}~\bibnamefont{Wiess}},
  \bibinfo{journal}{Phys. Lett. A} \textbf{\bibinfo{volume}{158}},
  \bibinfo{pages}{247} (\bibinfo{year}{1991}),
  \urlprefix\url{https://doi.org/10.1016/0375-9601(91)91008-2}.

\bibitem[{\citenamefont{Vil\'en et~al.}(2020)\citenamefont{Vil\'en, Canete,
  Cheal, Giatzoglou, {de Groote}, {de Roubin}, Eronen, Geldhof, Jokinen,
  Kankainen et~al.}}]{Vilen2020}
\bibinfo{author}{\bibfnamefont{M.}~\bibnamefont{Vil\'en}},
  \bibinfo{author}{\bibfnamefont{L.}~\bibnamefont{Canete}},
  \bibinfo{author}{\bibfnamefont{B.}~\bibnamefont{Cheal}},
  \bibinfo{author}{\bibfnamefont{A.}~\bibnamefont{Giatzoglou}},
  \bibinfo{author}{\bibfnamefont{R.}~\bibnamefont{{de Groote}}},
  \bibinfo{author}{\bibfnamefont{A.}~\bibnamefont{{de Roubin}}},
  \bibinfo{author}{\bibfnamefont{T.}~\bibnamefont{Eronen}},
  \bibinfo{author}{\bibfnamefont{S.}~\bibnamefont{Geldhof}},
  \bibinfo{author}{\bibfnamefont{A.}~\bibnamefont{Jokinen}},
  \bibinfo{author}{\bibfnamefont{A.}~\bibnamefont{Kankainen}},
  \bibnamefont{et~al.}, \bibinfo{journal}{Nuclear Instruments and Methods in
  Physics Research Section B: Beam Interactions with Materials and Atoms}
  \textbf{\bibinfo{volume}{463}}, \bibinfo{pages}{382} (\bibinfo{year}{2020}),
  ISSN \bibinfo{issn}{0168-583X},
  \urlprefix\url{https://www.sciencedirect.com/science/article/pii/S0168583X19302344}.

\bibitem[{\citenamefont{Gr\"aff et~al.}(1980)\citenamefont{Gr\"aff, Kalinowsky,
  and Traut}}]{Graff1980}
\bibinfo{author}{\bibfnamefont{G.}~\bibnamefont{Gr\"aff}},
  \bibinfo{author}{\bibfnamefont{H.}~\bibnamefont{Kalinowsky}},
  \bibnamefont{and} \bibinfo{author}{\bibfnamefont{J.}~\bibnamefont{Traut}},
  \bibinfo{journal}{Z Physik A} \textbf{\bibinfo{volume}{297}},
  \bibinfo{pages}{35} (\bibinfo{year}{1980}),
  \urlprefix\url{https://doi.org/10.1007/BF01414243}.

\bibitem[{\citenamefont{K{\"o}nig et~al.}(1995)\citenamefont{K{\"o}nig, Bollen,
  Kluge, Otto, and Szerypo}}]{Konig1995}
\bibinfo{author}{\bibfnamefont{M.}~\bibnamefont{K{\"o}nig}},
  \bibinfo{author}{\bibfnamefont{G.}~\bibnamefont{Bollen}},
  \bibinfo{author}{\bibfnamefont{H.-J.} \bibnamefont{Kluge}},
  \bibinfo{author}{\bibfnamefont{T.}~\bibnamefont{Otto}}, \bibnamefont{and}
  \bibinfo{author}{\bibfnamefont{J.}~\bibnamefont{Szerypo}},
  \bibinfo{journal}{Int. J. Mass Spectrom. Ion Process}
  \textbf{\bibinfo{volume}{142}}, \bibinfo{pages}{95} (\bibinfo{year}{1995}),
  \urlprefix\url{https://doi.org/10.1016/0168-1176(95)04146-C.}

\bibitem[{\citenamefont{Kretzschmar}(2007)}]{Kretzschmar2007}
\bibinfo{author}{\bibfnamefont{M.}~\bibnamefont{Kretzschmar}},
  \bibinfo{journal}{Int. J. Mass Spectrom.} \textbf{\bibinfo{volume}{264}},
  \bibinfo{pages}{122} (\bibinfo{year}{2007}),
  \urlprefix\url{https://doi.org/10.1016/j.ijms.2007.04.002}.

\bibitem[{\citenamefont{George et~al.}(2007)\citenamefont{George, Blaum,
  Herfurth, Herlert, Kretzschmar, Nagy, Schwarz, Schweikhard, and
  Yazidjian}}]{George2007}
\bibinfo{author}{\bibfnamefont{S.}~\bibnamefont{George}},
  \bibinfo{author}{\bibfnamefont{K.}~\bibnamefont{Blaum}},
  \bibinfo{author}{\bibfnamefont{F.}~\bibnamefont{Herfurth}},
  \bibinfo{author}{\bibfnamefont{A.}~\bibnamefont{Herlert}},
  \bibinfo{author}{\bibfnamefont{M.}~\bibnamefont{Kretzschmar}},
  \bibinfo{author}{\bibfnamefont{S.}~\bibnamefont{Nagy}},
  \bibinfo{author}{\bibfnamefont{S.}~\bibnamefont{Schwarz}},
  \bibinfo{author}{\bibfnamefont{L.}~\bibnamefont{Schweikhard}},
  \bibnamefont{and}
  \bibinfo{author}{\bibfnamefont{C.}~\bibnamefont{Yazidjian}},
  \bibinfo{journal}{Int. J. Mass Spectrom.} \textbf{\bibinfo{volume}{264}},
  \bibinfo{pages}{110} (\bibinfo{year}{2007}),
  \urlprefix\url{https://doi.org/10.1016/j.ijms.2007.04.003}.

\bibitem[{\citenamefont{Eronen et~al.}(2008)\citenamefont{Eronen, Elomaa,
  Hager, Hakala, Jokinen, Kankainen, Rahaman, Rissanen, Weber, and
  \"Ayst{\"o}}}]{Eronen2008}
\bibinfo{author}{\bibfnamefont{T.}~\bibnamefont{Eronen}},
  \bibinfo{author}{\bibfnamefont{V.-V.} \bibnamefont{Elomaa}},
  \bibinfo{author}{\bibfnamefont{U.}~\bibnamefont{Hager}},
  \bibinfo{author}{\bibfnamefont{J.}~\bibnamefont{Hakala}},
  \bibinfo{author}{\bibfnamefont{A.}~\bibnamefont{Jokinen}},
  \bibinfo{author}{\bibfnamefont{A.}~\bibnamefont{Kankainen}},
  \bibinfo{author}{\bibfnamefont{S.}~\bibnamefont{Rahaman}},
  \bibinfo{author}{\bibfnamefont{J.}~\bibnamefont{Rissanen}},
  \bibinfo{author}{\bibfnamefont{C.}~\bibnamefont{Weber}}, \bibnamefont{and}
  \bibinfo{author}{\bibfnamefont{J.}~\bibnamefont{\"Ayst{\"o}}},
  \bibinfo{journal}{Nucl. Inst. Meth. Phys. Sec. B}
  \textbf{\bibinfo{volume}{266}}, \bibinfo{pages}{4527} (\bibinfo{year}{2008}),
  \urlprefix\url{https://doi.org/10.1016/j.nimb.2008.05.076}.

\bibitem[{\citenamefont{Kellerbauer et~al.}(2003)\citenamefont{Kellerbauer,
  Blaum, Bollen, Herfurth, Kluge, Kuckein, Sauvan, Scheidenberger, and
  Schweikhard}}]{Kellerbauer2003}
\bibinfo{author}{\bibfnamefont{A.}~\bibnamefont{Kellerbauer}},
  \bibinfo{author}{\bibfnamefont{K.}~\bibnamefont{Blaum}},
  \bibinfo{author}{\bibfnamefont{G.}~\bibnamefont{Bollen}},
  \bibinfo{author}{\bibfnamefont{F.}~\bibnamefont{Herfurth}},
  \bibinfo{author}{\bibfnamefont{H.-J.} \bibnamefont{Kluge}},
  \bibinfo{author}{\bibfnamefont{M.}~\bibnamefont{Kuckein}},
  \bibinfo{author}{\bibfnamefont{E.}~\bibnamefont{Sauvan}},
  \bibinfo{author}{\bibfnamefont{C.}~\bibnamefont{Scheidenberger}},
  \bibnamefont{and}
  \bibinfo{author}{\bibfnamefont{L.}~\bibnamefont{Schweikhard}},
  \bibinfo{journal}{Eur. Phys. J. D} \textbf{\bibinfo{volume}{22}},
  \bibinfo{pages}{53} (\bibinfo{year}{2003}),
  \urlprefix\url{https://doi.org/10.1140/epjd/e2002-00222-0}.

\bibitem[{\citenamefont{Wang et~al.}(2021)\citenamefont{Wang, Huang, Kondev,
  Audi, and Naimi}}]{AME20}
\bibinfo{author}{\bibfnamefont{M.}~\bibnamefont{Wang}},
  \bibinfo{author}{\bibfnamefont{W.}~\bibnamefont{Huang}},
  \bibinfo{author}{\bibfnamefont{F.}~\bibnamefont{Kondev}},
  \bibinfo{author}{\bibfnamefont{G.}~\bibnamefont{Audi}}, \bibnamefont{and}
  \bibinfo{author}{\bibfnamefont{S.}~\bibnamefont{Naimi}},
  \bibinfo{journal}{Chinese Physics C} \textbf{\bibinfo{volume}{45}},
  \bibinfo{pages}{030003} (\bibinfo{year}{2021}),
  \urlprefix\url{https://doi.org/10.1088/1674-1137/abddaf}.

\bibitem[{\citenamefont{Kratz and Pfeiffer}()}]{Kratz2000}
\bibinfo{author}{\bibfnamefont{K.-L.} \bibnamefont{Kratz}} \bibnamefont{and}
  \bibinfo{author}{\bibfnamefont{B.}~\bibnamefont{Pfeiffer}},
  \bibinfo{note}{private communication, 2000}.

\bibitem[{\citenamefont{Lalkovski and Kondev}(2015)}]{ensdf112}
\bibinfo{author}{\bibfnamefont{S.}~\bibnamefont{Lalkovski}} \bibnamefont{and}
  \bibinfo{author}{\bibfnamefont{F.}~\bibnamefont{Kondev}},
  \bibinfo{journal}{Nuclear Data Sheets} \textbf{\bibinfo{volume}{124}},
  \bibinfo{pages}{157} (\bibinfo{year}{2015}), ISSN \bibinfo{issn}{0090-3752},
  \urlprefix\url{https://www.sciencedirect.com/science/article/pii/S0090375214007467}.

\bibitem[{\citenamefont{Blachot}(2012)}]{Blachot2012b}
\bibinfo{author}{\bibfnamefont{J.}~\bibnamefont{Blachot}},
  \bibinfo{journal}{Nuclear Data Sheets} \textbf{\bibinfo{volume}{113}},
  \bibinfo{pages}{515} (\bibinfo{year}{2012}), ISSN \bibinfo{issn}{0090-3752},
  \urlprefix\url{https://www.sciencedirect.com/science/article/pii/S0090375212000129}.

\bibitem[{\citenamefont{Kn{\"o}bel}()}]{Knoebel2008}
\bibinfo{author}{\bibnamefont{Kn{\"o}bel}}, \bibinfo{note}{phD thesis, {GSI},
  2008}.

\bibitem[{\citenamefont{Blachot}(2010)}]{Blachot2010}
\bibinfo{author}{\bibfnamefont{J.}~\bibnamefont{Blachot}},
  \bibinfo{journal}{Nuclear Data Sheets} \textbf{\bibinfo{volume}{111}},
  \bibinfo{pages}{717} (\bibinfo{year}{2010}), ISSN \bibinfo{issn}{0090-3752},
  \urlprefix\url{https://www.sciencedirect.com/science/article/pii/S0090375210000281}.

\bibitem[{\citenamefont{Matos}(2004)}]{Matos2004}
\bibinfo{author}{\bibfnamefont{M.}~\bibnamefont{Matos}}, Ph.D. thesis,
  \bibinfo{school}{Justus-Liebig-Universit\"at Giessen} (\bibinfo{year}{2004}).

\bibitem[{\citenamefont{Montes et~al.}(2006)\citenamefont{Montes, Estrade,
  Hosmer, Liddick, Mantica, Morton, Mueller, Ouellette, Pellegrini, Santi
  et~al.}}]{Montes2006}
\bibinfo{author}{\bibfnamefont{F.}~\bibnamefont{Montes}},
  \bibinfo{author}{\bibfnamefont{A.}~\bibnamefont{Estrade}},
  \bibinfo{author}{\bibfnamefont{P.~T.} \bibnamefont{Hosmer}},
  \bibinfo{author}{\bibfnamefont{S.~N.} \bibnamefont{Liddick}},
  \bibinfo{author}{\bibfnamefont{P.~F.} \bibnamefont{Mantica}},
  \bibinfo{author}{\bibfnamefont{A.~C.} \bibnamefont{Morton}},
  \bibinfo{author}{\bibfnamefont{W.~F.} \bibnamefont{Mueller}},
  \bibinfo{author}{\bibfnamefont{M.}~\bibnamefont{Ouellette}},
  \bibinfo{author}{\bibfnamefont{E.}~\bibnamefont{Pellegrini}},
  \bibinfo{author}{\bibfnamefont{P.}~\bibnamefont{Santi}},
  \bibnamefont{et~al.}, \bibinfo{journal}{Phys. Rev. C}
  \textbf{\bibinfo{volume}{73}}, \bibinfo{pages}{035801}
  (\bibinfo{year}{2006}),
  \urlprefix\url{https://link.aps.org/doi/10.1103/PhysRevC.73.035801}.

\bibitem[{\citenamefont{Lorusso et~al.}(2015)\citenamefont{Lorusso, Nishimura,
  Xu, Jungclaus, Shimizu, Simpson, S\"oderstr\"om, Watanabe, Browne, Doornenbal
  et~al.}}]{Lorusso2015}
\bibinfo{author}{\bibfnamefont{G.}~\bibnamefont{Lorusso}},
  \bibinfo{author}{\bibfnamefont{S.}~\bibnamefont{Nishimura}},
  \bibinfo{author}{\bibfnamefont{Z.~Y.} \bibnamefont{Xu}},
  \bibinfo{author}{\bibfnamefont{A.}~\bibnamefont{Jungclaus}},
  \bibinfo{author}{\bibfnamefont{Y.}~\bibnamefont{Shimizu}},
  \bibinfo{author}{\bibfnamefont{G.~S.} \bibnamefont{Simpson}},
  \bibinfo{author}{\bibfnamefont{P.-A.} \bibnamefont{S\"oderstr\"om}},
  \bibinfo{author}{\bibfnamefont{H.}~\bibnamefont{Watanabe}},
  \bibinfo{author}{\bibfnamefont{F.}~\bibnamefont{Browne}},
  \bibinfo{author}{\bibfnamefont{P.}~\bibnamefont{Doornenbal}},
  \bibnamefont{et~al.}, \bibinfo{journal}{Phys. Rev. Lett.}
  \textbf{\bibinfo{volume}{114}}, \bibinfo{pages}{192501}
  (\bibinfo{year}{2015}),
  \urlprefix\url{https://link.aps.org/doi/10.1103/PhysRevLett.114.192501}.

\bibitem[{\citenamefont{Hall et~al.}(2021)\citenamefont{Hall, Davinson,
  Estrade, Liu, Lorusso, Montes, Nishimura, Phong, Woods, Agramunt
  et~al.}}]{Hall2021}
\bibinfo{author}{\bibfnamefont{O.}~\bibnamefont{Hall}},
  \bibinfo{author}{\bibfnamefont{T.}~\bibnamefont{Davinson}},
  \bibinfo{author}{\bibfnamefont{A.}~\bibnamefont{Estrade}},
  \bibinfo{author}{\bibfnamefont{J.}~\bibnamefont{Liu}},
  \bibinfo{author}{\bibfnamefont{G.}~\bibnamefont{Lorusso}},
  \bibinfo{author}{\bibfnamefont{F.}~\bibnamefont{Montes}},
  \bibinfo{author}{\bibfnamefont{S.}~\bibnamefont{Nishimura}},
  \bibinfo{author}{\bibfnamefont{V.}~\bibnamefont{Phong}},
  \bibinfo{author}{\bibfnamefont{P.}~\bibnamefont{Woods}},
  \bibinfo{author}{\bibfnamefont{J.}~\bibnamefont{Agramunt}},
  \bibnamefont{et~al.}, \bibinfo{journal}{Phys. Lett. B}
  \textbf{\bibinfo{volume}{816}}, \bibinfo{pages}{136266}
  (\bibinfo{year}{2021}), ISSN \bibinfo{issn}{0370-2693},
  \urlprefix\url{https://www.sciencedirect.com/science/article/pii/S0370269321002069}.

\bibitem[{\citenamefont{Kameda et~al.}(2012)\citenamefont{Kameda, Kubo,
  Ohnishi, Kusaka, Yoshida, Yoshida, Ohtake, Fukuda, Takeda, Tanaka
  et~al.}}]{Kameda2012}
\bibinfo{author}{\bibfnamefont{D.}~\bibnamefont{Kameda}},
  \bibinfo{author}{\bibfnamefont{T.}~\bibnamefont{Kubo}},
  \bibinfo{author}{\bibfnamefont{T.}~\bibnamefont{Ohnishi}},
  \bibinfo{author}{\bibfnamefont{K.}~\bibnamefont{Kusaka}},
  \bibinfo{author}{\bibfnamefont{A.}~\bibnamefont{Yoshida}},
  \bibinfo{author}{\bibfnamefont{K.}~\bibnamefont{Yoshida}},
  \bibinfo{author}{\bibfnamefont{M.}~\bibnamefont{Ohtake}},
  \bibinfo{author}{\bibfnamefont{N.}~\bibnamefont{Fukuda}},
  \bibinfo{author}{\bibfnamefont{H.}~\bibnamefont{Takeda}},
  \bibinfo{author}{\bibfnamefont{K.}~\bibnamefont{Tanaka}},
  \bibnamefont{et~al.}, \bibinfo{journal}{Phys. Rev. C}
  \textbf{\bibinfo{volume}{86}}, \bibinfo{pages}{054319}
  (\bibinfo{year}{2012}),
  \urlprefix\url{https://link.aps.org/doi/10.1103/PhysRevC.86.054319}.

\bibitem[{\citenamefont{Erler et~al.}(2012)\citenamefont{Erler, Birge,
  Kortelainen, Nazarewicz, Olsen, Perhac, and Stoitsov}}]{Erler2012a}
\bibinfo{author}{\bibfnamefont{J.}~\bibnamefont{Erler}},
  \bibinfo{author}{\bibfnamefont{N.}~\bibnamefont{Birge}},
  \bibinfo{author}{\bibfnamefont{M.}~\bibnamefont{Kortelainen}},
  \bibinfo{author}{\bibfnamefont{W.}~\bibnamefont{Nazarewicz}},
  \bibinfo{author}{\bibfnamefont{E.}~\bibnamefont{Olsen}},
  \bibinfo{author}{\bibfnamefont{A.~M.} \bibnamefont{Perhac}},
  \bibnamefont{and} \bibinfo{author}{\bibfnamefont{M.}~\bibnamefont{Stoitsov}},
  \bibinfo{journal}{Nature} \textbf{\bibinfo{volume}{486}},
  \bibinfo{pages}{509} (\bibinfo{year}{2012}).

\bibitem[{\citenamefont{Agbemava et~al.}(2014)\citenamefont{Agbemava,
  Afanasjev, Ray, and Ring}}]{Agbemava2014a}
\bibinfo{author}{\bibfnamefont{S.~E.} \bibnamefont{Agbemava}},
  \bibinfo{author}{\bibfnamefont{A.~V.} \bibnamefont{Afanasjev}},
  \bibinfo{author}{\bibfnamefont{D.}~\bibnamefont{Ray}}, \bibnamefont{and}
  \bibinfo{author}{\bibfnamefont{P.}~\bibnamefont{Ring}},
  \bibinfo{journal}{Phys. Rev. C} \textbf{\bibinfo{volume}{89}},
  \bibinfo{pages}{054320} (\bibinfo{year}{2014}),
  \urlprefix\url{https://link.aps.org/doi/10.1103/PhysRevC.89.054320}.

\bibitem[{\citenamefont{Lu et~al.}(2015)\citenamefont{Lu, Li, Li, Yao, and
  Meng}}]{Lu2015}
\bibinfo{author}{\bibfnamefont{K.~Q.} \bibnamefont{Lu}},
  \bibinfo{author}{\bibfnamefont{Z.~X.} \bibnamefont{Li}},
  \bibinfo{author}{\bibfnamefont{Z.~P.} \bibnamefont{Li}},
  \bibinfo{author}{\bibfnamefont{J.~M.} \bibnamefont{Yao}}, \bibnamefont{and}
  \bibinfo{author}{\bibfnamefont{J.}~\bibnamefont{Meng}},
  \bibinfo{journal}{Phys. Rev. C} \textbf{\bibinfo{volume}{91}},
  \bibinfo{pages}{027304} (\bibinfo{year}{2015}),
  \urlprefix\url{https://link.aps.org/doi/10.1103/PhysRevC.91.027304}.

\bibitem[{\citenamefont{Yang et~al.}(2021)\citenamefont{Yang, Wang, Zhao, and
  Li}}]{Yang2021}
\bibinfo{author}{\bibfnamefont{Y.~L.} \bibnamefont{Yang}},
  \bibinfo{author}{\bibfnamefont{Y.~K.} \bibnamefont{Wang}},
  \bibinfo{author}{\bibfnamefont{P.~W.} \bibnamefont{Zhao}}, \bibnamefont{and}
  \bibinfo{author}{\bibfnamefont{Z.~P.} \bibnamefont{Li}},
  \bibinfo{journal}{Phys. Rev. C} \textbf{\bibinfo{volume}{104}},
  \bibinfo{pages}{054312} (\bibinfo{year}{2021}),
  \urlprefix\url{https://link.aps.org/doi/10.1103/PhysRevC.104.054312}.

\bibitem[{\citenamefont{Hilaire and Girod}(2007)}]{Hilaire2007a}
\bibinfo{author}{\bibfnamefont{S.}~\bibnamefont{Hilaire}} \bibnamefont{and}
  \bibinfo{author}{\bibfnamefont{M.}~\bibnamefont{Girod}},
  \bibinfo{journal}{Eur. Phys. J. A} \textbf{\bibinfo{volume}{33}},
  \bibinfo{pages}{237–241} (\bibinfo{year}{2007}).

\bibitem[{\citenamefont{Goriely et~al.}(2013)\citenamefont{Goriely, Chamel, and
  Pearson}}]{Goriely2013}
\bibinfo{author}{\bibfnamefont{S.}~\bibnamefont{Goriely}},
  \bibinfo{author}{\bibfnamefont{N.}~\bibnamefont{Chamel}}, \bibnamefont{and}
  \bibinfo{author}{\bibfnamefont{J.~M.} \bibnamefont{Pearson}},
  \bibinfo{journal}{Phys. Rev. C} \textbf{\bibinfo{volume}{88}},
  \bibinfo{pages}{061302} (\bibinfo{year}{2013}), ISSN
  \bibinfo{issn}{0556-2813}.

\bibitem[{\citenamefont{Ryssens}(2016)}]{RyssensPhD}
\bibinfo{author}{\bibfnamefont{W.}~\bibnamefont{Ryssens}}, Ph.D. thesis,
  \bibinfo{school}{Universit\'e Libre de Bruxelles} (\bibinfo{year}{2016}).

\bibitem[{\citenamefont{Ryssens et~al.}(2015)\citenamefont{Ryssens, Heenen, and
  Bender}}]{Ryssens15}
\bibinfo{author}{\bibfnamefont{W.}~\bibnamefont{Ryssens}},
  \bibinfo{author}{\bibfnamefont{P.-H.} \bibnamefont{Heenen}},
  \bibnamefont{and} \bibinfo{author}{\bibfnamefont{M.}~\bibnamefont{Bender}},
  \bibinfo{journal}{Phys. Rev. C} \textbf{\bibinfo{volume}{92}},
  \bibinfo{pages}{064318} (\bibinfo{year}{2015}), ISSN
  \bibinfo{issn}{0556-2813}, \eprint{1509.00252},
  \urlprefix\url{http://link.aps.org/doi/10.1103/PhysRevC.92.064318
  https://link.aps.org/doi/10.1103/PhysRevC.92.064318}.

\bibitem[{\citenamefont{Ryssens et~al.}(2019)\citenamefont{Ryssens, Bender, and
  Heenen}}]{Ryssens2019}
\bibinfo{author}{\bibfnamefont{W.}~\bibnamefont{Ryssens}},
  \bibinfo{author}{\bibfnamefont{M.}~\bibnamefont{Bender}}, \bibnamefont{and}
  \bibinfo{author}{\bibfnamefont{P.~H.} \bibnamefont{Heenen}},
  \bibinfo{journal}{Eur. Phys. J. A} \textbf{\bibinfo{volume}{55}},
  \bibinfo{pages}{93} (\bibinfo{year}{2019}), ISSN \bibinfo{issn}{1434-6001},
  \eprint{1812.08262}, \urlprefix\url{http://arxiv.org/abs/1812.08262
  http://link.springer.com/10.1140/epja/i2019-12766-6}.

\bibitem[{\citenamefont{Perez-Martin and Robledo}(2008)}]{Perez-Martin2008}
\bibinfo{author}{\bibfnamefont{S.}~\bibnamefont{Perez-Martin}}
  \bibnamefont{and} \bibinfo{author}{\bibfnamefont{L.~M.}
  \bibnamefont{Robledo}}, \bibinfo{journal}{Phys. Rev. C}
  \textbf{\bibinfo{volume}{78}}, \bibinfo{pages}{014304}
  (\bibinfo{year}{2008}), ISSN \bibinfo{issn}{0556-2813},
  \urlprefix\url{https://link.aps.org/doi/10.1103/PhysRevC.78.014304}.

\bibitem[{\citenamefont{M{\"o}ller et~al.}(2006)\citenamefont{M{\"o}ller,
  Bengtsson, Gillis~Carlsson, Olivius, and Ichikawa}}]{Moller2006}
\bibinfo{author}{\bibfnamefont{P.}~\bibnamefont{M{\"o}ller}},
  \bibinfo{author}{\bibfnamefont{R.}~\bibnamefont{Bengtsson}},
  \bibinfo{author}{\bibfnamefont{B.}~\bibnamefont{Gillis~Carlsson}},
  \bibinfo{author}{\bibfnamefont{P.}~\bibnamefont{Olivius}}, \bibnamefont{and}
  \bibinfo{author}{\bibfnamefont{T.}~\bibnamefont{Ichikawa}},
  \bibinfo{journal}{Phys. Rev. Lett.} \textbf{\bibinfo{volume}{97}},
  \bibinfo{pages}{162502} (\bibinfo{year}{2006}),
  \urlprefix\url{https://doi.org/10.1103/PhysRevLett.97.162502}.

\bibitem[{\citenamefont{Lunney et~al.}(2003)\citenamefont{Lunney, Pearson, and
  Thibault}}]{Lunney2003}
\bibinfo{author}{\bibfnamefont{D.}~\bibnamefont{Lunney}},
  \bibinfo{author}{\bibfnamefont{J.~M.} \bibnamefont{Pearson}},
  \bibnamefont{and} \bibinfo{author}{\bibfnamefont{C.}~\bibnamefont{Thibault}},
  \bibinfo{journal}{Rev. Mod. Phys.} \textbf{\bibinfo{volume}{75}},
  \bibinfo{pages}{1021} (\bibinfo{year}{2003}),
  \urlprefix\url{https://link.aps.org/doi/10.1103/RevModPhys.75.1021}.

\bibitem[{\citenamefont{Bender et~al.}(2008)\citenamefont{Bender, Bertsch, and
  Heenen}}]{Bender08a}
\bibinfo{author}{\bibfnamefont{M.}~\bibnamefont{Bender}},
  \bibinfo{author}{\bibfnamefont{G.~F.} \bibnamefont{Bertsch}},
  \bibnamefont{and} \bibinfo{author}{\bibfnamefont{P.-H.}
  \bibnamefont{Heenen}}, \bibinfo{journal}{Phys. Rev. C}
  \textbf{\bibinfo{volume}{78}}, \bibinfo{pages}{054312}
  (\bibinfo{year}{2008}),
  \urlprefix\url{https://link.aps.org/doi/10.1103/PhysRevC.78.054312}.

\bibitem[{\citenamefont{{Kn\"obel, R.} et~al.}(2016)\citenamefont{{Kn\"obel,
  R.}, {Diwisch, M.}, {Geissel, H.}, {Litvinov, Yu. A.}, {Patyk, Z.},
  {Pla\ss{}, W. R.}, {Scheidenberger, C.}, {Sun, B.}, {Weick, H.}, {Bosch, F.}
  et~al.}}]{Knoebel2016}
\bibinfo{author}{\bibnamefont{{Kn\"obel, R.}}},
  \bibinfo{author}{\bibnamefont{{Diwisch, M.}}},
  \bibinfo{author}{\bibnamefont{{Geissel, H.}}},
  \bibinfo{author}{\bibnamefont{{Litvinov, Yu. A.}}},
  \bibinfo{author}{\bibnamefont{{Patyk, Z.}}},
  \bibinfo{author}{\bibnamefont{{Pla\ss{}, W. R.}}},
  \bibinfo{author}{\bibnamefont{{Scheidenberger, C.}}},
  \bibinfo{author}{\bibnamefont{{Sun, B.}}},
  \bibinfo{author}{\bibnamefont{{Weick, H.}}},
  \bibinfo{author}{\bibnamefont{{Bosch, F.}}}, \bibnamefont{et~al.},
  \bibinfo{journal}{Eur. Phys. J. A} \textbf{\bibinfo{volume}{52}},
  \bibinfo{pages}{138} (\bibinfo{year}{2016}),
  \urlprefix\url{https://doi.org/10.1140/epja/i2016-16138-6}.

\bibitem[{\citenamefont{Bender et~al.}(2000)\citenamefont{Bender, Rutz,
  Reinhard, and A.}}]{Bender00a}
\bibinfo{author}{\bibfnamefont{M.}~\bibnamefont{Bender}},
  \bibinfo{author}{\bibfnamefont{K.}~\bibnamefont{Rutz}},
  \bibinfo{author}{\bibfnamefont{P.-G.} \bibnamefont{Reinhard}},
  \bibnamefont{and} \bibinfo{author}{\bibfnamefont{.~M.~J.} \bibnamefont{A.}},
  \bibinfo{journal}{Eur. Phys. J A} \textbf{\bibinfo{volume}{8}},
  \bibinfo{pages}{59–75} (\bibinfo{year}{2000}),
  \urlprefix\url{https://doi.org/10.1007/s10050-000-4504-z}.

\bibitem[{\citenamefont{Duguet et~al.}(2001{\natexlab{a}})\citenamefont{Duguet,
  Bonche, Heenen, and Meyer}}]{Duguet01a}
\bibinfo{author}{\bibfnamefont{T.}~\bibnamefont{Duguet}},
  \bibinfo{author}{\bibfnamefont{P.}~\bibnamefont{Bonche}},
  \bibinfo{author}{\bibfnamefont{P.-H.} \bibnamefont{Heenen}},
  \bibnamefont{and} \bibinfo{author}{\bibfnamefont{J.}~\bibnamefont{Meyer}},
  \bibinfo{journal}{Phys. Rev. C} \textbf{\bibinfo{volume}{65}},
  \bibinfo{pages}{014311} (\bibinfo{year}{2001}{\natexlab{a}}),
  \urlprefix\url{https://link.aps.org/doi/10.1103/PhysRevC.65.014311}.

\bibitem[{\citenamefont{Wu et~al.}(2016)\citenamefont{Wu, Changizi, and
  Qi}}]{Wu16a}
\bibinfo{author}{\bibfnamefont{Z.}~\bibnamefont{Wu}},
  \bibinfo{author}{\bibfnamefont{S.~A.} \bibnamefont{Changizi}},
  \bibnamefont{and} \bibinfo{author}{\bibfnamefont{C.}~\bibnamefont{Qi}},
  \bibinfo{journal}{Phys. Rev. C} \textbf{\bibinfo{volume}{93}},
  \bibinfo{pages}{034334} (\bibinfo{year}{2016}),
  \urlprefix\url{https://link.aps.org/doi/10.1103/PhysRevC.93.034334}.

\bibitem[{\citenamefont{Kumar}(1972)}]{Kumar1972}
\bibinfo{author}{\bibfnamefont{K.}~\bibnamefont{Kumar}},
  \bibinfo{journal}{Phys. Rev. Lett.} \textbf{\bibinfo{volume}{28}},
  \bibinfo{pages}{249} (\bibinfo{year}{1972}), ISSN \bibinfo{issn}{0031-9007}.

\bibitem[{\citenamefont{Cline}(1986)}]{Cline1986}
\bibinfo{author}{\bibfnamefont{D.}~\bibnamefont{Cline}},
  \bibinfo{journal}{Annu. Rev. Nucl. Part. Sci.} \textbf{\bibinfo{volume}{36}},
  \bibinfo{pages}{683} (\bibinfo{year}{1986}), ISSN \bibinfo{issn}{00664243},
  \urlprefix\url{http://nucl.annualreviews.org/cgi/doi/10.1146/annurev.nucl.36.1.683}.

\bibitem[{\citenamefont{Liu et~al.}(2013)\citenamefont{Liu, Hamilton, Ramayya,
  Zhu, Shi, Xu, Batchelder, Brewer, Hwang, Luo et~al.}}]{Liu2013}
\bibinfo{author}{\bibfnamefont{S.~H.} \bibnamefont{Liu}},
  \bibinfo{author}{\bibfnamefont{J.~H.} \bibnamefont{Hamilton}},
  \bibinfo{author}{\bibfnamefont{A.~V.} \bibnamefont{Ramayya}},
  \bibinfo{author}{\bibfnamefont{S.~J.} \bibnamefont{Zhu}},
  \bibinfo{author}{\bibfnamefont{Y.}~\bibnamefont{Shi}},
  \bibinfo{author}{\bibfnamefont{F.~R.} \bibnamefont{Xu}},
  \bibinfo{author}{\bibfnamefont{J.~C.} \bibnamefont{Batchelder}},
  \bibinfo{author}{\bibfnamefont{N.~T.} \bibnamefont{Brewer}},
  \bibinfo{author}{\bibfnamefont{J.~K.} \bibnamefont{Hwang}},
  \bibinfo{author}{\bibfnamefont{Y.~X.} \bibnamefont{Luo}},
  \bibnamefont{et~al.}, \bibinfo{journal}{Phys. Rev. C}
  \textbf{\bibinfo{volume}{87}}, \bibinfo{pages}{057302}
  (\bibinfo{year}{2013}),
  \urlprefix\url{https://link.aps.org/doi/10.1103/PhysRevC.87.057302}.

\bibitem[{\citenamefont{Fotiades et~al.}(2003)\citenamefont{Fotiades, Cizewski,
  Kr\"ucken, McNabb, Becker, Bernstein, Younes, Clark, Fallon, Lee
  et~al.}}]{Fotiades2003}
\bibinfo{author}{\bibfnamefont{N.}~\bibnamefont{Fotiades}},
  \bibinfo{author}{\bibfnamefont{J.~A.} \bibnamefont{Cizewski}},
  \bibinfo{author}{\bibfnamefont{R.}~\bibnamefont{Kr\"ucken}},
  \bibinfo{author}{\bibfnamefont{D.~P.} \bibnamefont{McNabb}},
  \bibinfo{author}{\bibfnamefont{J.~A.} \bibnamefont{Becker}},
  \bibinfo{author}{\bibfnamefont{L.~A.} \bibnamefont{Bernstein}},
  \bibinfo{author}{\bibfnamefont{W.}~\bibnamefont{Younes}},
  \bibinfo{author}{\bibfnamefont{R.~M.} \bibnamefont{Clark}},
  \bibinfo{author}{\bibfnamefont{P.}~\bibnamefont{Fallon}},
  \bibinfo{author}{\bibfnamefont{I.~Y.} \bibnamefont{Lee}},
  \bibnamefont{et~al.}, \bibinfo{journal}{Phys. Rev. C}
  \textbf{\bibinfo{volume}{67}}, \bibinfo{pages}{064304}
  (\bibinfo{year}{2003}),
  \urlprefix\url{https://link.aps.org/doi/10.1103/PhysRevC.67.064304}.

\bibitem[{\citenamefont{S{\"{o}}derstr{\"{o}}m
  et~al.}(2013)\citenamefont{S{\"{o}}derstr{\"{o}}m, Lorusso, Watanabe,
  Nishimura, Doornenbal, Thiamova, Browne, Gey, Jung, Sumikama
  et~al.}}]{Soderstrom2013}
\bibinfo{author}{\bibfnamefont{P.-A.} \bibnamefont{S{\"{o}}derstr{\"{o}}m}},
  \bibinfo{author}{\bibfnamefont{G.}~\bibnamefont{Lorusso}},
  \bibinfo{author}{\bibfnamefont{H.}~\bibnamefont{Watanabe}},
  \bibinfo{author}{\bibfnamefont{S.}~\bibnamefont{Nishimura}},
  \bibinfo{author}{\bibfnamefont{P.}~\bibnamefont{Doornenbal}},
  \bibinfo{author}{\bibfnamefont{G.}~\bibnamefont{Thiamova}},
  \bibinfo{author}{\bibfnamefont{F.}~\bibnamefont{Browne}},
  \bibinfo{author}{\bibfnamefont{G.}~\bibnamefont{Gey}},
  \bibinfo{author}{\bibfnamefont{H.~S.} \bibnamefont{Jung}},
  \bibinfo{author}{\bibfnamefont{T.}~\bibnamefont{Sumikama}},
  \bibnamefont{et~al.}, \bibinfo{journal}{Phys. Rev. C}
  \textbf{\bibinfo{volume}{88}}, \bibinfo{pages}{024301}
  (\bibinfo{year}{2013}), ISSN \bibinfo{issn}{0556-2813},
  \urlprefix\url{https://link.aps.org/doi/10.1103/PhysRevC.88.024301}.

\bibitem[{\citenamefont{Luo et~al.}(2004)\citenamefont{Luo, Wu, Gilat,
  Rasmussen, Hamilton, Ramayya, Hwang, Beyer, Zhu, Kormicki et~al.}}]{Luo2004}
\bibinfo{author}{\bibfnamefont{Y.~X.} \bibnamefont{Luo}},
  \bibinfo{author}{\bibfnamefont{S.~C.} \bibnamefont{Wu}},
  \bibinfo{author}{\bibfnamefont{J.}~\bibnamefont{Gilat}},
  \bibinfo{author}{\bibfnamefont{J.~O.} \bibnamefont{Rasmussen}},
  \bibinfo{author}{\bibfnamefont{J.~H.} \bibnamefont{Hamilton}},
  \bibinfo{author}{\bibfnamefont{A.~V.} \bibnamefont{Ramayya}},
  \bibinfo{author}{\bibfnamefont{J.~K.} \bibnamefont{Hwang}},
  \bibinfo{author}{\bibfnamefont{C.~J.} \bibnamefont{Beyer}},
  \bibinfo{author}{\bibfnamefont{S.~J.} \bibnamefont{Zhu}},
  \bibinfo{author}{\bibfnamefont{J.}~\bibnamefont{Kormicki}},
  \bibnamefont{et~al.}, \bibinfo{journal}{Phys. Rev. C}
  \textbf{\bibinfo{volume}{69}}, \bibinfo{pages}{024315}
  (\bibinfo{year}{2004}),
  \urlprefix\url{https://link.aps.org/doi/10.1103/PhysRevC.69.024315}.

\bibitem[{\citenamefont{Hagen et~al.}(2018)\citenamefont{Hagen, G{\"{o}}rgen,
  Korten, Grente, Salsac, Farget, Braunroth, Bruyneel, Celikovic, Cl{\'{e}}ment
  et~al.}}]{Hagen2018}
\bibinfo{author}{\bibfnamefont{T.~W.} \bibnamefont{Hagen}},
  \bibinfo{author}{\bibfnamefont{A.}~\bibnamefont{G{\"{o}}rgen}},
  \bibinfo{author}{\bibfnamefont{W.}~\bibnamefont{Korten}},
  \bibinfo{author}{\bibfnamefont{L.}~\bibnamefont{Grente}},
  \bibinfo{author}{\bibfnamefont{M.~D.} \bibnamefont{Salsac}},
  \bibinfo{author}{\bibfnamefont{F.}~\bibnamefont{Farget}},
  \bibinfo{author}{\bibfnamefont{T.}~\bibnamefont{Braunroth}},
  \bibinfo{author}{\bibfnamefont{B.}~\bibnamefont{Bruyneel}},
  \bibinfo{author}{\bibfnamefont{I.}~\bibnamefont{Celikovic}},
  \bibinfo{author}{\bibfnamefont{E.}~\bibnamefont{Cl{\'{e}}ment}},
  \bibnamefont{et~al.}, \bibinfo{journal}{Eur. Phys. J. A}
  \textbf{\bibinfo{volume}{54}} (\bibinfo{year}{2018}), ISSN
  \bibinfo{issn}{1434601X}.

\bibitem[{\citenamefont{Ring and Schuck}(1980)}]{RingSchuck}
\bibinfo{author}{\bibfnamefont{P.}~\bibnamefont{Ring}} \bibnamefont{and}
  \bibinfo{author}{\bibfnamefont{P.}~\bibnamefont{Schuck}},
  \emph{\bibinfo{title}{The Nuclear Many-Body problem}}
  (\bibinfo{publisher}{Springer-Verlag Berlin Heidelberg},
  \bibinfo{year}{1980}).

\bibitem[{\citenamefont{Duguet et~al.}(2001{\natexlab{b}})\citenamefont{Duguet,
  Bonche, Heenen, and Meyer}}]{Duguet2001a}
\bibinfo{author}{\bibfnamefont{T.}~\bibnamefont{Duguet}},
  \bibinfo{author}{\bibfnamefont{P.}~\bibnamefont{Bonche}},
  \bibinfo{author}{\bibfnamefont{P.-H.} \bibnamefont{Heenen}},
  \bibnamefont{and} \bibinfo{author}{\bibfnamefont{J.}~\bibnamefont{Meyer}},
  \bibinfo{journal}{Physical Review C} \textbf{\bibinfo{volume}{65}},
  \bibinfo{pages}{014310} (\bibinfo{year}{2001}{\natexlab{b}}), ISSN
  \bibinfo{issn}{0556-2813, 1089-490X},
  \urlprefix\url{https://link.aps.org/doi/10.1103/PhysRevC.65.014310}.

\bibitem[{\citenamefont{Duguet et~al.}(2001{\natexlab{c}})\citenamefont{Duguet,
  Bonche, Heenen, and Meyer}}]{Duguet2001b}
\bibinfo{author}{\bibfnamefont{T.}~\bibnamefont{Duguet}},
  \bibinfo{author}{\bibfnamefont{P.}~\bibnamefont{Bonche}},
  \bibinfo{author}{\bibfnamefont{P.-H.} \bibnamefont{Heenen}},
  \bibnamefont{and} \bibinfo{author}{\bibfnamefont{J.}~\bibnamefont{Meyer}},
  \bibinfo{journal}{Physical Review C} \textbf{\bibinfo{volume}{65}},
  \bibinfo{pages}{014311} (\bibinfo{year}{2001}{\natexlab{c}}), ISSN
  \bibinfo{issn}{0556-2813, 1089-490X},
  \urlprefix\url{https://link.aps.org/doi/10.1103/PhysRevC.65.014311}.

\bibitem[{\citenamefont{Strutinsky}(1967)}]{Strutinsky1967}
\bibinfo{author}{\bibfnamefont{V.~M.} \bibnamefont{Strutinsky}},
  \bibinfo{journal}{Nucl. Phys. A} \textbf{\bibinfo{volume}{95}},
  \bibinfo{pages}{420} (\bibinfo{year}{1967}), ISSN \bibinfo{issn}{0375-9474},
  \urlprefix\url{https://www.sciencedirect.com/science/article/pii/0375947467905106}.

\bibitem[{\citenamefont{Brack et~al.}(1972)\citenamefont{Brack, Damgaard,
  Jensen, Pauli, Strutinsky, and Wong}}]{Brack1972}
\bibinfo{author}{\bibfnamefont{M.}~\bibnamefont{Brack}},
  \bibinfo{author}{\bibfnamefont{J.}~\bibnamefont{Damgaard}},
  \bibinfo{author}{\bibfnamefont{A.~S.} \bibnamefont{Jensen}},
  \bibinfo{author}{\bibfnamefont{H.~C.} \bibnamefont{Pauli}},
  \bibinfo{author}{\bibfnamefont{V.~M.} \bibnamefont{Strutinsky}},
  \bibnamefont{and} \bibinfo{author}{\bibfnamefont{C.~Y.} \bibnamefont{Wong}},
  \bibinfo{journal}{Reviews of Modern Physics} \textbf{\bibinfo{volume}{44}},
  \bibinfo{pages}{320} (\bibinfo{year}{1972}), ISSN \bibinfo{issn}{0034-6861}.

\bibitem[{\citenamefont{Gallagher and Moszkowski}(1958)}]{Gallagher1958}
\bibinfo{author}{\bibfnamefont{C.}~\bibnamefont{Gallagher}} \bibnamefont{and}
  \bibinfo{author}{\bibfnamefont{S.~A.} \bibnamefont{Moszkowski}},
  \bibinfo{journal}{Phys. Rev.} \textbf{\bibinfo{volume}{111}},
  \bibinfo{pages}{1282} (\bibinfo{year}{1958}),
  \urlprefix\url{https://doi.org/10.1103/PhysRev.111.1282}.

\bibitem[{\citenamefont{Robledo et~al.}(2014)\citenamefont{Robledo, Bernard,
  and Bertsch}}]{Robledo2014}
\bibinfo{author}{\bibfnamefont{L.~M.} \bibnamefont{Robledo}},
  \bibinfo{author}{\bibfnamefont{R.~N.} \bibnamefont{Bernard}},
  \bibnamefont{and} \bibinfo{author}{\bibfnamefont{G.~F.}
  \bibnamefont{Bertsch}}, \bibinfo{journal}{Physical Review C}
  \textbf{\bibinfo{volume}{89}}, \bibinfo{pages}{021303}
  (\bibinfo{year}{2014}), ISSN \bibinfo{issn}{0556-2813, 1089-490X}.

\end{thebibliography}

\end{document}